\documentclass[a4paper]{article}
\pdfoutput=1
\usepackage[utf8]{inputenc}
\usepackage{geometry}
\usepackage{color}
\usepackage{mathtools}
\usepackage{amsmath}
\usepackage{amssymb}
\usepackage{setspace}
\usepackage{float}
\usepackage{jheppub}
\usepackage{subfig}
\usepackage{xcolor}
\usepackage{braket}
\usepackage{mathrsfs}
\usepackage{lineno}
\usepackage{lmodern}
\usepackage{tikz}
\usepackage[most]{tcolorbox}
\allowdisplaybreaks

\newcommand{\reduced}{Fused }
\newcommand{\reducedWeb}{Fused-Web }
\newcommand{\reducedWebs}{Fused-Webs }
\newcommand{\reduceds}{Fused-diagrams }

\def\beq{\begin{eqnarray}}
\def\eeq{\end{eqnarray}}

\def\bra#1{%
	\left\langle\smash{#1}{\vphantom1}\right|}
\def\ket#1{%
	\left|\smash{#1}{\vphantom1}\right\rangle}

\newcommand{\e}{\epsilon}

\newcommand{\as}{\alpha_s}
\newcommand{\warning}[1]{\textcolor{black}{#1}}

\title {Building blocks of Cwebs in multiparton scattering amplitudes}
\author[a]{Neelima Agarwal,}
\author[b]{Sourav Pal,}
\author[b]{Aditya Srivastav,}
\author[b]{Anurag Tripathi}
\affiliation[a]{Department of Physics, Chaitanya Bharathi Institute of Technology, \\ Gandipet, Hyderabad, Telangana State 500075, India}
\affiliation[b]{Department of Physics, Indian Institute of Technology Hyderabad, \\ Kandi, Sangareddy, Telangana State 502284, India}
\emailAdd{neelimaagarwal$\_$physics@cbit.ac.in}
\emailAdd{spalexam@gmail.com}
\emailAdd{shrivastavadi333@gmail.com}
\emailAdd{tripathi@phy.iith.ac.in}
\abstract{The correlators of Wilson-line operators in non-abelian gauge theories are known to exponentiate, and their logarithms can be organised in terms of the collections of Feynman diagrams called Cwebs. The colour factors that appear in the logarithm correspond to completely connected diagrams and are determined by the web mixing matrices. In this article we introduce several new concepts: (a) {Normal ordering} of the diagrams of a Cweb, (b) Fused-Webs (c) Basis and Family of Cwebs. We use these ideas together with a Uniqueness theorem that we prove to arrive at an understanding of the diagonal blocks, and several null matrices that appear in  the mixing matrices. 
We demonstrate using our formalism that, once the basis Cwebs present upto order $\alpha_{s}^{n}$ are determined, the number of exponentiated colour factors for several classes of Cwebs starting at order $\alpha_{s}^{n+1}$ can be predicted. We further provide complete results for the mixing matrices, to all orders in perturbation theory, for two special classes of Cwebs using our framework.}

\begin{document}
	\maketitle

\section{Introduction}

The infrared (IR) structure of scattering amplitudes in gauge  theories is an important object of study, and has a long history spanning almost a century~\cite{Bloch:1937pw,Sudakov:1954sw,Yennie:1961ad,Kinoshita:1962ur,Lee:1964is,Grammer:1973db,Mueller:1979ih,Collins:1980ih,Sen:1981sd,
	Sen:1982bt,Korchemsky:1987wg,Korchemsky:1988hd,Magnea:1990zb,
	Dixon:2008gr,Gardi:2009qi,Becher:2009qa,Feige:2014wja}. These structures are universal, that is they are independent of the hard scattering processes. The universality of these structures gives us remarkable all order insights of the perturbation theory. 
	A recent review on the subject can be found in \cite{Agarwal:2021ais}. 
	These studies have practical applications in the study of high energy scattering {experiments} at different colliders.
	The IR singularities that appear in the intermediate stages of calculations of the observables such as cross-sections cancel when the contributions of real emissions and virtual corrections are added. However, these singularities often leave their imprints in the form of large logarithms of kinematic invariants, which
	 can make a fixed order result lose predictive power in certain kinematical regions. It is the universality of the IR structure which enables a 
	  summation of  these large logarithms to all orders in perturbation theory and allows us to recover the predictive power in those kinematical regions~\cite{Sterman:1995fz,Laenen:2004pm,Luisoni:2015xha}. 
	 Furthermore, a knowledge of these structures is also very helpful in organizing fixed order calculations. The cancellation of these IR singularities for complicated observables in colliders is not  a trivial task, and using the universality of  the IR singularities several efficient subtraction procedures for this purpose have been developed  ~\cite{GehrmannDeRidder:2005cm,Somogyi:2005xz,Catani:2007vq,
	Czakon:2010td,Boughezal:2015dva,Sborlini:2016hat,Caola:2017dug,
	Herzog:2018ily,Magnea:2018hab,Magnea:2018ebr,Capatti:2020xjc, Magnea:2020trj, TorresBobadilla:2020ekr}.        

The factorization property of QCD in the IR limit enables us in  studying these singular parts efficiently, without calculating the  complicated hard parts. The soft function, that controls the IR singular parts in a scattering process, can be expressed in terms of matrix elements of Wilson line correlators  \cite{Erdogan:2014gha,Falcioni:2019nxk}.  These matrix elements also play an important role in QCD based effective theories ~\cite{Manohar:2000dt,
	Brambilla:2004jw,Becher:2014oda}. The Wilson-line operators $\Phi ( \gamma )$  evaluated on smooth space-time contours $\gamma$ are defined as,
\begin{align}
\Phi \left(  \gamma \right) \, \equiv \, \mathcal{P} \exp \left[ {\rm i} g \!
\int_\gamma d x \cdot {\bf A} (x) \right] \, .
\label{genWL}
\end{align}
where ${\bf A}^\mu (x) = A^\mu_a (x) \, {\bf T}^a$ is a non-abelian gauge field, 
and ${\bf T}^a$ is a generator of the gauge algebra, which can be taken to belong
to any desired representation, and $ \mathcal{P} $ denotes path ordering of the gauge fields.
If we restrict ourselves to multi-particle scattering amplitudes in gauge theories, then, 
we can write the soft function as, 
 \begin{align}
{\cal S}_n \Big( \beta_i \cdot \beta_j, \as (\mu^2), \e \Big) \, \equiv \, 
\bra{0} \prod_{k = 1}^n \Phi_{\beta_k} \left( \infty, 0 \right) \ket{0} , \quad
\Phi_\beta \left( \infty, 0 \right) \, \equiv \, \mathcal{P} \exp \left[ {\rm i} g \!
\int_0^\infty d \lambda \, \beta \cdot {\bf A} (\lambda \beta) \right] .
\label{softWLC}
\end{align}
{Here the Wilson lines are semi-infinite and point} along the direction of the hard particle, that is, the smooth contours run along $ \beta_k $, the velocities of the particles involved in this scattering have limit from origin to $ \infty $.

 $ \mathcal{S}_{n}$ suffers from both ultra-violet (UV) and IR (soft) singularities, and requires renormalization. In dimensional regularization $ \mathcal{S}_{n} $ vanishes as it involves only scaleless integrals and thus, after renormalization it is given by its  UV counterterms. 
The renormalized soft function obeys a renormalization group equation which leads to the following exponentiation:
\begin{align}
\mathcal{S}_n \Big( \beta_i \cdot \beta_j, \as (\mu^2), \e \Big) \, = \, 
\mathcal{P} \exp \left[ - \frac{1}{2} \int_{0}^{\mu^2} \frac{d \lambda^2}  
{\lambda^2} \, {\bf \Gamma}_n \Big( \beta_i \cdot \beta_j, \alpha_s (\lambda^2), 
\e \Big) \right]  \, ,
\label{softmatr}
\end{align}    
where $ {\bf \Gamma}_n $ is known as the soft anomalous dimension. In case of processes involving multi-parton scatterings, the soft anomalous dimension is a matrix, which is an important object of study on which we will focus in this article. The perturbative calculation of soft-anomalous dimension using renormalization group approach has a history of more than twenty years. ${\bf \Gamma}_n$ was computed at one loop in~\cite{Kidonakis-1998} 
(see also~\cite{Korchemskaya:1994qp}); at two loops in the massless case 
in~\cite{Aybat:2006wq,Aybat:2006mz}, and in the massive case in~\cite{Mitov:2009sv,
	Ferroglia:2009ep,Ferroglia:2009ii,Kidonakis:2009ev,Chien:2011wz}; finally, at three 
loops in the massless case in~\cite{Almelid:2015jia,Almelid:2017qju}. The calculation of soft anomalous dimension at four loops is an ongoing effort, and preliminary results are presented in  ~\cite{Becher:2019avh,
	Falcioni:2020lvv,Falcioni:2021buo,Vernazza:2021oqa,Catani:2019nqv,Moch:2017uml,Ahrens:2012qz,Moch:2018wjh,Chetyrkin:2017bjc,vonManteuffel:2020vjv,Henn:2019swt}.

The scale invariance of the soft function
	puts strong constraints on the form of $ {\bf\Gamma}_n $. It was shown in \cite{Gardi:2009qi,Becher:2009cu,Becher:2009qa,Gardi-Magnea,Magnea:2021fvy} that  ${\bf \Gamma}_n$ can only involve dipole correlations between the Wilson
	lines upto two loops; beyond 
	two loops, quadrupole correlations can arise, which must depend on
	scale-invariant conformal cross ratios of the form $\rho_{ijkl} \equiv (\beta_i 
	\cdot  \beta_j  \beta_k \cdot  \beta_l)/(\beta_i \cdot  \beta_k  \beta_j \cdot  \beta_l)$:
	the first such correlations arise at three loops, with at least four Wilson lines,
	and were computed in~\cite{Almelid:2015jia,Almelid:2017qju}; further correlations may 
	arise only in association with higher-order Casimir operators.

An alternative approach to determine the exponent of the soft function is through diagrammatic exponentiation. In terms of Feynman diagrams, the soft function has the form, 
\begin{align}
{\cal S}_n \left( \gamma_i \right) \, = \, \exp \Big[ {\cal W}_n \left( \gamma_i \right) 
\Big]  \, ,
\label{diaxp}
\end{align}
where $ {\cal W}_n \left( \gamma_i \right) $ are known as \textit{webs}, and can be directly computed using Feynman diagrams. Webs are defined as connected photon sub-diagrams in the abelian gauge theory, while in non-abelian gauge theory, for two Wilson line processes, webs are defined as two-line irreducible diagrams, that is, diagrams that remain connected upon cutting the Wilson lines~\cite{Sterman-1981,
	Gatheral,Frenkel-1984}.  In case of multi-parton
scattering process, the webs in non-abelian gauge theory are defined as sets of diagrams that differ from each other by the order of gluon attachments on each Wilson line~\cite{Mitov:2010rp,Gardi:2010rn}. The kinematics and the colour factors of a diagram in a web mix among themselves, through a web mixing matrix, which can be determined using a replica trick algorithm~\cite{Gardi:2010rn,Laenen:2008gt}. 

Cwebs --- a generalization of webs --- 
 which are a set of skeleton diagrams built out of connected gluon correlators attached to Wilson lines, and are closed under shuffles of the gluon attachments to each Wilson line were introduced in \cite{Agarwal:2020nyc,Agarwal:2021him}. 
The mixing between diagrams of a web does not get altered upon replacing the diagrams by the corresponding skeleton diagrams. That is, the same mixing matrix describes the mixing of diagrams of a web and its corresponding Cweb.
The Cweb mixing matrices are central objects in the study of non-abelian exponentiation. An alternative approach of generating functionals was developed in \cite{Vladimirov:2015fea,Vladimirov:2014wga,Vladimirov:2017ksc}. 
There have been attempts to determine the web mixing matrices bypassing the replica trick algorithm. Combinatoric ideas such as  partial order sets
have been employed in constructing matrices for {certain classes} of Cwebs  \cite{Dukes:2013gea, Dukes:2013wa,Dukes:2016ger}. All prime dimensional mixing matrices were also constructed directly without using replica trick algorithm in \cite{Agarwal:2021him}.

In this article we have introduced several new ideas such as: (a) {Normal {o}rdering} of the diagrams of a Cweb (b) \reducedWebs and (c) Basis and Family of Cwebs which prove extremely useful in making the structures present in the mixing matrices very transparent.
We prove a  {\it Uniqueness theorem} which together with the above ideas helps to determine the diagonal blocks of the mixing matrices of a Cweb. 
These ideas provide us with an ability to predict the rank of the mixing matrices or equivalently the number of independent exponentiated colour factors for several Cwebs.  

This paper is structured as follows. In section \ref{sec:conje}, we review the known properties of the mixing matrices, and provide a Uniqueness theorem for Cweb mixing matrices. In section \ref{sec:\reduced}, we define an ordering among the diagrams of a Cweb, and describe the construction of \reduced diagrams and Fused-Webs. These entities shed light on the different blocks of the mixing matrices.  Further, we calculate the explicit forms of mixing matrices for two classes of Cwebs without using the replica trick in section \ref{sec:explicit}. In section \ref{sec:Direct-cons}, we use \reducedWebs to calculate the diagonal blocks for three classes of Cwebs. 
 Finally, we conclude our findings in section \ref{sec:conclu}. Appendices \ref{sec:repl}, \ref{sec:table-of-av-webs} and \ref{sec:basis} describe the replica trick, the application of \reducedWebs to provide the rank of mixing matrices, and the mixing matrices for the basis Cwebs present up to four loops, respectively.

\section{Cweb mixing matrices: Properties and a Uniqueness theorem}
\label{sec:conje}
We begin with the definition   \cite{Agarwal:2020nyc,Agarwal:2021him} 
of a  {\it Correlator Web }({Cweb}). It is  a set of skeleton diagrams, built out of 
connected gluon correlators attached to Wilson lines, and closed under shuffle
of the gluon attachments to each Wilson line. 
Cwebs are not fixed-order quantities, but 
admit their own perturbative expansion in powers of the gauge coupling $g$. 
Below, we will use the notation $W_n^{(c_2, \ldots , c_p)} (k_1, \ldots  , k_n)$ 
for a Cweb constructed out of $c_m$ $m$-point connected gluon correlators 
($m = 2, \ldots, p$).
For the sake of classification we choose the ordering $k_1 \leq k_2 
\leq \ldots \leq k_n$, where $k_{i}$ denotes attachments on different Wilson lines.
Taking into account the fact that the 
perturbative expansion for an $m$-point connected gluon correlator starts 
at ${\cal O} (g^{m - 2})$, while each attachment to a Wilson line carries a 
further power of $g$, the perturbative expansion for a Cweb can be written as
\beq
W_n^{(c_2, \ldots , c_p)} (k_1, \ldots  , k_n)  \, = \, 
g^{\, \sum_{i = 1}^n k_i \, + \,  \sum_{r = 2}^p c_r (r - 2)} \, \sum_{j = 0}^\infty \,
W_{n, \, j}^{(c_2, \ldots , c_p)} (k_1, \ldots  , k_n) \, g^{2 j} \, ,
\label{pertCweb}
\eeq
which defines the perturbative coefficients $W_{n, \, j}^{(c_2, \ldots , c_p)} 
(k_1, \ldots  , k_n)$. 
The perturbative order of Cwebs is defined as the order at which they receive their lowest order contributions, which is given by the prefactor $ g^{\, \sum_{i = 1}^n k_i \, + \,  \sum_{r = 2}^p c_r (r - 2)} $. The remaining powers of $ g $ in eq.~\eqref{pertCweb} arise from the attachments within the \textit{blobs} and they do not enter into the counting of order of Cwebs.

Cwebs are the proper building blocks of the logarithm of Soft function; and are also useful in the organisation and counting of diagrammatic contributions at higher perturbative orders. 
The logarithm of the Soft function is a sum over all the Cwebs at each perturbative order:
\begin{align}
{\cal S} \, = \, \exp \left[ \sum_w
\sum_{d,d' \in  w} {\cal K} (d) \, R_w (d, d') \, C (d')
\right] \, .
\label{Snwebs}
\end{align}
The $d$ here denotes a diagram in a  Cweb $w$ and its corresponding kinematic and colour factor are denoted by $ {\cal K} (d) $ and $C(d)$. The action of web mixing matrix $R_w$ on the colour of a  diagram $ d $ generates its exponentiated colour factor $\widetilde{C}$,  
\begin{align}
\widetilde{C} (d)  \, = \, \sum_{d'\in w} R_w (d, d') \, C(d') \, .
\label{eq:ecf}
\end{align}

In addition to the kinematic and colour factors, we also define a  weight factor $ s(d) $ for a given diagram $ d $. Given a diagram $d$,
consider the set of its gluon correlators (after  
removing the Wilson lines), $\{ d_{\rm c}^i \subset d  \}$; we say that a 
gluon correlator $d_{\rm c}^i$ can be shrunk to the common origin 
of the Wilson lines if all the vertices connecting the gluon correlator to the Wilson 
lines can be moved to the origin without encountering vertices associated
with other gluon correlators. Then, the weight factor $s(d)$ for a diagram $ d $ is defined as the number of different ways in which the gluon correlators 
$d_{\rm c}^i$ of $d$ can be {\it sequentially} shrunk to their common origin. 
We can construct a column weight vector out of these $ s $-factors for a Cweb with $ n $ diagrams as 
\begin{align} 
S=\{s(d_1),s(d_2), \ldots,s(d_n)\}.
\end{align}
Now, we will classify the diagrams based on their $s$ values.
\subsection*{Classification of diagrams}
\noindent {\it Reducible diagram}: $d$ is a reducible diagram if $s(d) \neq 0$, \vspace{0.1cm}\\  
\noindent {\it Irreducible diagram}: $d$ is an irreducible diagram if $s(d) = 0$.
\\ \\
\noindent We further classify irreducible diagrams into the following two categories: \\  \\ 
\noindent {\it Completely entangled diagram}: An irreducible diagram in which all the gluon correlators are entangled and thus none of the gluon correlators can be independently shrunk to the origin.  \vspace{0.1cm}\\ 
\noindent {\it Partially entangled diagram}: An irreducible diagram which has at least one gluon correlator which is not entangled with the other correlators. 

\begin{figure}[t]
	\centering
	\subfloat[][]{\includegraphics[height=4cm,width=4cm]{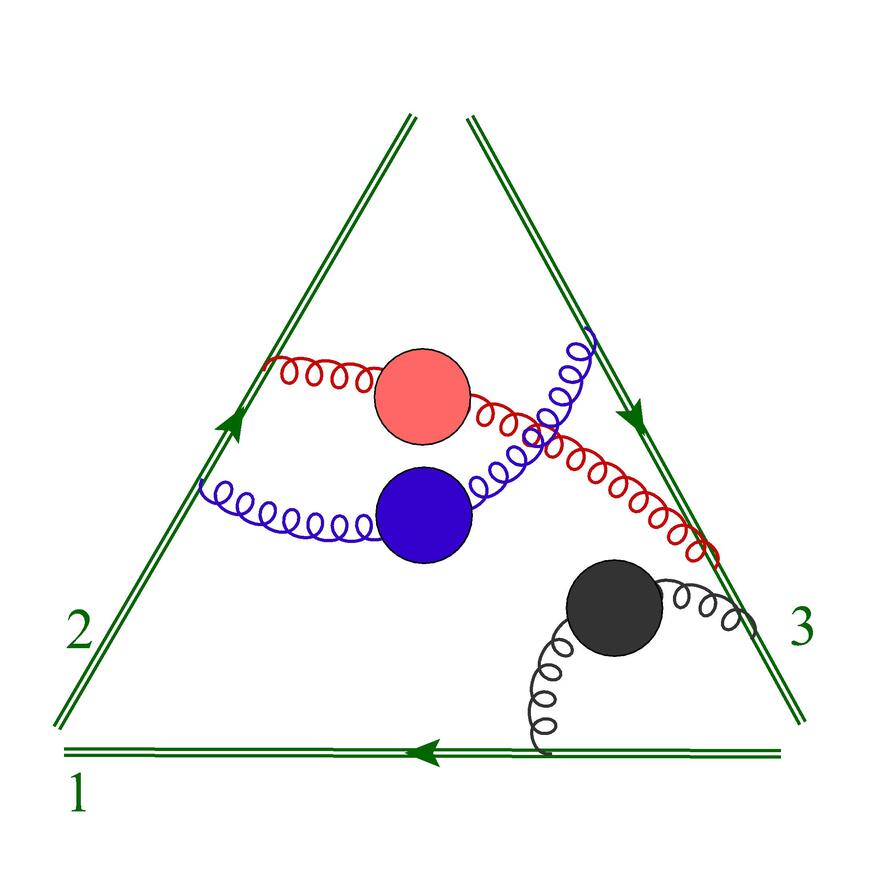} }
	\qquad 
	\subfloat[][]{\includegraphics[height=4cm,width=4cm]{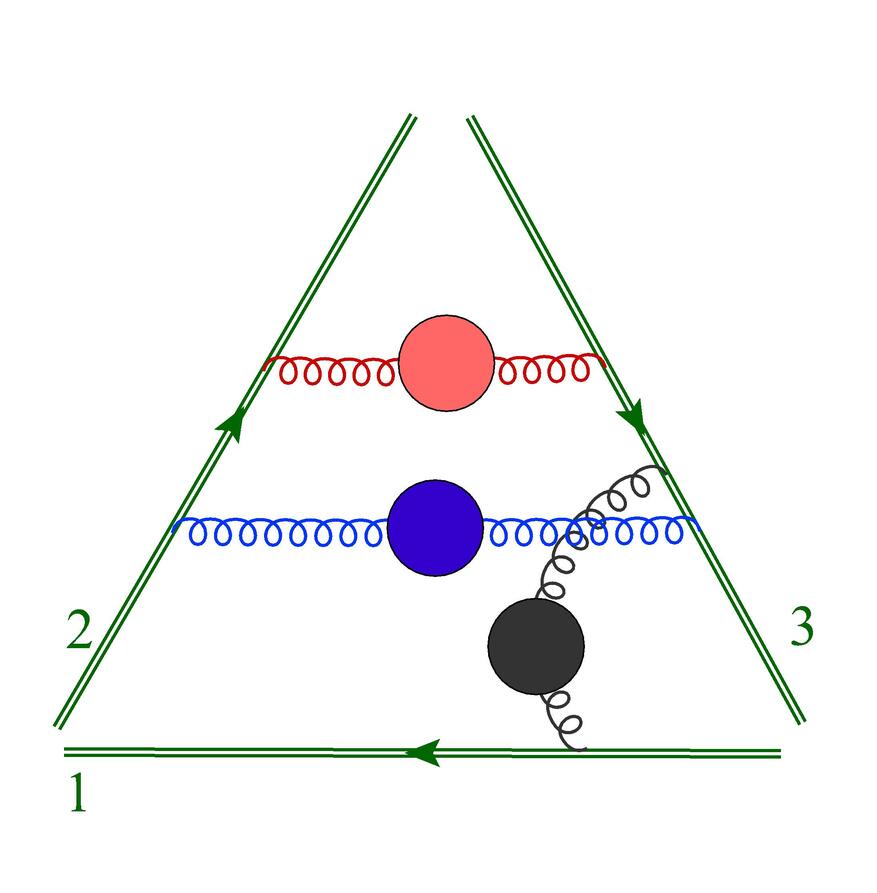} }	
	\qquad 
	\subfloat[][]{\includegraphics[height=4cm,width=4cm]{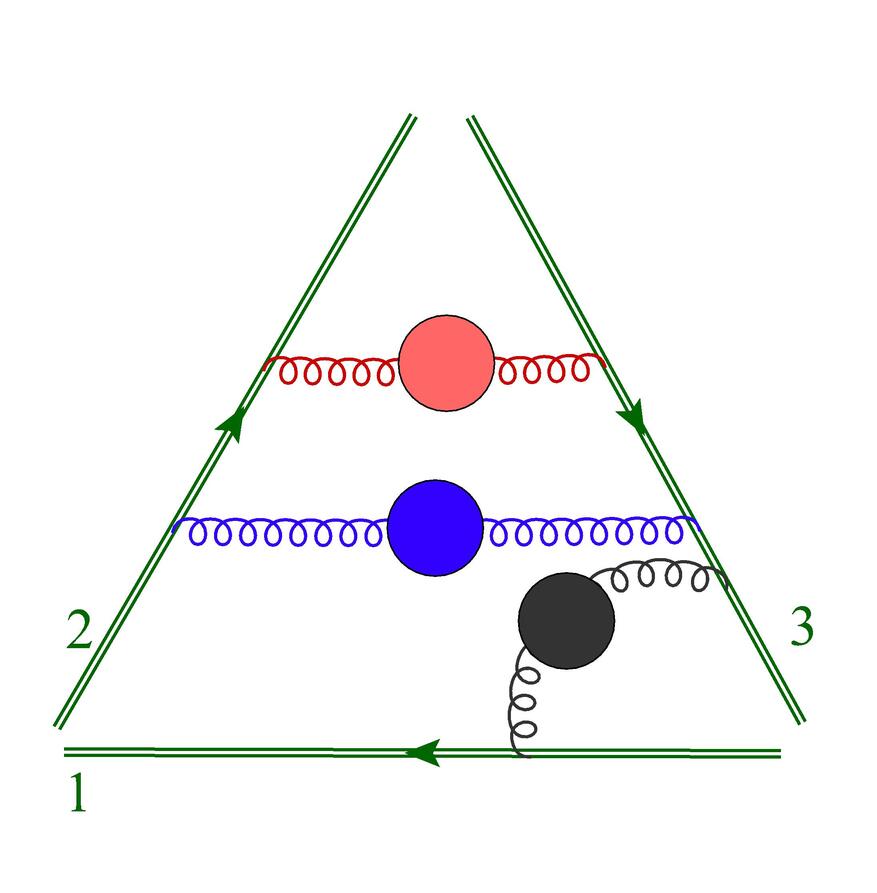} }	
	
	\caption{Diagram (a) is a reducible diagram; diagrams (b) and (c) are completely, and partially entangled diagrams respectively.}
	\label{fig:irreducible}
\end{figure}

In this article we draw Wilson lines with an arrow whose tails are at the origin.
The diagram in  fig.~(\ref{fig:irreducible}\textcolor{blue}{a})  is a reducible diagram. This diagram has three correlators, all of which can be sequentially shrunk to the origin in only one possible way and thus has $ s=1 $. The diagrams in  fig.~(\ref{fig:irreducible}\textcolor{blue}{b}), and (\ref{fig:irreducible}\textcolor{blue}{c}) are examples of  irreducible diagrams; for these diagrams there is no possible way to sequentially shrink all the correlators to the origin independently, that is, they both have $ s=0 $. The diagram in fig.  (\ref{fig:irreducible}\textcolor{blue}{b}) is a completely entangled diagram because, if any of the three correlators is dragged to the origin, all of them get dragged simultaneously.
In diagram (\ref{fig:irreducible}\textcolor{blue}{c}), the black correlator can be independently shrunk to the origin after shrinking the blue and red correlators simultaneously. Thus, this diagram is a partially entangled diagram. 

\noindent A Cweb with multiple diagrams having the same $ s $-factors has the column weight vector of the form, $S = \{s_1, \ldots, s_1, \ldots, s_i, \ldots, s_i, \ldots, s_l, \ldots s_l\},$
where, $ s_{1} < s_{2} < \cdots < s_{l}$. Here we have ordered the diagrams according to their $ s $-factors. 
$ S $ can be written in a compact form
\begin{align}\label{eq:S-definition}
S& = \left\{ (s_{1})_{{k_{1}}}, (s_{2})_{{k_{2}}}  , \ldots,  (s_{l})_{{k_{l}}}  \right\}\,,
\end{align}
if first $ k_1 $ diagrams in the Cweb have weight $ s_1 $, followed by the  $k_{2}$ diagrams with weight $s_{2}$ and so forth. 
We denote the corresponding mixing matrix for the Cweb as,
\begin{align}
R\Big(  (s_{1})_{{k_{1}}}, (s_{2})_{{k_{2}}}  , \ldots,  (s_{l})_{{k_{l}}}  \Big).
\end{align}

\noindent The web mixing matrices are essential quantities for the determination of Wilson line correlators, and thus, the soft anomalous dimension matrix, and we now turn our focus on them.
\subsection{Properties of mixing matrices}
General all order properties of the mixing matrices were first observed in \cite{Gardi:2010rn}, and were proven in \cite{Gardi:2011wa}. Further, a  conjecture regarding the columns of the mixing matrices was proposed in  \cite{Gardi:2011yz}. Below we list down these properties.

\begin{enumerate}
	\item {\it Idempotence:} These matrices are idempotent and act as projection operators:
	\begin{align}
	R^2 \, = \,  R \, .
	\label{eq:idempo} 
	\end{align}
	Thus their eigenvalues can only be either 0 or 1, which further implies that their trace is  equal to their rank.
	\item {\it Non-abelian exponentiation:} The general non-abelian exponentiation theorem~\cite{Gardi:2013ita} states that the colour factors that survive the above projection by $ R$ are the ones that are associated with a diagram that has only one gluon correlator. 
	\item {\it Row sum rule:} The elements of web mixing matrices obey the row sum rule 
	\begin{align}
	\sum_{d'} R (d, d') \, = \, 0 \,.
	\label{eq:rowsum}
	\end{align}
	\item {\it Column sum rule:}  
	The mixing matrices obey the following column sum conjecture:
	\begin{align}
	\sum_d s(d) R(d, d')\,=0\,.
	\label{eq:column-sum} 
	\end{align}
\end{enumerate}
The idempotence of mixing matrices implies that $R$ projects onto only those combinations of kinematic factors that do not contain ultraviolet sub-divergences\footnote{We refer here to UV divergences arising from sub-diagrams involving the Wilson lines: interactions away from the Wilson lines will still involve the usual gauge-theory UV divergences, which are dealt with by means of ordinary
	renormalization techniques.}. For the case of two Wilson lines, the absence of sub-divergences was proved in \cite{Gatheral,Frenkel-1984,Sterman-1981}. However, for more than two Wilson lines, the all order proof for column sum is not available, although this property has been verified upto four loops~\cite{Gardi:2011yz,Gardi:2013ita,Agarwal:2020nyc,Agarwal:2021him}.
The connection of the column-sum rule to UV sub-divergences is explained in a coordinate-space
picture in ~\cite{Erdogan:2014gha}. In coordinate-space, UV divergences 
arise from short distances between interaction vertices, and, thus the `shrinkable' 
correlators are naturally associated to UV sub-divergences,  eq.~(\ref{eq:column-sum}) 
guarantees that these correlators are projected out of the webs.

\subsection{Uniqueness Theorem}\label{sec:Uniqueness-theorem}

A careful survey of the elements of the large number of web mixing matrices available in the literature --- at two \cite{Gardi:2010rn}, three \cite{Gardi:2013ita}, and four loops \cite{Agarwal:2020nyc,Agarwal:2021him} --- gives an impression of
repeating structures. One is tempted to find some organising principle that 
could remove the veil from these structures and possibly find the building blocks of these matrices. Towards this goal we prove the 
following Uniqueness theorem:
\begin{itemize}
\item[] {\it Uniqueness}:  For a given column weight vector $S=\{s(d_1),s(d_2), \ldots,s(d_n)\}$
 with all $s(d_{i}) \neq 0$, the mixing matrix is unique.
\end{itemize}
An important consequence of uniqueness is that, if mixing matrix of a Cweb  at some order, that has a column weight vector $S$ with only non-zero entries, is known, then we can without any further work write down the mixing matrix of another Cweb that appears at the same or higher perturbative order if it has the same weight vector $S$. 

\subsection*{The proof} 
\begin{figure}[H]
	\vspace{0.5cm}
	\captionsetup[subfloat]{labelformat=empty}
	\centering	
	\subfloat[][]{\includegraphics[scale=0.23]{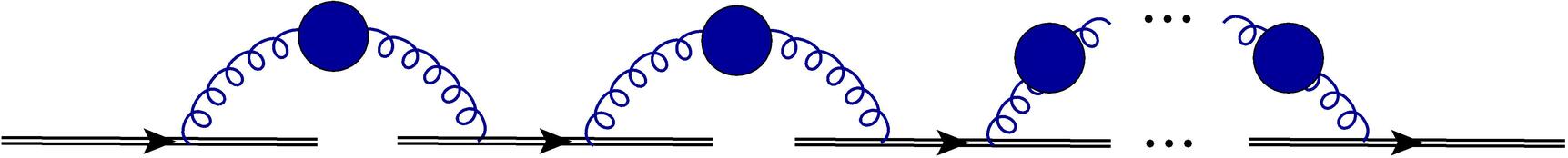} }
	\caption{A Cweb constructed out of $n$ two-point correlators connecting $n+1$ Wilson lines. This is one of the basis Cwebs at order $\alpha_{s}^{n}$.}
	\label{fig:MaximalCon}
\end{figure}

\begin{figure}[b]
	\centering
	\subfloat[][$ \text{W}_3^{(2)}(1,1,2) $]{\includegraphics[height=4.0cm,width=4.0cm]{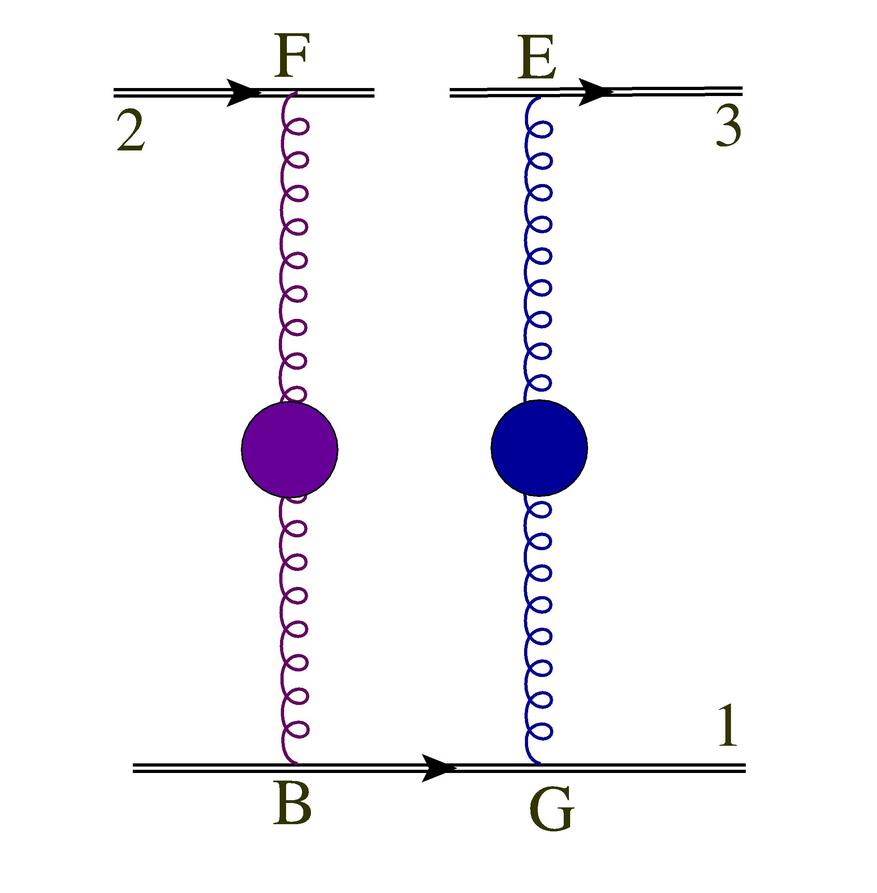} }
	\qquad 	
	\subfloat[][$ \text{W}_4^{(3)}(1,1,1,3) $]{\includegraphics[height=4.0cm,width=4.0cm]{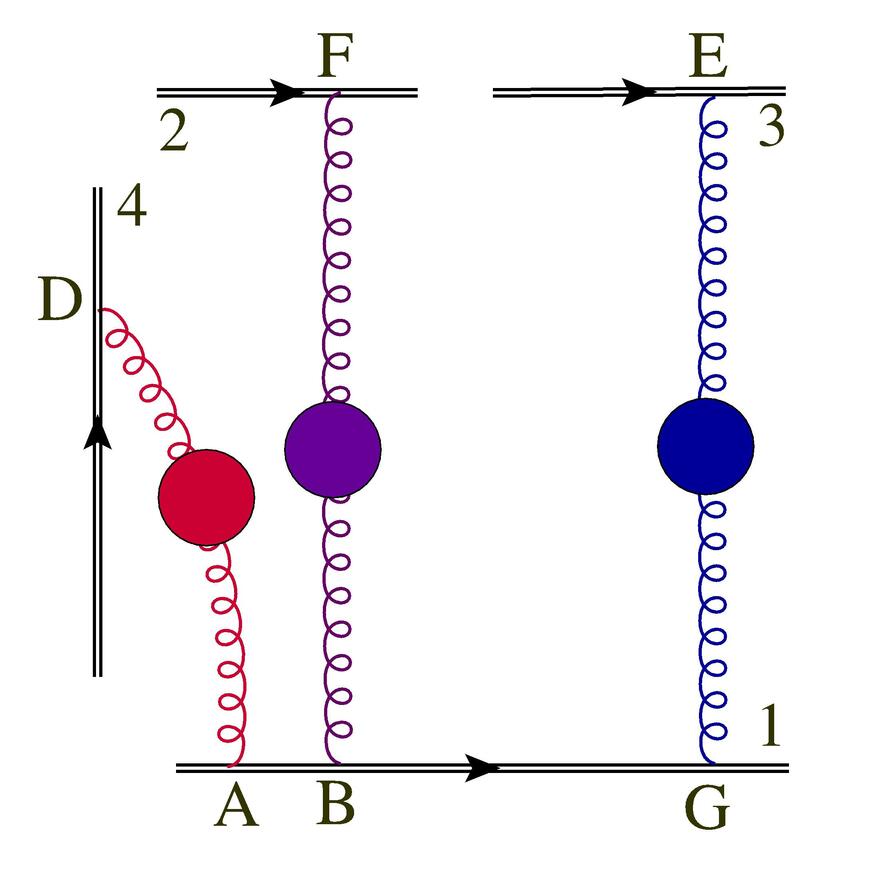} }
	\qquad 	
	\subfloat[][$ \text{W}_4^{(3)}(1,1,2,2) $]{\includegraphics[height=4.0cm,width=4.0cm]{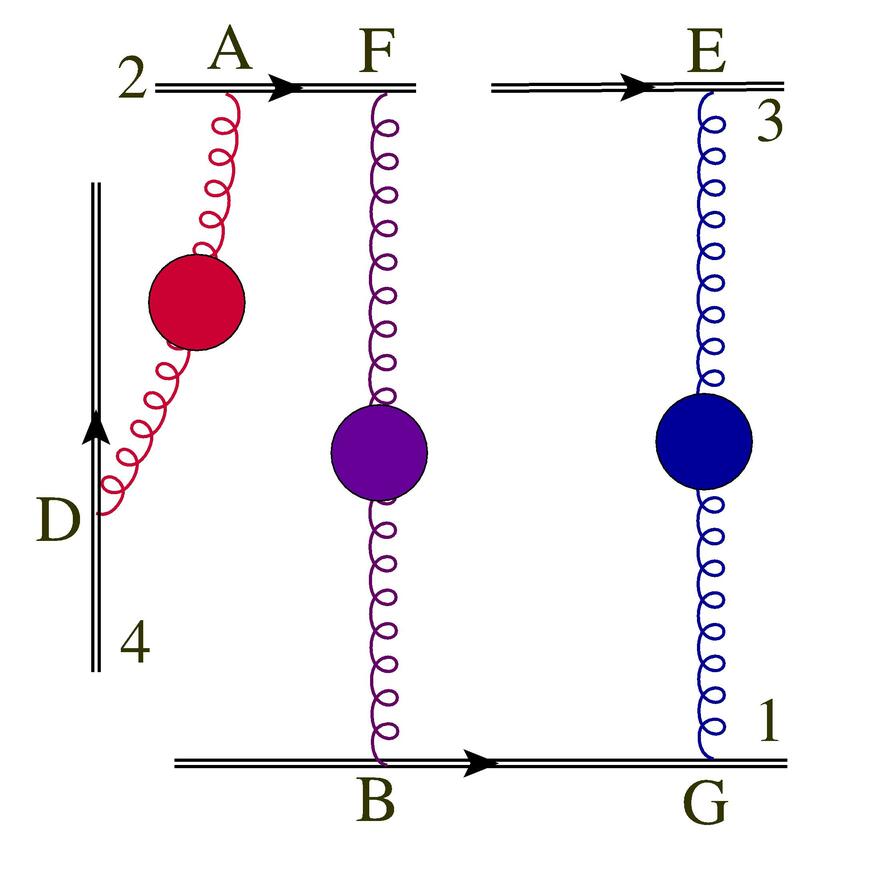} }
	\caption{Basis Cwebs at two and three loop order}
	\label{fig:basisWebs}
\end{figure}

In the diagram, shown in fig.~(\ref{fig:MaximalCon})  $n$ two-point gluon correlators connect $n+1$ Wilson lines. This diagram is reducible. Any rearrangement of the attachments on and across the Wilson lines, keeping the number of Wilson lines fixed, gives another diagram which is still reducible. 
\noindent{Any} rearrangement that lead{s} to an irreducible diagram cannot connect all the $n+1$ Wilson lines. Based on this observation we introduce,
\\ \\
\noindent {\it Basis Cwebs}:  Cwebs formed by connecting $n$ two-point gluon correlators to $n+1$ Wilson lines.
\\
\\
The basis Cwebs that are present at two  and three loops\footnote{These Cwebs were studied in detail in \cite{Falcioni:2014pka}, using the results of \cite{Gardi:2013saa}.} are shown in the fig.~(\ref{fig:basisWebs}).  
Now we start at one loop -- the Cusp, where one two-point gluon correlator connects two Wilson lines and apply the recursive algorithm presented in \cite{Agarwal:2020nyc}
to generate higher order Cwebs. Keeping in mind the possibility of adding a Wilson line with no attachments at lower orders, the steps of the algorithm are:
\begin{enumerate}
	\item Connect any two Wilson lines by introducing a two-point gluon correlator.
	\item Connect any existing $m$-point gluon correlator to a Wilson line, 
	turning it into an  $(m+1)$-point gluon correlator.
	\item Connect an existing $m$-point gluon correlator to an existing 
	$n$-point gluon correlator by a single gluon, resulting in an $(n+m)$-point 
	correlator.
	\item Discard Cwebs that are given by the product of two or more
	disconnected lower-order webs.
	\item In a massless theory, discard all self-energy Cwebs, where all 
	gluon lines attach to the same Wilson line, as they vanish as a 
	consequence of the eikonal Feynman rules. 
	\item Discard Cwebs that have been generated by the procedure more 
	than once.
\end{enumerate}

\begin{figure}[b]
	\centering
	\subfloat[][$ \text{W}_3^{(2)}(1,1,2) $]{\includegraphics[height=4cm,width=4cm]{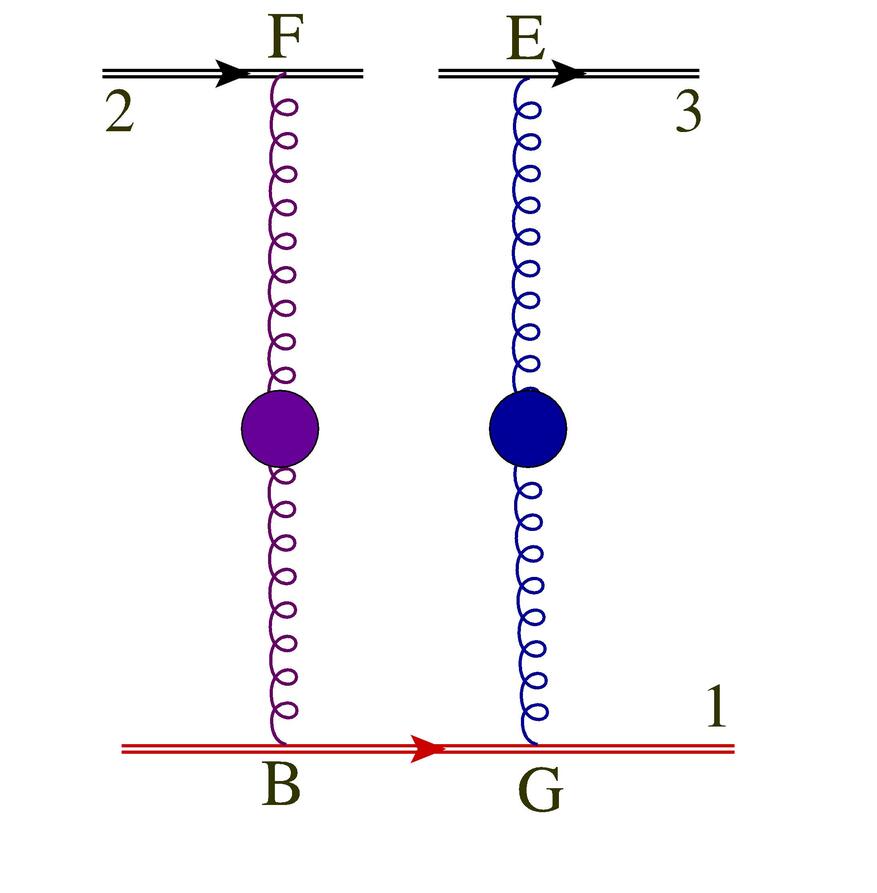} }
	\quad
	\subfloat[][$ \text{W}_5^{(0,2)}(1,1,1,1,2) $]{\includegraphics[height=4cm,width=4cm]{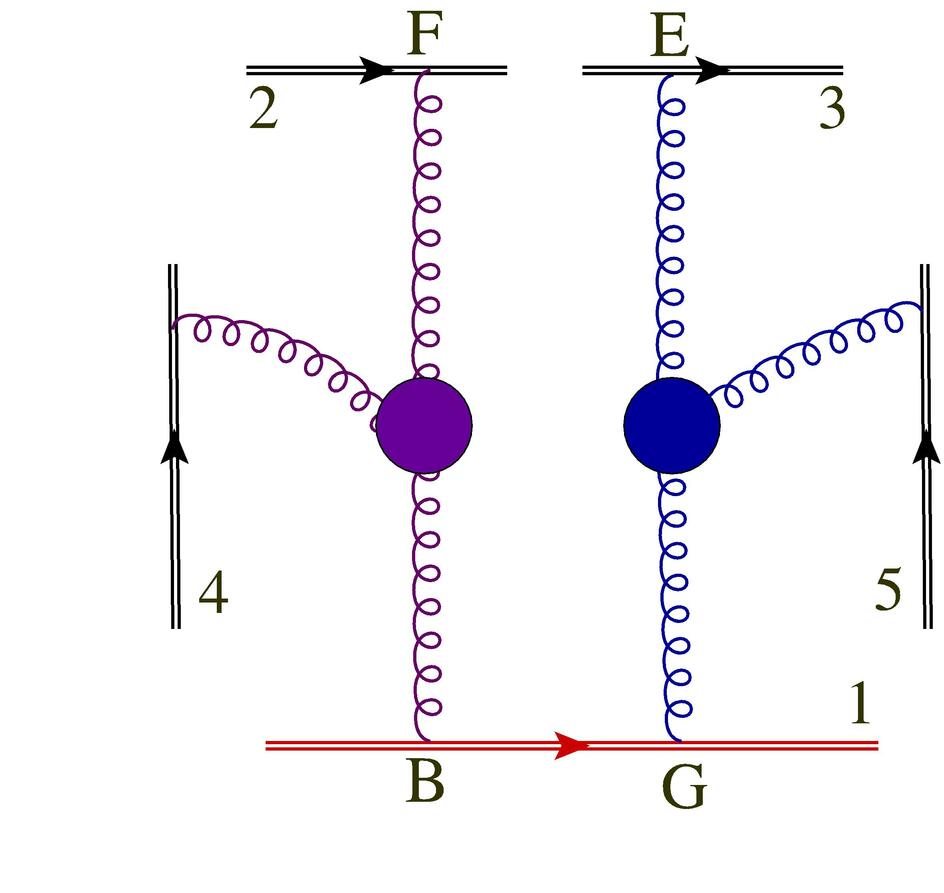} }
	\qquad 
	\subfloat[][$ \text{W}_5^{(0,0,2)}(1,1,2,2,2) $]{\includegraphics[height=4cm,width=4cm]{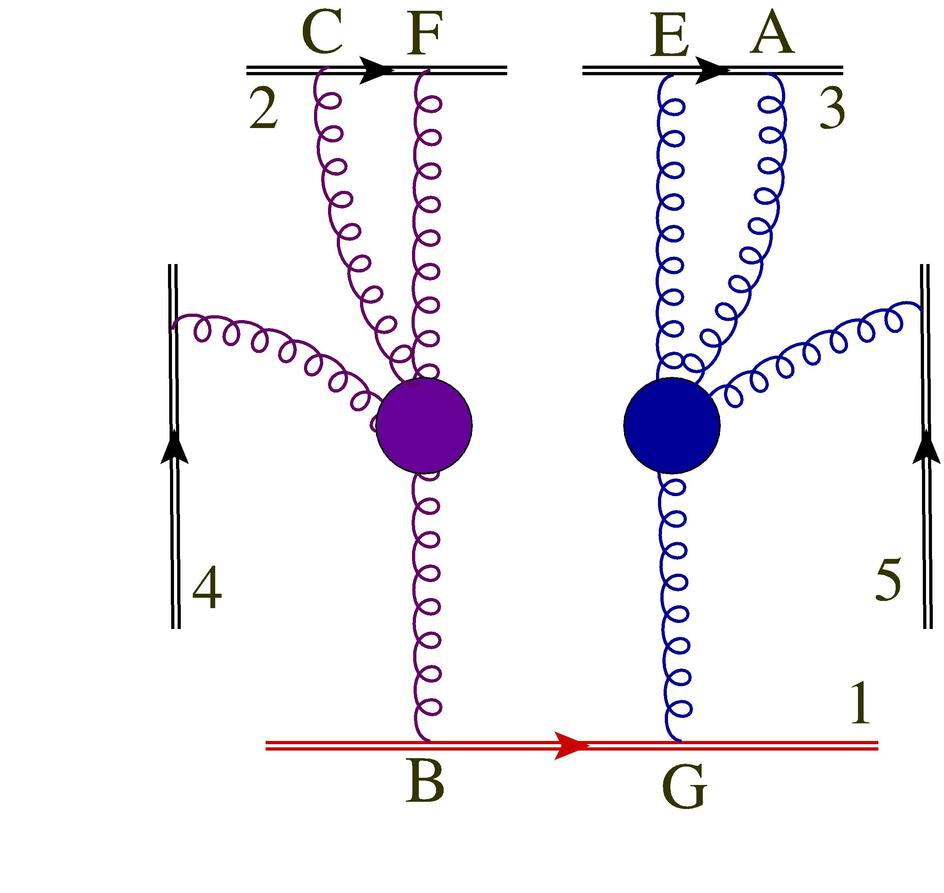} }
	\caption{Three Cwebs of family $ f(1_2) $}
	\label{fig:family-Ex}
\end{figure}

Cwebs generated from the above algorithm that admit same possible shuffles of the attachments on each of the Wilson lines present in the parent diagram, and no possible shuffles on any additional Wilson line that has been introduced, form a {\it family}, $f$. Further, we call a Cweb that belongs to a  family a  {\it member} of that family.  The weight vector $ S $ is the same for all members of the family as shuffles on the Wilson lines remain the same. Thus a family\footnote{A family is not uniquely specified by $ f(S) $ for the cases where $ S $ contains zero entries.} will be denoted by $f(S)$.
 Furthermore,
since the mixing matrix of a Cweb is completely determined by the shuffles on the Wilson lines, there is a unique mixing matrix for every family.

\begin{figure}
	\centering
	\subfloat[][$ \text{W}_4^{(3)}(1,1,1,3) $]{\includegraphics[height=4cm,width=4cm]{Converse-Conjecture-2b} }
	\qquad
	\subfloat[][$ \text{W}_4^{(1,2)}(1,2,2,3) $]{\includegraphics[height=4cm,width=4cm]{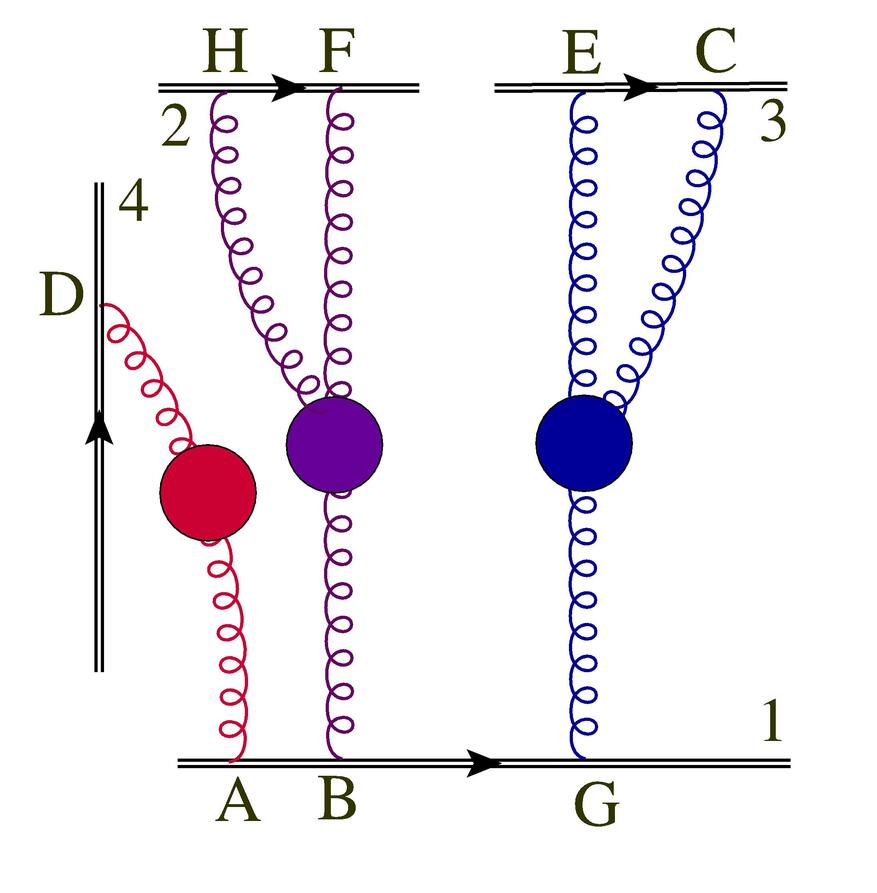} }
	\qquad 	
	\subfloat[][$ \text{W}_4^{(2,1)}(1,1,2,3) $]{\includegraphics[height=4cm,width=4cm]{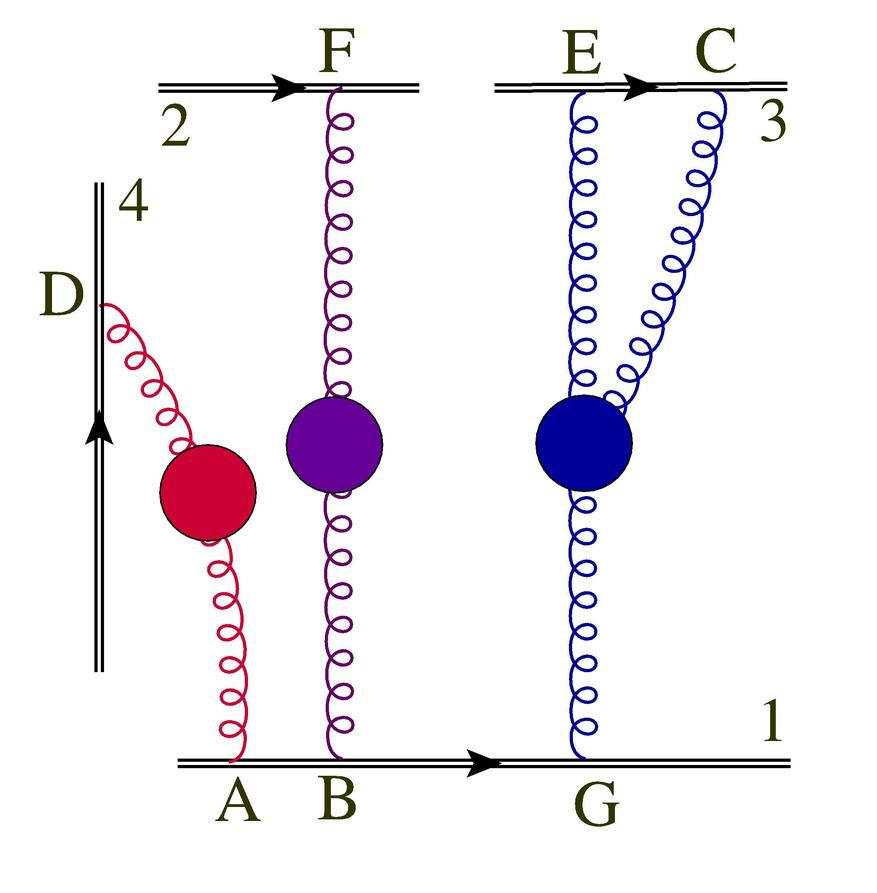} }
	\caption{Two Cwebs (\textcolor{blue}{b}) and (\textcolor{blue}{c}) of family $ f(1_6) $  are generated from basis Cweb (\textcolor{blue}{a}) 
	}
	\label{fig:Converse-Conjecture-1}
\end{figure}

For example, the three Cwebs shown in fig.~(\ref{fig:family-Ex})  belong to the same family $ f(1_2) $.  
The shuffle of the attachments on the Wilson line $ 1 $ --- \{BG\} and \{GB\} --- for each of these {Cwebs}, generates two diagrams $d_{1}$ and $d_{2}$ {of} the respective Cweb. Both $d_{1}$ and $d_{2}$ have $s=1$.
Similarly all the Cwebs shown in fig.~(\ref{fig:Converse-Conjecture-1})  have same shuffles possible on each of the Wilson lines and thus belong to the same family $ f(1_6) $.

Starting from a Cweb with irreducible ($ s=0 $) diagram one can not generate a Cweb with only reducible ($ s\neq0 $) diagrams. Since uniqueness theorem concerns itself with Cwebs that have only reducible diagrams, below we will consider only such Cwebs as \textit{parent} Cwebs. 

\begin{figure}[h]
	\centering
	\subfloat[][$ \text{W}_3^{(1,1)}(1,2,2) $]{\includegraphics[height=4cm,width=4cm]{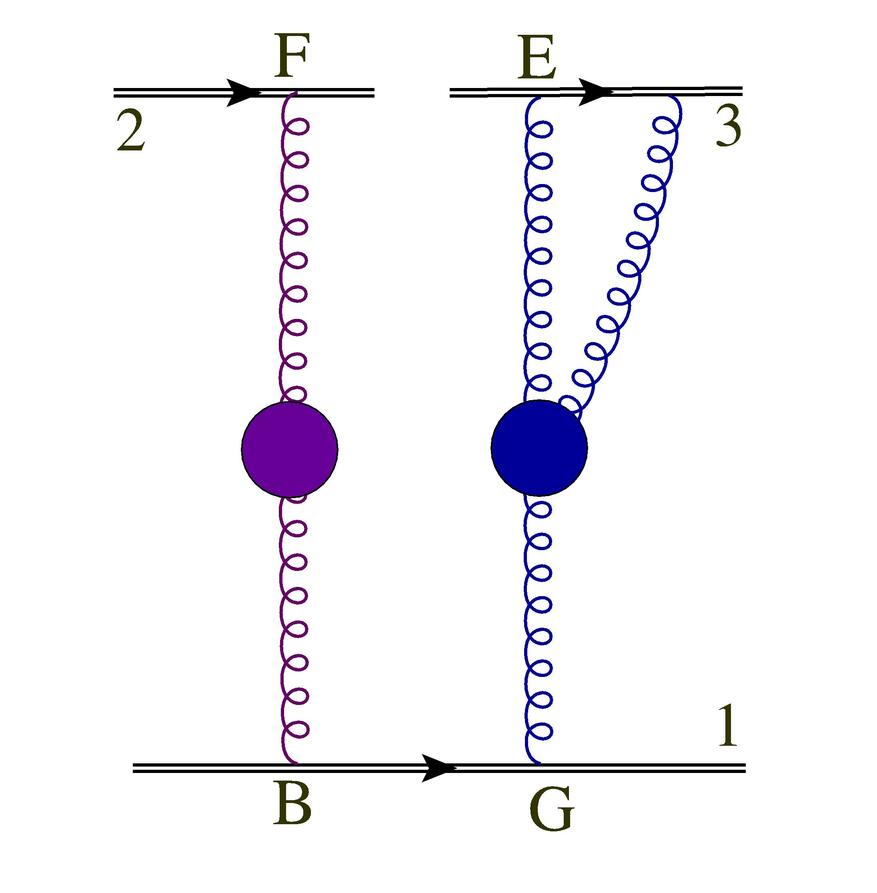} }
	\qquad 	
	\subfloat[][$ \text{W}_4^{(2,1)}(1,1,2,3) $]{\includegraphics[height=4cm,width=4cm]{Converse-Conjecture-1b} }
	\qquad 	
	\subfloat[][$ \text{W}_3^{(2,1)}(1,3,3) $]{\includegraphics[height=4cm,width=4cm]{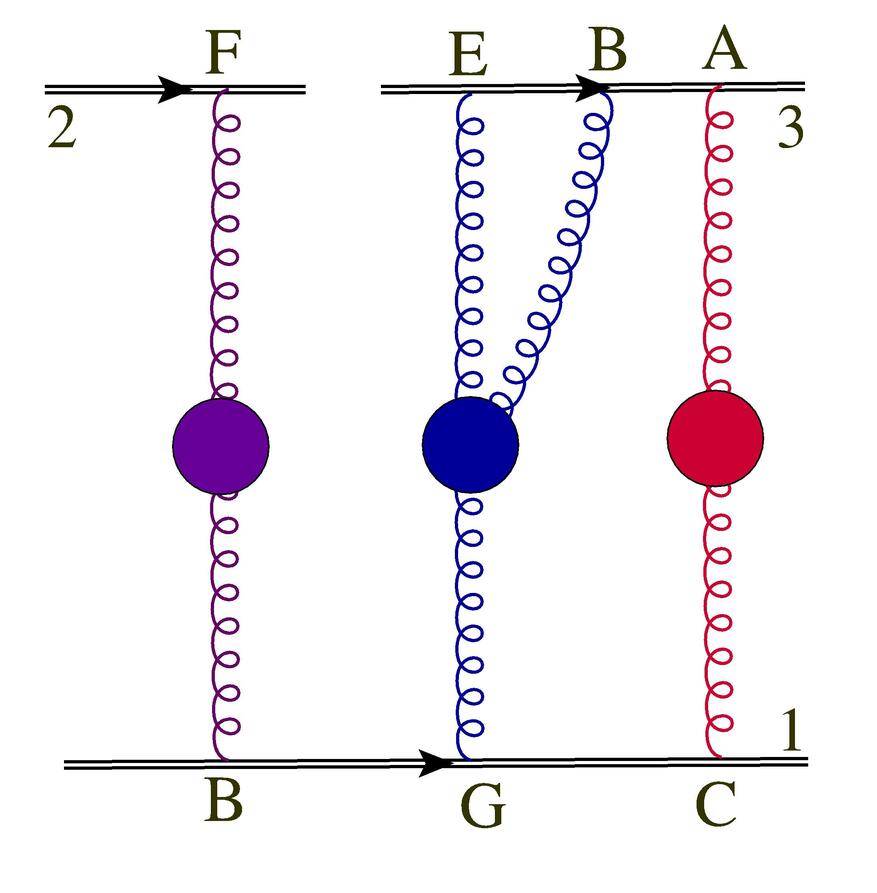} }
	\caption{Two Cwebs (\textcolor{blue}{b}) and (\textcolor{blue}{c}) generated from Cweb (\textcolor{blue}{a}) using first step of the algorithm 
	}
	\label{fig:first-category-conjec}
\end{figure}

The first step of the algorithm on a {parent} Cweb at some order in $\alpha_{s}$ with only reducible diagrams can generate a Cweb that has some irreducible diagrams, 
or it can generate a member of a family that has appeared at a lower order in $\alpha_s$, or it can produce a Cweb which belongs to a new family.
For example, the three-loop Cweb shown in fig.~(\ref{fig:first-category-conjec}\textcolor{blue}{a}), which has $ S = \{1_2\} $, generates Cwebs shown in fig.~(\ref{fig:first-category-conjec}\textcolor{blue}{b}), and (\ref{fig:first-category-conjec}\textcolor{blue}{c}) upon including the red two-point gluon correlator. The order of shuffle in the  Cwebs (\ref{fig:first-category-conjec}\textcolor{blue}{b}),  and (\ref{fig:first-category-conjec}\textcolor{blue}{c}) are different from that of its parent Cweb, and so are their column weight vectors $ S $. However, we are interested in the Cwebs for which the column weight vectors do not contain any zero entries. We find that only Cweb $ \text{W}^{(2,1)}_{4} (1,1,2,3) $, shown in fig.~(\ref{fig:first-category-conjec}\textcolor{blue}{b}), satisfies this criterion,  and it belongs to the family $ f(1_6) $ that appears
at three loops, and thus has the mixing matrix $ R(1_6) $. 

\begin{figure}[H]
	\vspace*{-0.9cm}
	\centering
	\subfloat[][$ \text{W}_4^{(3)}(1,1,1,3) $]{\includegraphics[height=3.5cm,width=3.5cm]{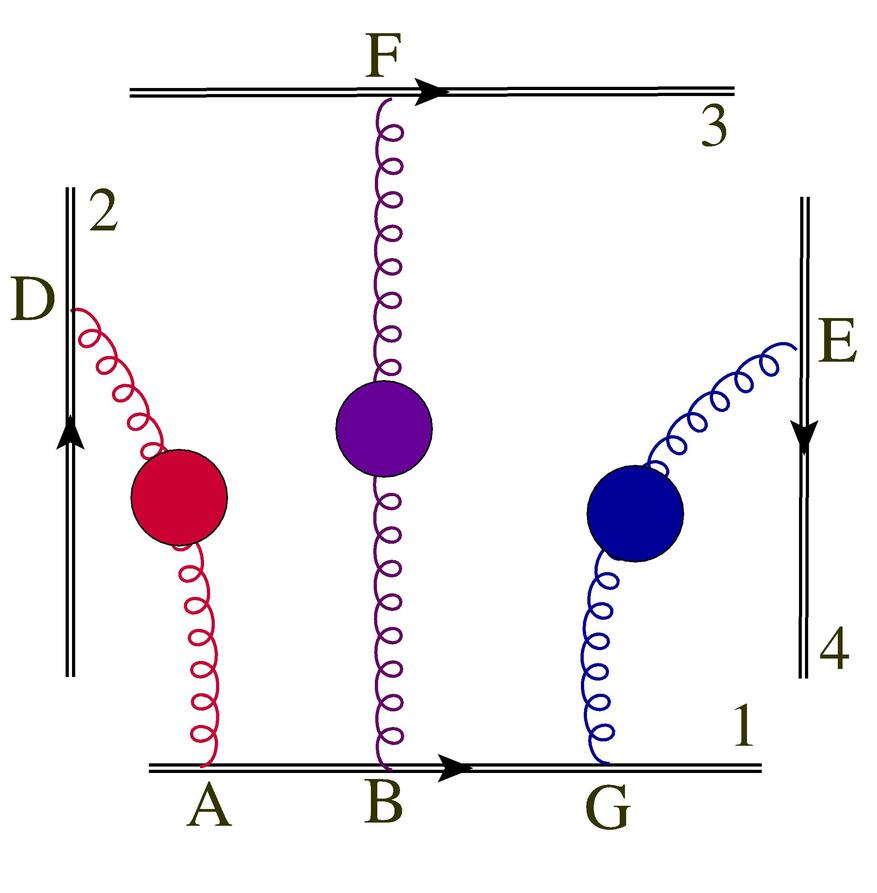} }
	\qquad
	\subfloat[][$ \text{W}_4^{(2,1)}(1,1,1,4) $]{\includegraphics[height=3.5cm,width=3.5cm]{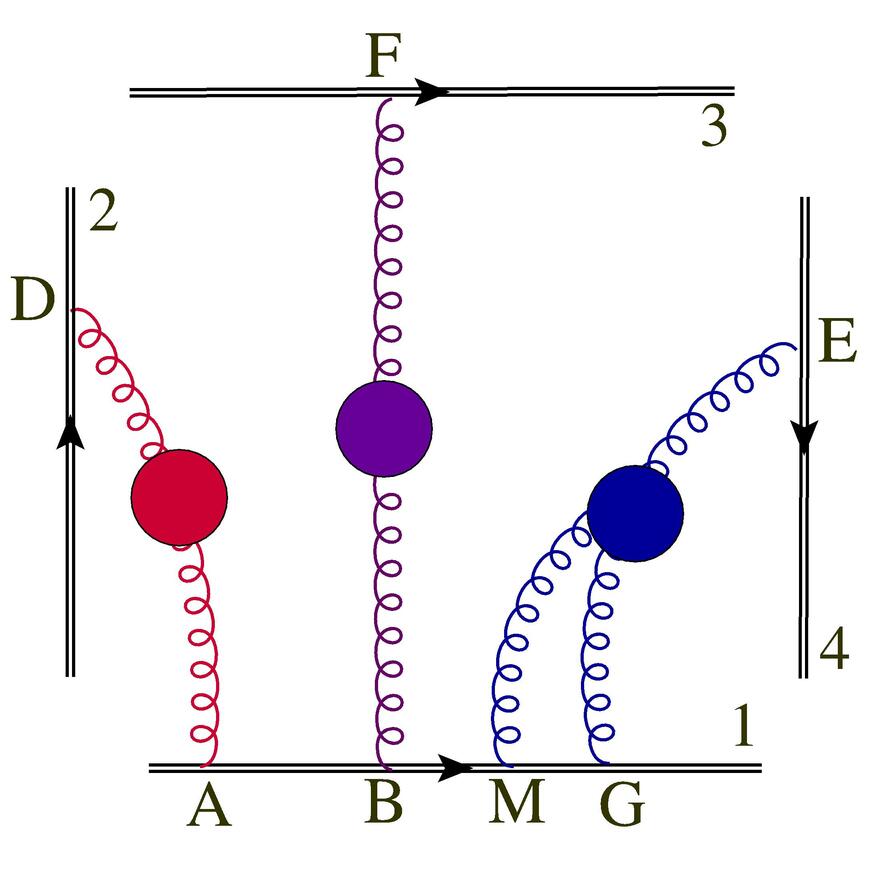} }
	\qquad 	
	\subfloat[][$ \text{W}_{4,\,\text{I}}^{(2,1)}(1,1,2,3) $]{\includegraphics[height=3.5cm,width=3.5cm]{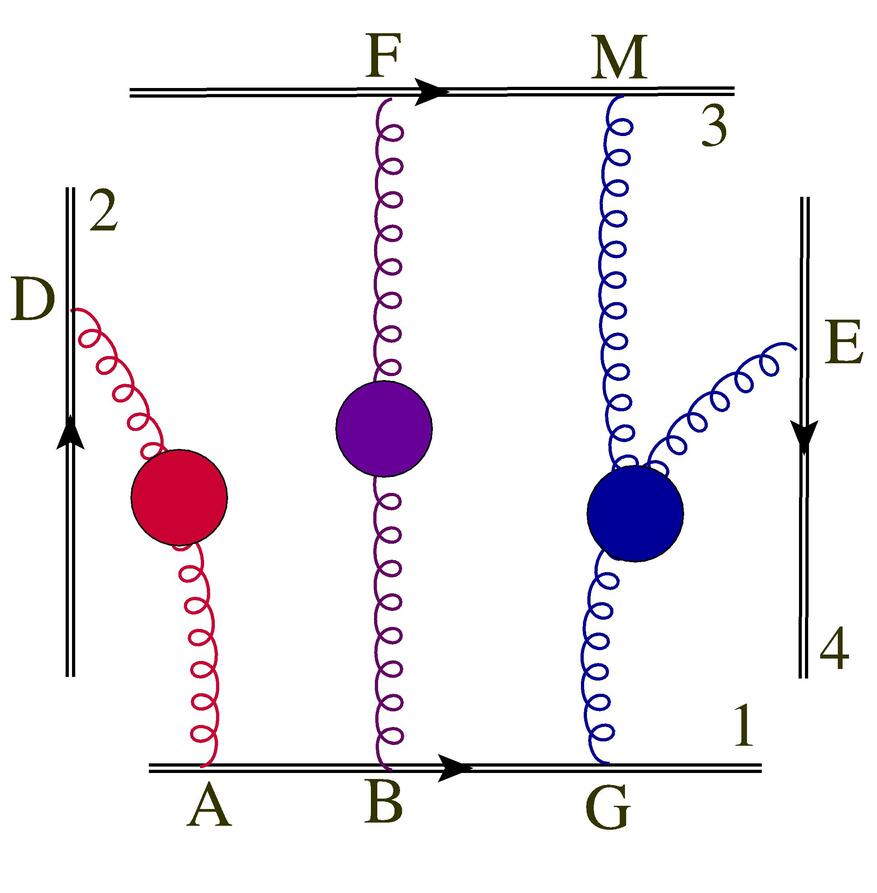} }
	\qquad 	
	\subfloat[][$ \text{W}_{4,\,\text{II}}^{(2,1)}(1,1,2,3) $]{\includegraphics[height=3.5cm,width=3.5cm]{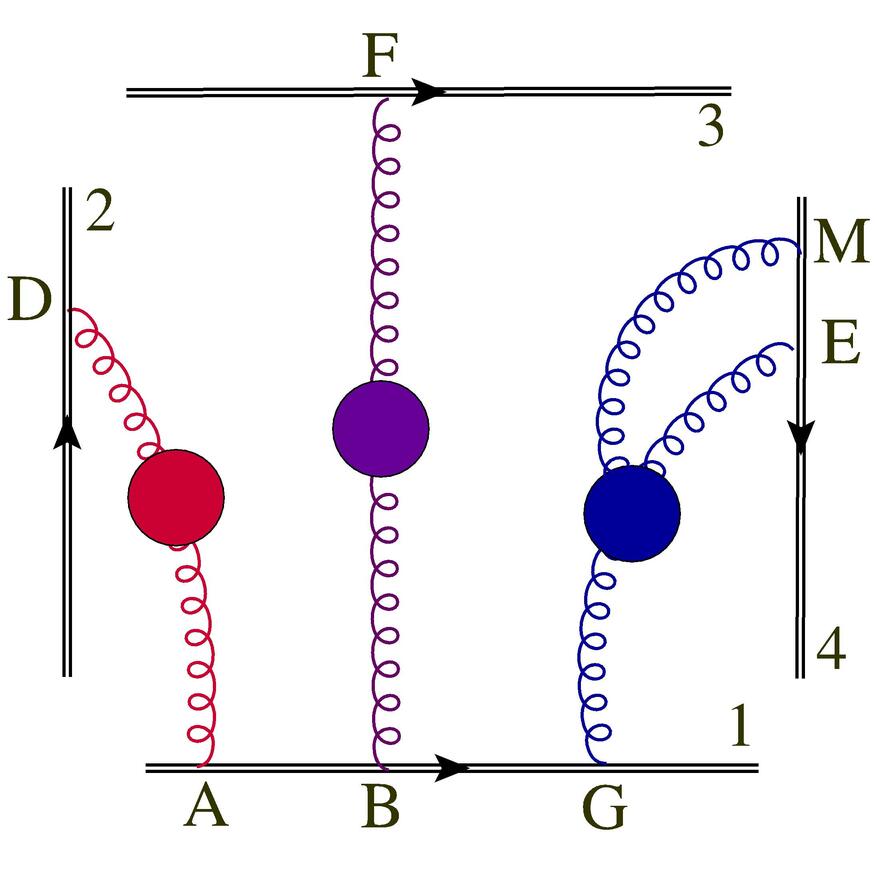} }
	\qquad 	
	\subfloat[][$ \text{W}_{4,\,\text{III}}^{(2,1)}(1,1,2,3) $]{\includegraphics[height=3.5cm,width=3.5cm]{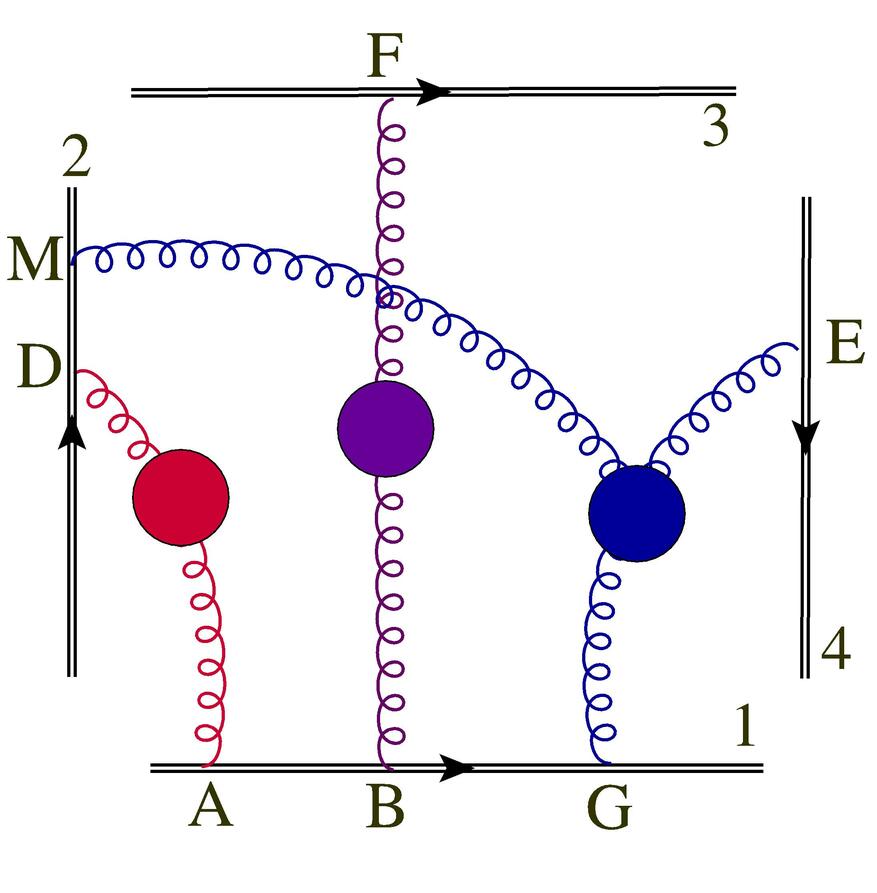} }
	\qquad 	
	\subfloat[][$ \text{W}_5^{(1,2)}(1,1,1,1,3) $]{\includegraphics[height=3.5cm,width=3.5cm]{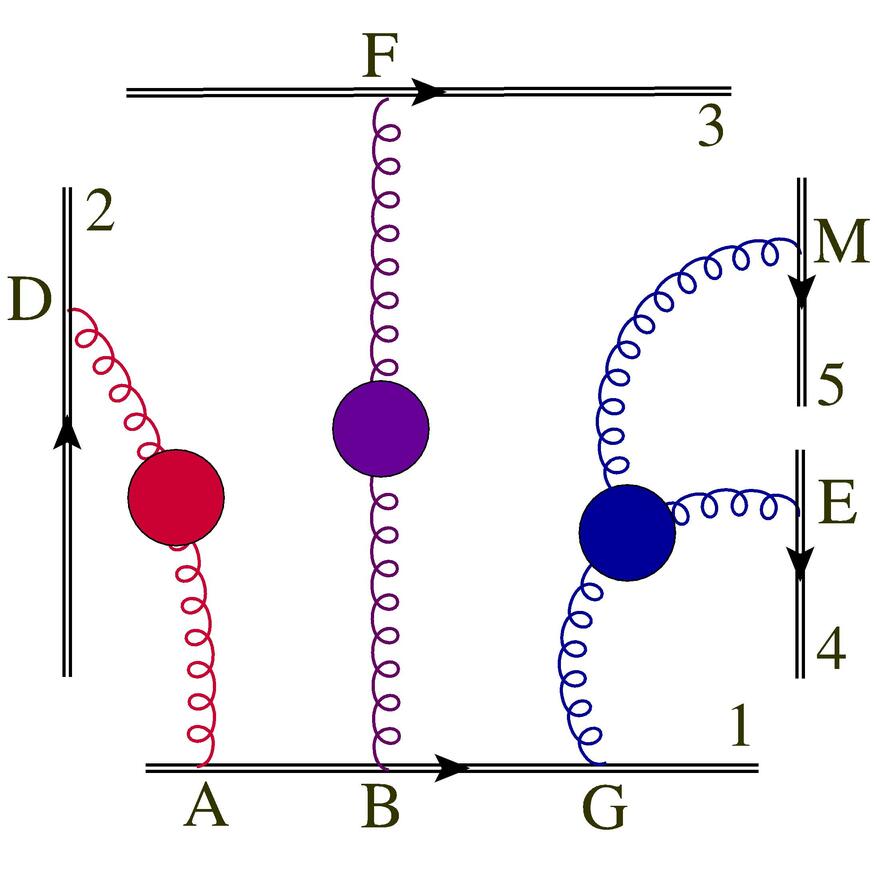} }
	\caption{The five Cwebs (\textcolor{blue}{b}), (\textcolor{blue}{c}), (\textcolor{blue}{d}), (\textcolor{blue}{e}), and (\textcolor{blue}{f}) are generated from Cweb (\textcolor{blue}{a}) using second step of the algorithm}
	\label{fig:Converse-Conjecture-3}
\end{figure}

The second step when applied to a parent Cweb with only reducible diagrams can generate either a Cweb that contains at least one irreducible diagram, or a Cweb that is a member of the same family. 
We can illustrate this by starting with the Cweb $ \text{W}^{(3)}_{4} (1,1,1,3) $, at three loops. This is a basis Cweb that has six diagrams. A representative diagram of the Cweb, shown in fig.~(\ref{fig:Converse-Conjecture-3}\textcolor{blue}{a}), whose possible shuffles and  $ s $-factors are given in table \ref{tab:Parent-Cweb-step-two-conj}. Let us, now  generate Cwebs  by turning the blue correlator into a three point correlator;  this generates the five Cwebs shown in fig.~(\ref{fig:Converse-Conjecture-3}).

We note that the Cwebs shown in figs.~(\ref{fig:Converse-Conjecture-3}\textcolor{blue}{b}), (\ref{fig:Converse-Conjecture-3}\textcolor{blue}{c}) and (\ref{fig:Converse-Conjecture-3}\textcolor{blue}{e}) have different shuffles in comparison to their parent Cweb shown in fig.~(\ref{fig:Converse-Conjecture-3}\textcolor{blue}{a}). Further, they are also the only Cwebs which contain irreducible diagrams. Unlike these, the Cwebs shown in figs.~(\ref{fig:Converse-Conjecture-3}\textcolor{blue}{d}) and (\ref{fig:Converse-Conjecture-3}\textcolor{blue}{f}) do not have any irreducible ($s=0$)  diagram and have exactly the same shuffle content as their parent Cweb shown in table \ref{tab:Parent-Cweb-step-two-conj}. 
\warning{Note that the second step generates Cwebs with at least one $ n $-point gluon correlator ($ n>2 $), thus, by definition these cannot be basis Cwebs.}
Hence  we conclude that the second step of algorithm does not generate any new basis Cweb if we start from a  Cweb that contains only reducible ($s\neq 0$) diagrams, 
but it generates Cwebs at one order higher that belong to the same family.

\begin{figure}[H]
	\centering
	\subfloat[][$ \text{W}_3^{(2)}(1,1,2) $]{\includegraphics[height=3.5cm,width=3.5cm]{ConjectureWebsBasis} }
	\qquad
	\subfloat[][$ \text{W}_4^{(3)}(1,1,1,3) $]{\includegraphics[height=3.5cm,width=3.5cm]{Converse-Conjecture-2b} }
	\qquad 	
	\subfloat[][$ \text{W}_{4}^{(1,1)}(1,1,1,3) $]{\includegraphics[height=3.5cm,width=3.5cm]{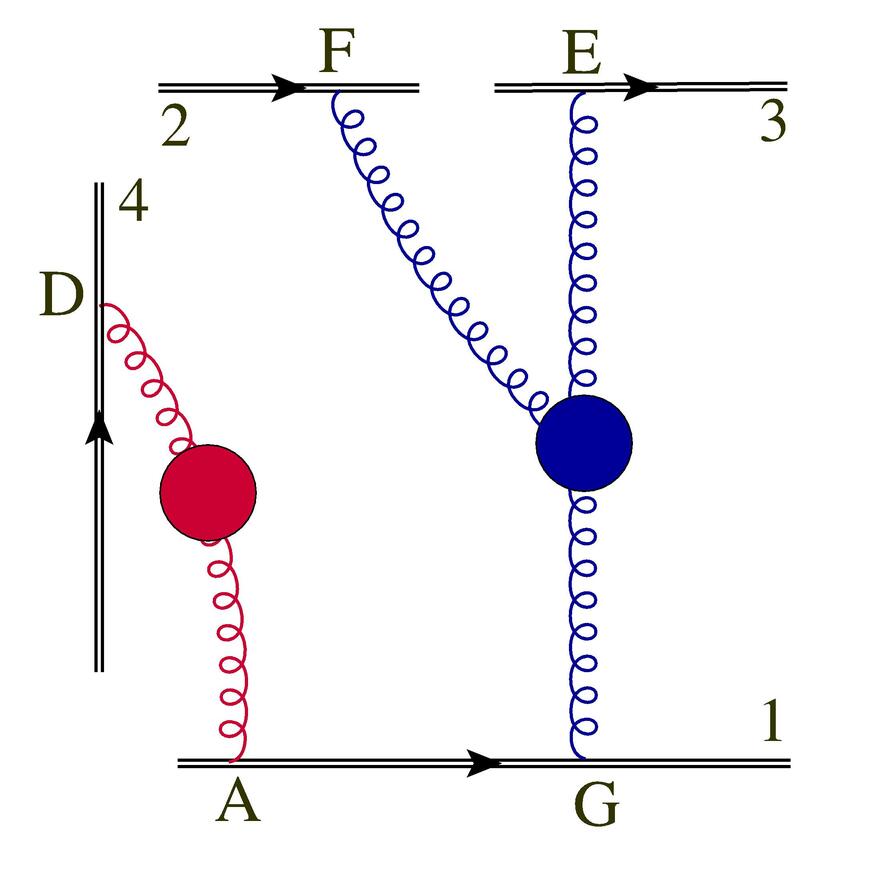} }
	\qquad 	
	\subfloat[][$ \text{W}_{3}^{(0,0,1)}(1,1,2) $]{\includegraphics[height=3.5cm,width=3.5cm]{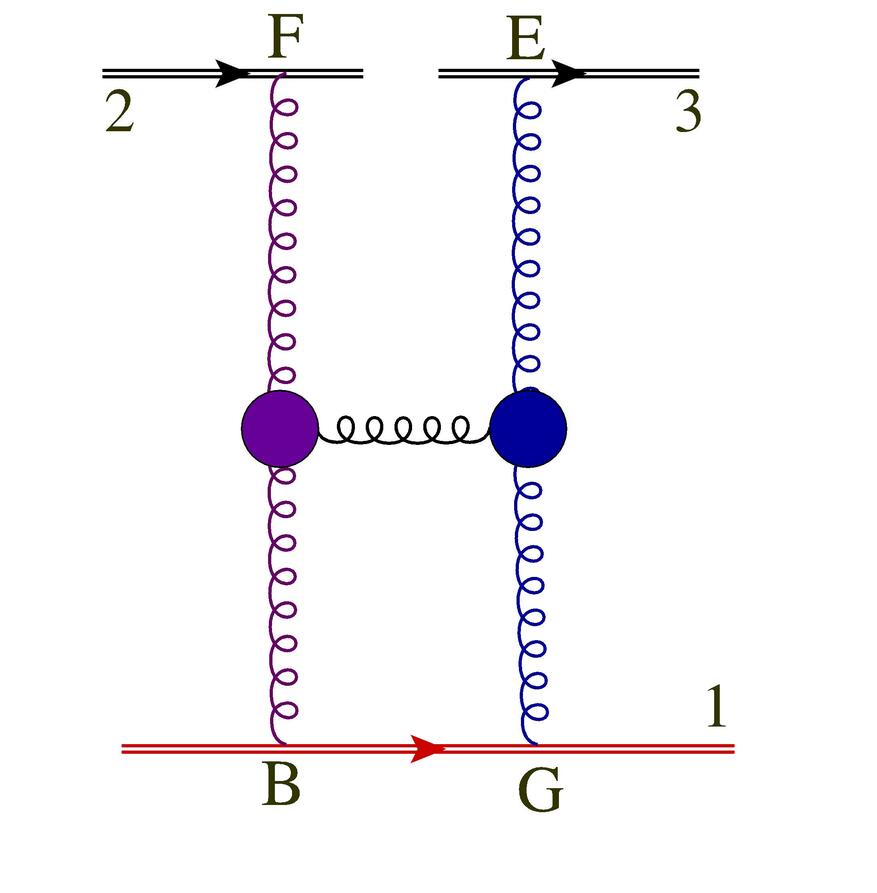} }
	\qquad 	
	\subfloat[][$ \text{W}_{4}^{(1,0,1)}(1,1,1,3) $]{\includegraphics[height=3.5cm,width=3.5cm]{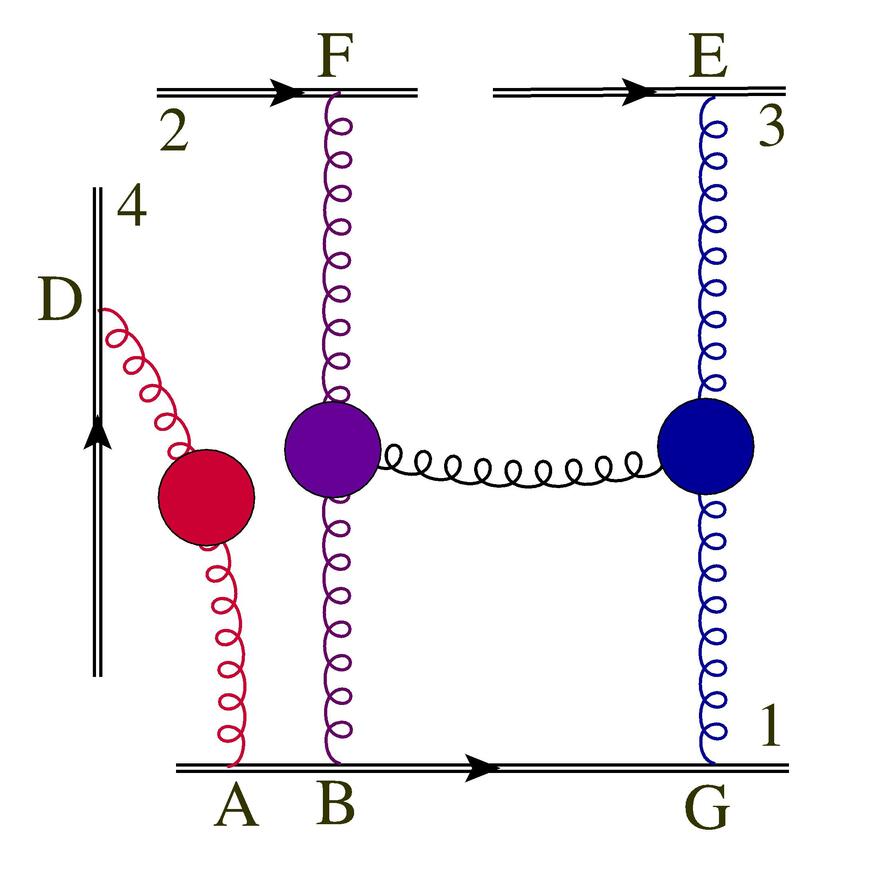} }
	\qquad 	
	\subfloat[][$ \text{W}_4^{(1,0,1)}(1,1,1,3) $]{\includegraphics[height=3.5cm,width=3.5cm]{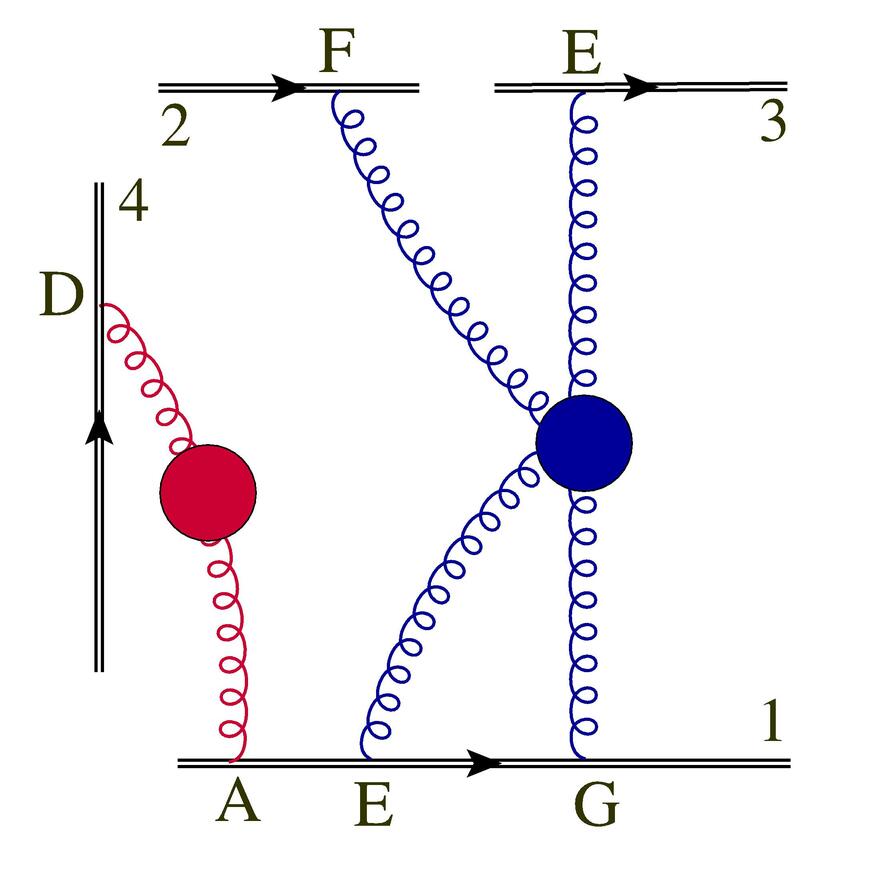} }
	\caption{The two Cwebs (\textcolor{blue}{d}) and (\textcolor{blue}{e}) are generated using the third step on the parent Cwebs  (\textcolor{blue}{a}) and  (\textcolor{blue}{b}). The Cweb (\textcolor{blue}{f}) is generated from Cweb (\textcolor{blue}{c}) using second step. The Cwebs (\textcolor{blue}{e}) and  (\textcolor{blue}{f}) are exactly the same.}
	\label{fig:Converse-Conjecture-Algo-3}
\end{figure}

The third step will produce Cwebs that contain only one diagram, or have been already generated by first and second steps of the algorithm.
For example the Cweb shown in fig.~(\ref{fig:Converse-Conjecture-Algo-3}\textcolor{blue}{e}) is same as that in fig.~(\ref{fig:Converse-Conjecture-Algo-3}\textcolor{blue}{f}). Therefore, for the proof of theorem it is sufficient to analyse only the first two categories.

We conclude, thus, that only the first step of the algorithm can produce a new basis Cweb.  Crucially, it  is evident that if the steps of the algorithm generate a Cweb with a given weight vector $S$ with all non-zero entries, then it is a member of a {\it unique} family. \warning{Recall that each family has a unique mixing matrix} and this completes the proof of the Uniqueness theorem.

\begin{table}\centering
		\begin{tabular}{|c|c|c|}
			\hline 
			\textbf{Diagrams}  & \textbf{Sequences}  & \textbf{s-factors}  \\ 
			\hline
			$ d_1 $ &$ \{ABG\} $&1\\ \hline
			$ d_2 $ &$ \{AGB\} $&1\\ \hline
			$ d_3 $ &$ \{GAB\} $&1\\ \hline
			$ d_4 $ &$ \{GBA\} $&1\\ \hline
			$ d_5 $ &$ \{BAG\} $&1\\ \hline
			$ d_6 $ &$ \{BGA\} $&1\\ \hline
		\end{tabular}
\caption{Diagrams and $ s $-factors of parent Cweb $ \text{W}^{(3)}_{4} (1,1,1,3) $ }
\label{tab:Parent-Cweb-step-two-conj}
\end{table}

\subsection*{Application at four loops}
We will now show the utility of the Uniqueness theorem for Cwebs that appear at four loops.
In table~\ref{tab:conjecture-table-two-three}, we list down all the  Cwebs that belong to one of the families of basis Cwebs upto three loops.  
Also a reordering of diagrams with same s-factors does not change the mixing matrix of a Cweb with S having all non-zero entries.
The explicit expressions of the  matrices ${ R(1_2) }$, $ R(1_2,2_2) $ and $ R(1_6) $ are given by
\begin{align}
R\,(1_2)\,=\,&\frac{1}{2}\left(\begin{array}{cc}
1 & -1 \\
-1 & 1
\end{array}\right), \, R(1_2,2_2)\,=\,\frac{1}{6} \left(
\begin{array}{cccc}
2 & 2 & -2 & -2 \\
2 & 2 & -2 & -2 \\
-1 & -1 & 1 & 1 \\
-1 & -1 & 1 & 1 \\
\end{array}
\right), \,
R(1_6) =\frac{1}{6} \left(
\begin{array}{cccccc}
2 & -1 & -1 & -1 & -1 & 2 \\
-1 & 2 & -1 & 2 & -1 & -1 \\
-1 & -1 & 2 & -1 & 2 & -1 \\
-1 & 2 & -1 & 2 & -1 & -1 \\
-1 & -1 & 2 & -1 & 2 & -1 \\
2 & -1 & -1 & -1 & -1 & 2 \\
\end{array}
\right).
\label{eq:basis-text} 
\end{align} 
From Uniqueness theorem it is clear that at four loops the Cwebs with all reducible diagrams, listed in table~\ref{tab:conjecture-table} are either a new basis or a member of a known family already present at two or three loops\footnote{The $ s $-factors for $ \text{W}^{(4)}_{5}(1,1,1,2,2)  $, $ \text{W}^{(4)}_{5}(1,1,1,2,3) $ had typos in~\cite{Agarwal:2020nyc} which have been corrected in this article.}. From this table, we note that there are total fourteen Cwebs with all reducible diagrams out of which first eleven Cwebs belong to known families and remaining three Cwebs are new basis appearing at four loops.
Thus, we can write the explicit form of eleven matrices which is $ 78\% $  of all the Cwebs that have reducible diagrams, without any further work. 
The explicit forms of the mixing matrices for new basis webs that appear at four loops can be found in the appendix \ref{sec:basis}.
\begin{table}[t]
	\begin{center}
		\begin{tabular}{|l|c|c|c|}
			\hline
			Name of Cweb & loop order & Column weight vector $ S $   & Web mixing matrix \\ \hline
			&&&\\
			$ \text{W}^{(2)}_{3}(1,1,2) $  & 2 & $ \{1_2\} $ & $ R(1_2) $ \\ \hline
			&&&\\
			
			$ \text{W}^{(1,1)}_{4}(1,1,1,2) $  & 3 & $ \{1_2\} $ & $ R(1_2) $ \\ \hline
			&&&\\
			$ \text{W}^{(1,1)}_{3}(1,2,2) $  & 3 & $ \{1_2\} $ & $ R(1_2) $ \\ \hline
			&&&\\
			$ \text{W}^{(3)}_{4}(1,1,1,3) $  & 3 & $ \{1_6\} $ & $ R(1_6) $ \\ \hline &&& \\ 
			$ \text{W}^{(3)}_{4}(1,1,2,2) $  & 3 & $ \{1_2,2_2\} $ & $ R(1_2,2_2) $ \\ \hline
		\end{tabular}	
	\end{center}
	\caption{Cwebs upto three loops that have only non-zero entries in $S$ and their mixing matrices.}
	\label{tab:conjecture-table-two-three}
\end{table}
\begin{table}[t]
	\begin{center}
		\begin{tabular}{|l|c|c|c|}
			\hline
			Name of Cweb & loop order & Column weight vector $ S $   & Web mixing matrix \\ \hline &&& \\
			$ \text{W}^{(0,2)}_{5}(1,1,1,1,2) $  & 4 & $ \{1_2\} $ & $ R(1_2) $ \\ \hline
			
			&&& \\
			$ \text{W}^{(1,0,1)}_{5}(1,1,1,1,2) $  & 4 & $ \{1_2\} $ & $ R(1_2) $ \\ \hline
			&&&\\
			$ \text{W}^{(1,0,1)}_{4,\text{II}}(1,1,2,2) $  & 4 & $ \{1_2\} $ & $ R(1_2) $ \\ \hline
			&&&\\
			$ \text{W}^{(0,2)}_{4,\text{I}}(1,1,2,2) $  & 4 & $ \{1_2\} $ & $ R(1_2) $ \\ \hline
			&&&\\
			$ \text{W}^{(0,2)}_{3,\text{I}}(2,2,2) $  & 4 & $ \{1_2\} $ & $ R(1_2) $ \\ \hline &&& \\
			$ \text{W}^{(1,0,1)}_{3,\text{II}}(1,2,3) $  & 4 & $ \{1_2\} $ & $ R(1_2) $ \\ \hline 	&&&\\
			$ \text{W}^{(2,1)}_{4,\text{III}}(1,1,2,3) $  & 4 & $ \{1_6\} $ & $ R(1_6) $ \\ \hline
			&&&\\
			$ \text{W}^{(2,1)}_{5}(1,1,1,1,3) $  & 4 & $ \{1_6\} $ & $ R(1_6) $ \\ \hline 
			&&& \\	
			$ \text{W}^{(2,1)}_{4,\text{I}}(1,2,2,2) $  & 4 & $ \{1_2,2_2\} $ & $ R(1_2,2_2) $ \\ \hline
			&&& \\
			$ \text{W}^{(2,1)}_{5,\text{I}}(1,1,1,2,2) $  & 4 & $ \{1_2,2_2\} $ & $ R(1_2,2_2) $ \\ \hline
			&&&\\
			$ \text{W}^{(2,1)}_{5,\text{II}}(1,1,1,2,2) $  & 4 & $ \{1_2,2_2\} $ & $ R(1_2,2_2) $ \\ \hline 
			&&& \\
			$ \text{W}^{(4)}_{5}(1,1,1,2,2) $  & 4 & $ \{1_2,\warning{2_4},4_2\} $ & $ R(1_2,2_4,4_2) $ \\ \hline
			&&&\\
			$ \text{W}^{(4)}_{5}(1,1,1,2,3) $  & 4 & $ \{1_4,\warning{2_8\}} $ & $ R(1_4,2_8) $ \\ \hline
			&&&\\
			$ \text{W}^{(4)}_{5}(1,1,1,1,4) $  & 4 & $ \{1_{24}\} $ & $ R(1_{24}) $ \\ \hline
		\end{tabular}	
	\end{center}
	\caption{Cwebs at four loops that have only non-zero entries in $S$ and their mixing matrices}
	\label{tab:conjecture-table}
\end{table}
\begin{figure}[b]
	\centering \includegraphics[height=4cm,width=4cm]{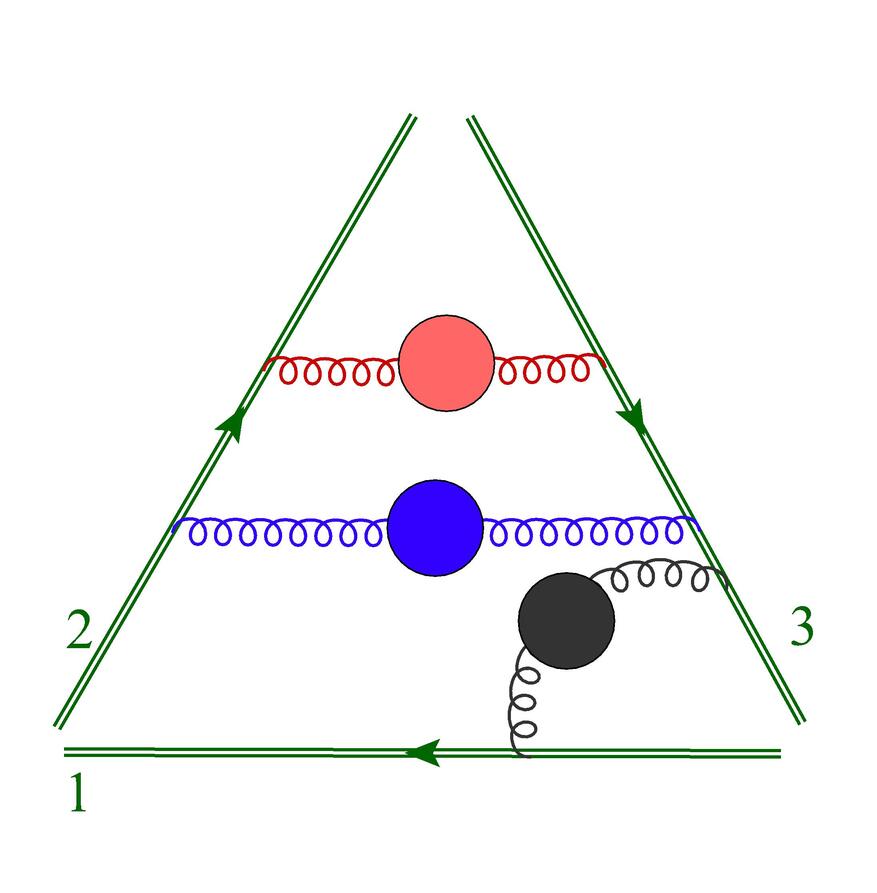}
	\caption{A Cweb which generates multiple duplicates.}
	\label{fig:dcweb}
\end{figure}

We will see in later sections that the matrices of basis Cwebs form the building blocks of the matrices corresponding to those Cwebs that contain irreducible diagrams. We would like to emphasise that there is a certain class of Cwebs in which the mixing matrices are not made up of these building blocks. In this class, identical copies of two or more correlators are attached to the 
same set of Wilson lines in an identical manner. For this class, the shuffle of correlators on each Wilson line generates multiple duplicates of each diagram in a Cweb, which we need to discard before applying the replica trick algorithm. As the replica trick algorithm for these Cwebs is only applied to a subset of shuffles, thus, in general the building blocks of mixing matrices for them cannot be constructed using the matrices of table \ref{tab:conjecture-table}. This class of Cwebs and its complexities while applying the replica trick is mentioned in detail in section 3 of \cite{Agarwal:2021him}. 
An example of this class is shown in fig.~(\ref{fig:dcweb}). The two point correlators attached between Wilson lines 2 and 3 are the reason behind the generation of multiple duplicates while doing shuffle of correlators. The shuffle generates twelve diagrams, out of which six are duplicates, and are discarded from the list of  diagrams for this Cweb. None of the ideas of later sections are applicable to this class except that of Normal ordering to be discussed in the next section.

\section{Normal ordering and Fused-Webs}
\label{sec:\reduced}
Having said all that we had for the Cwebs which contain only reducible diagrams in the previous section, we now turn our attention to those Cwebs which have one or more irreducible diagrams. We will describe here how the basis matrices present upto order $ \alpha_s^n $ appear in a general mixing matrix of 
these Cwebs, at orders higher than $ \alpha_s^n $.
\subsection{Normal ordering}
We begin by ordering the diagrams of a Cweb in such a way that the irreducible diagrams appear before the reducible diagrams. Using the fact
that the exponentiated colour factors of reducible diagrams are independent of the irreducible diagrams in a Cweb \cite{Gardi:2010rn},
we immediately find that upon ordering in this manner,  the general structure of a $ n\times n $ mixing matrix for a Cweb with $ l $ diagrams having $ s=0 $, and $ m $  diagrams having $ s \neq 0$  becomes, \\
\begin{align}
R\,=\,\left(\begin{array}{cccccccccc}
a_{11} & a_{12} & a_{13} &\ldots & a_{1l} & b_{11} & b_{12} & b_{13} &\ldots & b_{1m}\\
a_{21} & a_{22} & a_{23}  &\ldots & a_{2l} & b_{21} & b_{22} & b_{23} &\ldots & b_{2m}\\
a_{31} & a_{32} & a_{33}  &\ldots & a_{3l} & b_{31} & b_{32} & b_{33} &\ldots & b_{3m} \\
\vdots & \vdots & \vdots  &\vdots & \vdots & \vdots & \vdots & \vdots &\vdots & \vdots \\
a_{l1} & a_{l2} & a_{l3}  &\ldots & a_{ll} & b_{l1} & b_{l2} & b_{l3} &\ldots & b_{lm} \\
0 & 0 & 0 & \ldots  & 0  & d_{11} & d_{12} & d_{13} &\ldots & d_{1m}\\
0 & 0 & 0 & \ldots  & 0  & d_{21} & d_{22} & d_{23} &\ldots & d_{2m} \\
0 & 0 & 0 & \ldots  & 0  & d_{31} & d_{32} & d_{33} &\ldots & d_{3m} \\
0 & 0 & 0 & \ldots  & 0  & d_{m1} & d_{m2} & d_{m3} &\ldots & d_{mm} \\
\end{array}\right)\,,
\label{eq:R-gen-form}
\end{align}
where $ l+m=n \,$.  We write this more compactly as
\begin{align}
R=\left(\begin{array}{cc}
A_{l\times l}  & B_{l\times m} \\
O_{m \times l} & D_{m\times m}
\end{array}\right)\,,
\label{eq:R-block}
\end{align}
where  $ A $ and $ D $ are square matrices of order $ l\times l $, and $ m\times m $ corresponding to irreducible and reducible diagrams respectively; $ O $ is null matrix of order $m \times l$, and $ B $ is a matrix of order $ l\times m $. More layers of structure get unveiled if the diagrams are further ordered as 
\begin{align}
\Big\{ d_{1}, \ldots,  d_{k}, \, d_{k+1}, \ldots, d_{l}, \,  d_{l+1}, \ldots, d_{n} \Big\}\,, \quad \text{where},\quad&\, d_{1}, \ldots d_{k} \in \text{completely entangled diagrams},  \nonumber \\ 
&d_{k+1}, \ldots d_{l} \in \text{partially entangled diagrams}, \nonumber \\
 \quad & d_{l+1}, \ldots d_{n} \in \text{reducible diagrams, and }\nonumber\\
& s(d_{l+1})\leq s(d_{{l+2}}) \leq \cdots \leq s(d_{l+m=n}). 
\label{diaglist}
\end{align}
That is we  arrange the diagrams such that the first $k$ diagrams are completely entangled, followed by $(l-k)$
partially entangled, further followed by reducible diagrams which appear in ascending order of their $s$-factors. We define this order of diagrams of a Cweb as \textit{Normal order}, and the corresponding mixing matrix to be Normal ordered.

\begin{figure}
	\captionsetup[subfloat]{labelformat=empty}
	\centering
	\vspace{-3mm}
	\subfloat[][$ d_1 $]{\includegraphics[height=4cm,width=4cm]{W123-1} }
	\qquad 
	\subfloat[][$ d_2 $]{\includegraphics[height=4cm,width=4cm]{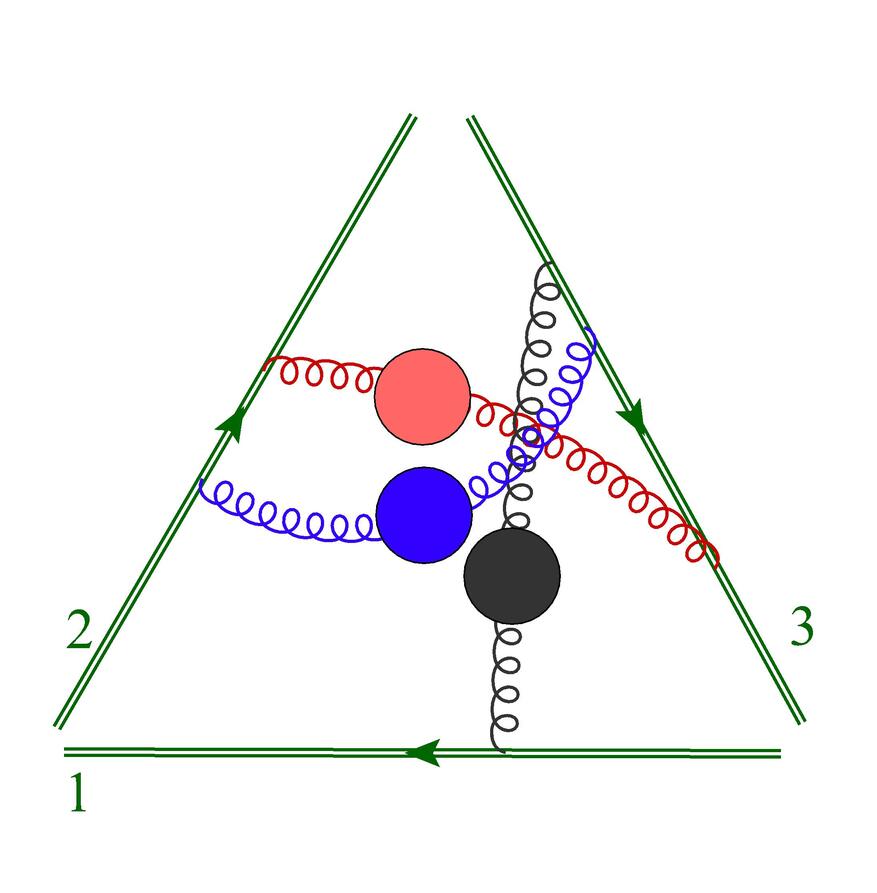} }
	\qquad 	
	\subfloat[][$ d_3 $]{\includegraphics[height=4cm,width=4cm]{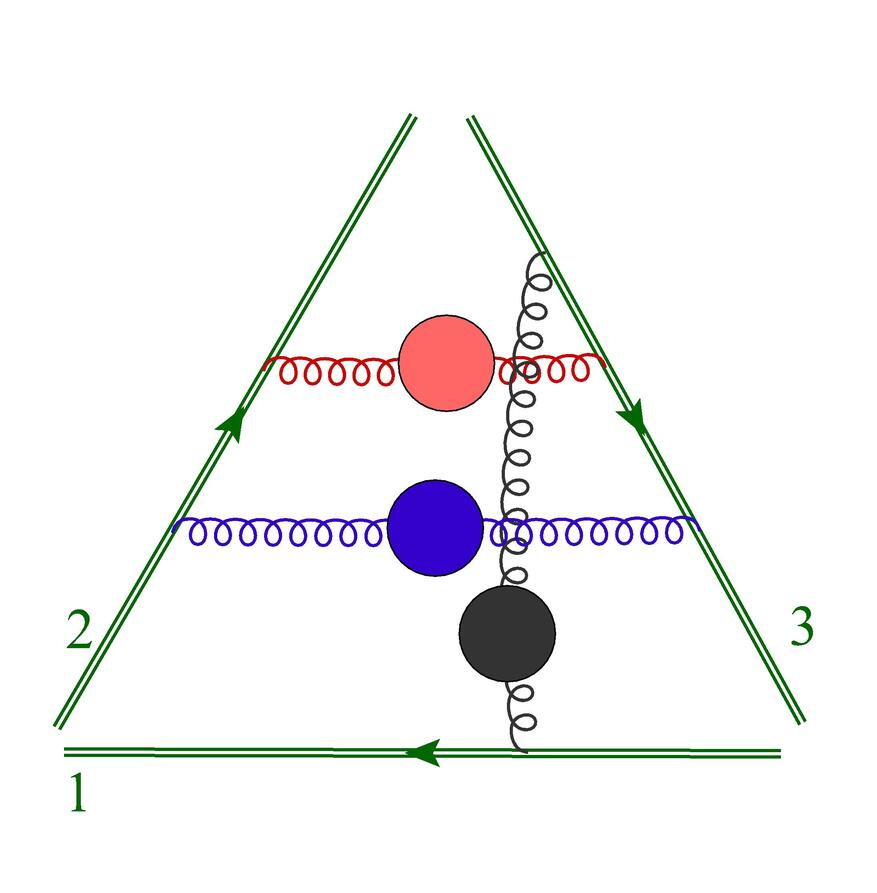} }
	\qquad 	
	\subfloat[][$ d_4 $]{\includegraphics[height=4cm,width=4cm]{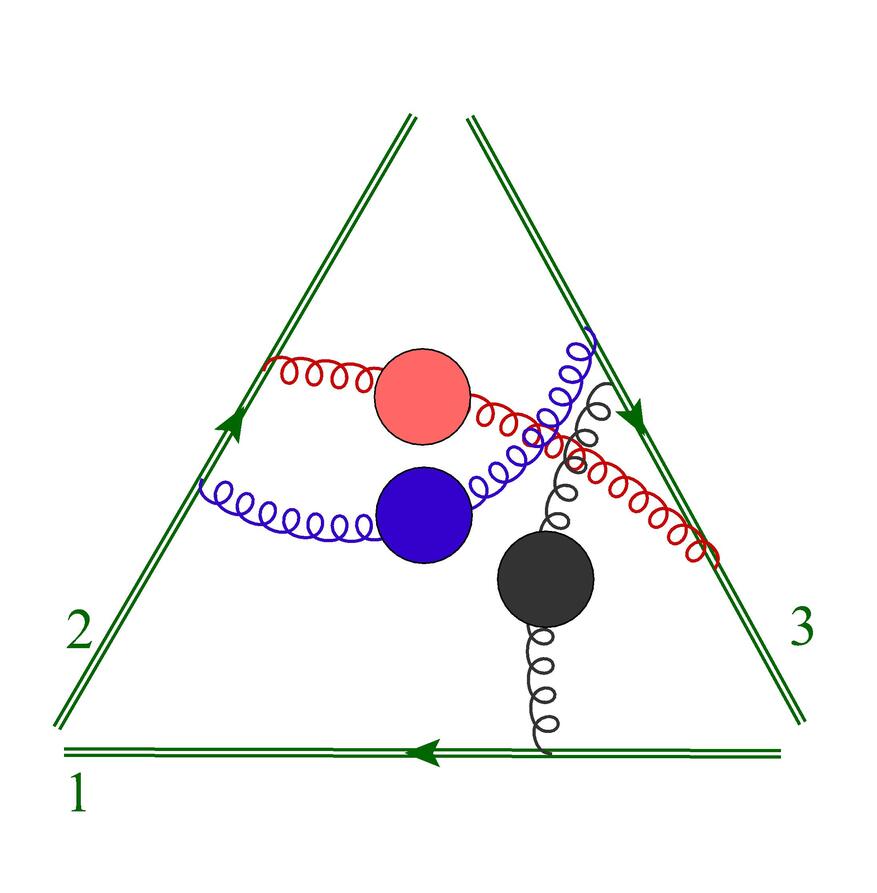} }	
	\qquad
	\subfloat[][$ d_5 $]{\includegraphics[height=4cm,width=4cm]{W123-5} }	
	\qquad 
	\subfloat[][$ d_6 $]{\includegraphics[height=4cm,width=4cm]{W123-6} }	
	
	\caption{Diagrams of the Cweb $ \text{W}_3^{(3)}(1,2,3) $.}
	\label{fig:order-example}
\end{figure}

Consider $ \text{W}_3^{(3)}(1,2,3) $ that has six diagrams shown in fig.~(\ref{fig:order-example}) of which $ d_1 $, $ d_3 $, and $ d_6 $ are irreducible, and the remaining three diagrams are reducible. Among three  irreducible diagrams  $ d_3 $ and $ d_6 $ are partially entangled as the black correlator can be shrunk to the origin independent of the other correlators. After Normal ordering, $ \lbrace d_1,d_2,d_3,d_4,d_5,d_6 \rbrace $ becomes $ \lbrace d_1,d_3,d_6,d_2,d_4,d_5 \rbrace $. 

The replica trick algorithm which determines the explicit form of the mixing matrices works in a fashion that it can disentangle two entangled correlators, however, the converse is not possible. 
This implies that the ECF of $d_{1}$ in (\ref{diaglist}), for example, is of the following form (see eq.~\eqref{expocolf})
\begin{align}
\widetilde{C}(d_{1}) = C(d_{1}) + 0 \cdot C(d_{2}) + \ldots + 0 \cdot C(d_{k}) + \sum_{j=k+1}^{{n}} \alpha_{j} C(d_{j})\,.
\end{align}
In general for a completely entangled diagram $ d_i $ in (\ref{diaglist}), the ECF is given by, 
\begin{align}
\widetilde{C}(d_{i}) = C(d_{i})  + \sum_{j=k+1}^{n} \alpha_{ij}\, C(d_{j})\,,  \qquad 1 \leq i \leq k\,.
\end{align}
That is, the ECF for a completely entangled diagram consists of colour of the partially entangled diagrams and reducible diagrams along with its own colour. 
A careful examination of the replica trick algorithm reveals that the ECF of a partially entangled diagram will not contain the colours of the completely entangled diagrams.   
\begin{align}
\widetilde{C}(d_{i}) =  \sum_{j=k+1}^{n} \beta_{ij}\, C(d_{j})\,,  \qquad k+1 \leq i \leq l\,,
\end{align}
Thus, if a Cweb contains $ l $ irreducible diagrams, out of which $ k $ are completely entangled, and ($ l-k $) are partially entangled, then the Normal ordering puts the matrix $ A $ in the following general form 

\begin{align}
A=\left(\begin{array}{cc}
\hspace*{-0.5cm}{I}_{k \times k} & (A_U)_{k\times (l-k)} \\
{O}_{(l-k)\times k} & \quad \, \, (A_L)_{(l-k) \times (l-k)}
\end{array}\right),
\label{eq:A-gen-form}
\end{align}
where $ I $ is an identity matrix of order $ k $ that corresponds to completely entangled diagrams, $ A_L $ is a square matrix of order $ (l-k) $ corresponding to partially entangled diagrams, $ O $ is null matrix of order $ (l-k)\times k $, and $ A_U $ is a matrix of order $ k \times (l-k) $. 

\noindent From the foregoing discussion we conclude that the general structure of the mixing matrix $ R $ after Normal ordering is
\begin{align}
R=\left(\begin{array}{c|c}
\begin{array}{cc}
\hspace*{-0.5cm}{I}_{k \times k} & (A_U)_{k\times (l-k)} \\
{O}_{(l-k)\times k} & \quad \, \, (A_L)_{(l-k) \times (l-k)}
\end{array} & B_{l\times m} \\ 
\hline
{O}_{m\times l } & D_{m\times m}
\end{array}\right)\,. 
\label{eq:R-gen-big-form}
\end{align}

\noindent We further explore the structure of matrices $ A $ and $ D $ in the next subsections.

\subsection{Diagonal block $ D $ of a mixing matrix $ R $}\label{sec:D-properties}
{Recall that the matrix $ D $ gives mixing between the reducible diagrams of a Cweb. Now, we prove that $ D $ obeys all the properties of the mixing matrices that are known to date:} 
\begin{itemize}
	\item {Idempotence:} Using $ R^2=R $, eq.~(\ref{eq:R-block}) gives
	\begin{align}
	D^2=D\,.
	\end{align} 
	That is, $ D $ is idempotent. 
	\item Row sum rule: Applying row sum rule to the general form given in eq.~(\ref{eq:R-gen-form}), the rows associated with the reducible diagrams give,
	\begin{align}
	\sum_{j=1}^{m} d_{ij} = 0, \quad 1\leq i\leq m,
	\end{align}
	where $ d_{ij} $ are the elements of matrix $ D $ given in eq.~(\ref{eq:R-gen-form}).  
	That is the block $ D $ also satisfies zero row-sum rule. 
	\item Column sum rule:	Applying column sum rule on the explicit form of the mixing matrix given in eq.~(\ref{eq:R-gen-form}) and remembering that \warning{the first $ l $ diagrams have $ s=0 $, the columns corresponding to the reducible diagrams give,}
	\begin{align}
	\sum_{i=1}^m s(d_{l+i})\, d_{ij}\,=0\,, \quad  1 \leq j \leq m  
	\label{eq:column-sum1} 
	\end{align}
	Thus, the matrix $ D $ satisfies the zero column-sum rule. 
\end{itemize}

\noindent Hence, we conclude that $ D $ satisfies the known properties of a web mixing matrix with $S_D=\{ s_{l+1} \ldots s_{l+m}\} $. 
The diagrams present in $ D $ form a closed set under the action of replica ordering operator \textbf{R} (see appendix \ref{sec:repl}) in the same way as that of a Cweb with  $S_D=\{ s_{l+1} \ldots s_{l+m}\} $. Then mixing between the diagrams of $ D $ is same as that of the Cweb, thus, they have the same mixing matrix $ R(S_D) $. 
It is possible, however, that $ D $ does not correspond to any actual Cweb. For example, if
$ D $ is associated with three diagrams, having $ S_D=\{1,1,1\} $, then although $ D $ satisfies all the known properties, it is not a mixing matrix
as there are no Cwebs with $ S=\{1,1,1\} $~\cite{Agarwal:2021him}.

As we have proved that $ D $ satisfies the properties of a mixing matrix, and has its own column weight vector $ S_D $, we can use the Uniqueness theorem stated in section \ref{sec:conje} to write the explicit form of $ D $, provided $ S_D $ is column weight vector for a known family of Cwebs $ f(S_D) $. 

\subsection{Structure of the block $A$:  Fused-Webs }\label{sec:avatar-webs}

\begin{figure}
	\centering
	\subfloat[][]{\includegraphics[height=4cm,width=4cm]{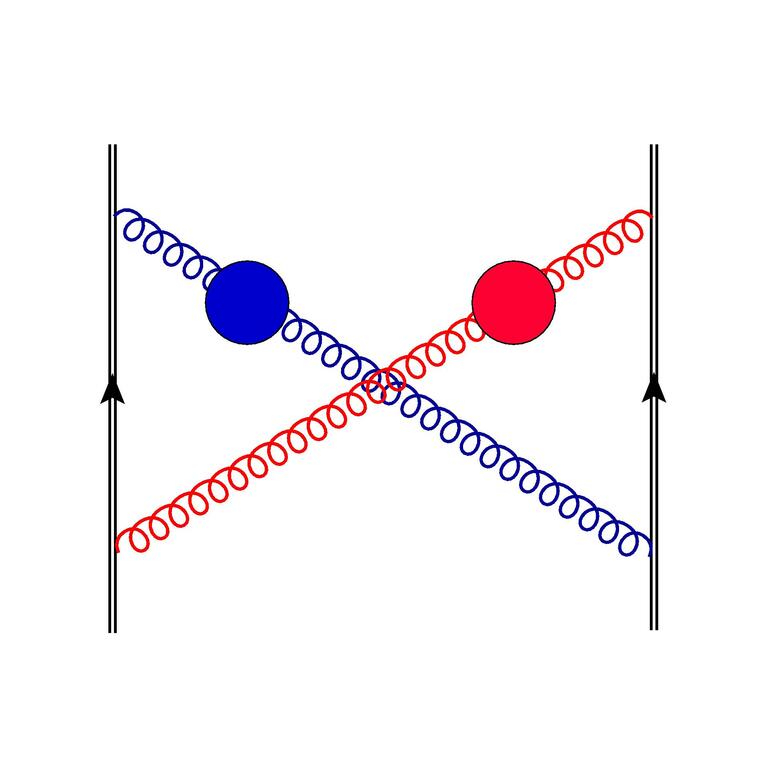} }
	\qquad 
	\subfloat[][]{\includegraphics[height=4cm,width=4cm]{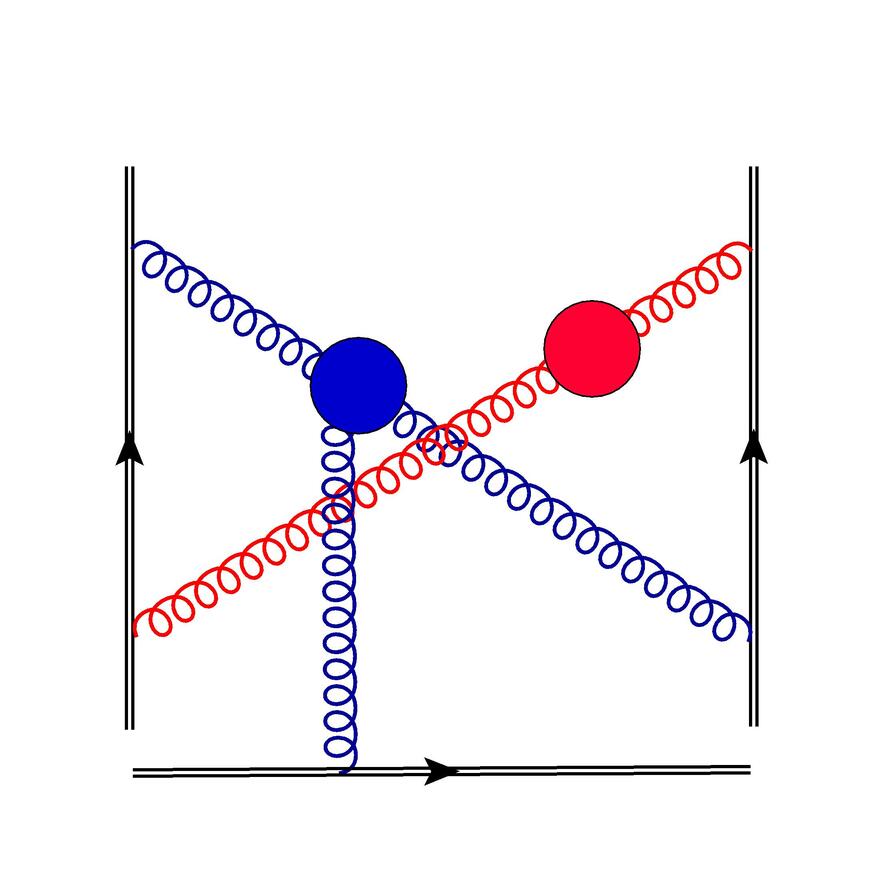} }
	\qquad 
	\subfloat[][]{\includegraphics[height=4cm,width=4cm]{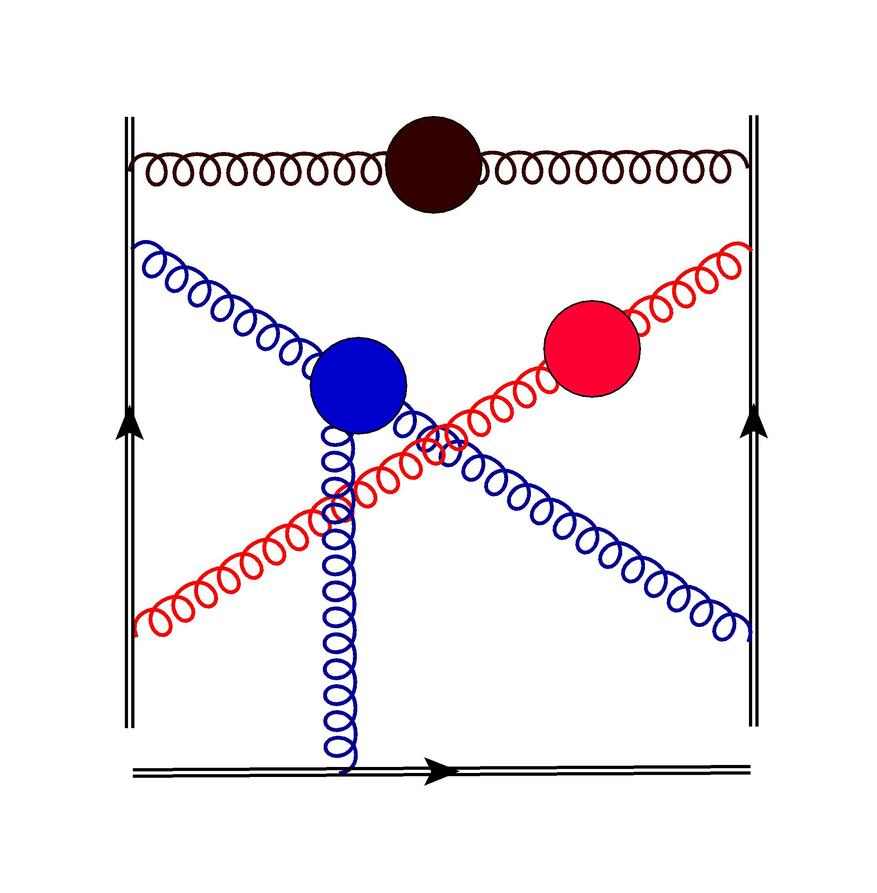} }	
	\quad 
	\subfloat[][]{\includegraphics[height=4cm,width=4cm]{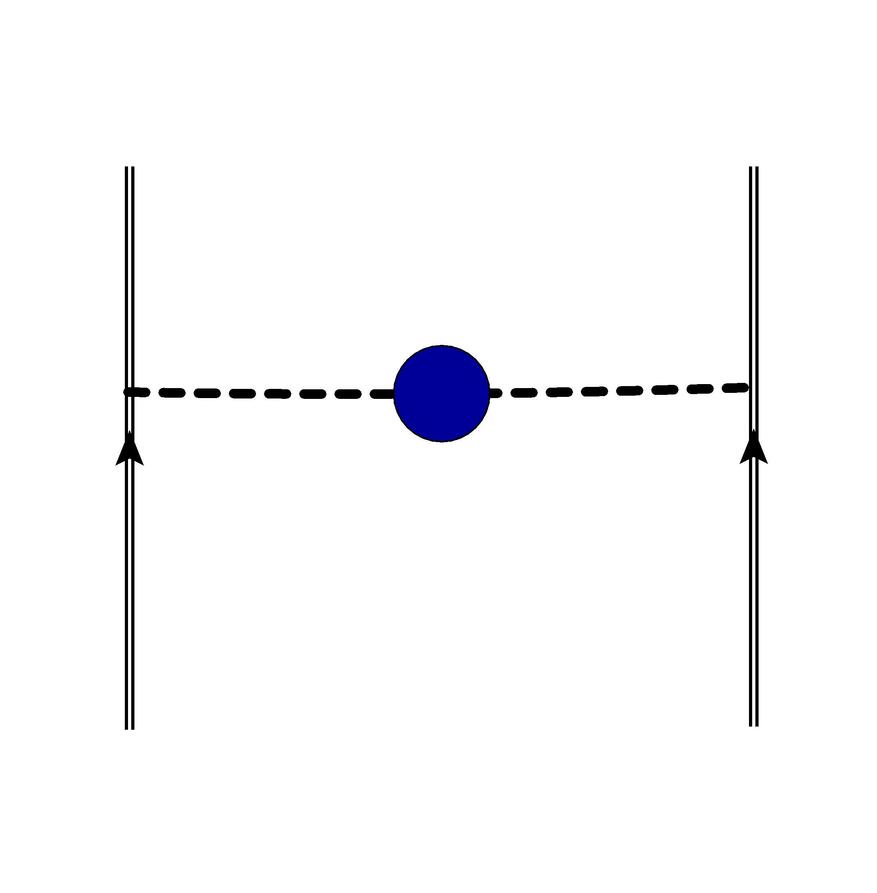} }
	\qquad 
	\subfloat[][]{\includegraphics[height=4cm,width=4cm]{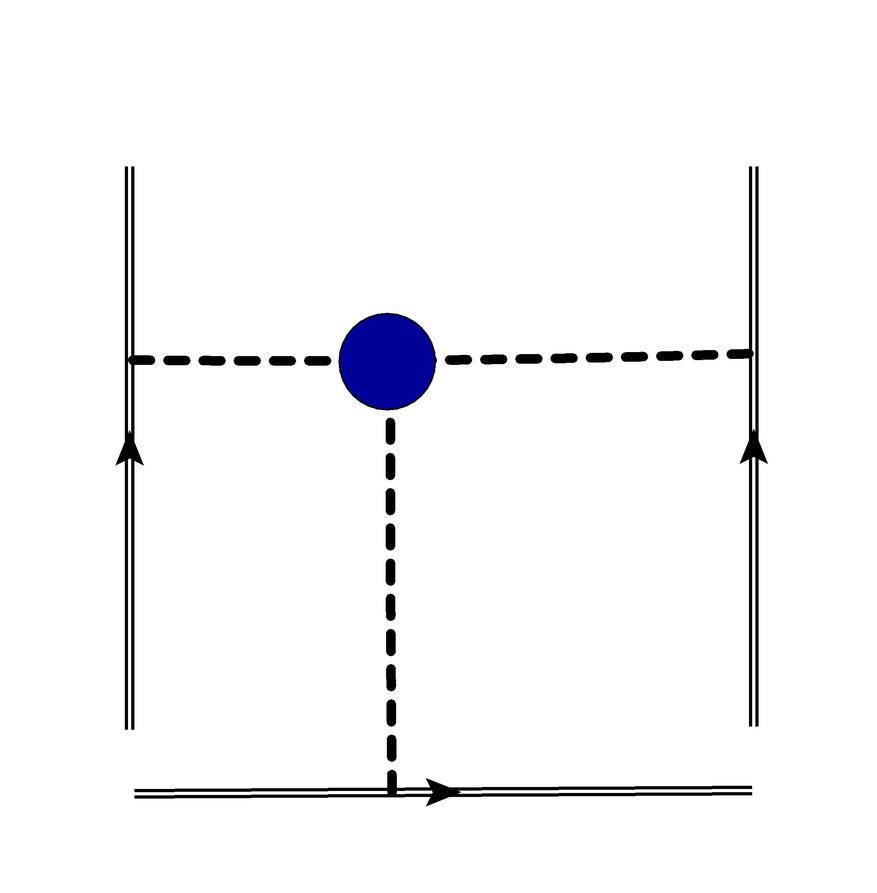} }
	\qquad  
	\subfloat[][]{\includegraphics[height=4cm,width=4cm]{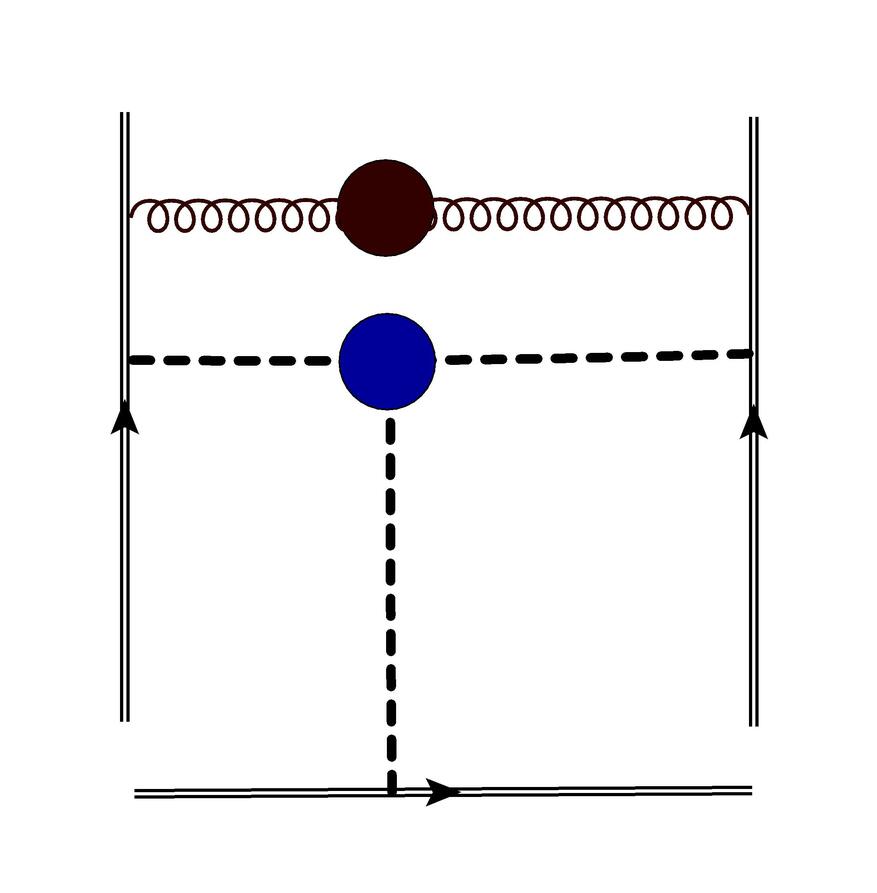} }
	\caption{Replacing an entangled piece by $ n $-point correlators.     
		Here, the diagrams in~(\textcolor{blue}{a}), (\textcolor{blue}{b}) and (\textcolor{blue}{c}) have an entangled piece made up of the blue and the red gluon correlators involving two and three Wilson lines. Therefore, replacing the entangled piece of diagram (\textcolor{blue}{a}) with a two point \reduced correlator; and that of diagram (\textcolor{blue}{b}), and (\textcolor{blue}{c}) with a three point \reduced correlator, provides the  \reduced diagrams (\textcolor{blue}{d}), (\textcolor{blue}{e}), and (\textcolor{blue}{f}) respectively. With this replacement, the diagrams (\textcolor{blue}{a}), (\textcolor{blue}{b}), and (\textcolor{blue}{c}), having $ s=0 $ correspond to  \reduced diagrams (\textcolor{blue}{d}), (\textcolor{blue}{e}), and (\textcolor{blue}{f}), with $ s=1 $, respectively. The colour factors of (\textcolor{blue}{d}), (\textcolor{blue}{e}) and (\textcolor{blue}{f}) are same as those of (\textcolor{blue}{a}), (\textcolor{blue}{b}) and (\textcolor{blue}{c}) respectively. 
	}
	\label{fig:irreducible-h1}
\end{figure}

In this section, we explore the structure of the block $ A $ of mixing matrices that corresponds to mixing of the irreducible diagrams of a Cweb. In each of these irreducible diagrams, there are at least two correlators which are entangled. We call, an \textit{entangled piece}, the group of correlators for which dragging  any correlator will drag all the other correlators to the origin. Next we introduce the concepts of Fused diagrams and Fused-Webs which will prove to be very useful in understanding the structure of block $ A $.
\\ \\
\textit{\reduced diagram:} The diagram generated by replacing each entangled piece of an irreducible diagram involving $ n $-Wilson lines, with an $ n $-point correlator connecting the same $ n $-lines, is defined as \textit{\reduced diagram} associated with the irreducible diagram. This new $ n $-point correlator is called \textit{\reduced correlator}, which is denoted by the dotted lines\footnote{Denoting these new correlators by gluon lines introduces spurious problem of counting in the number of diagrams for a Cweb.}. These \reduced correlators are essentially gluon correlators. {However,} note that an $ n $-point \reduced correlator does not have the colour structure of an $ n $-point gluon correlator. The illustration of obtaining \reduced diagrams is shown in fig.~(\ref{fig:irreducible-h1}). 
\\ \\
We define $ s $-factor, for the \reduced diagram by following the usual definition of $ s $-factors, as the number of ways in which all the correlators, including the \reduced correlators can be sequentially shrunk to the hard interaction vertex (origin). 

Starting from a \reduced diagram we can generate a {\it fictitious} Cweb in the usual way by shuffling all the attachments on each Wilson line. This Cweb has its own mixing matrix $R_{\text{fict}}$ which can be put in the form of  eq.~(\ref{eq:R-block}) by reordering the diagrams. The mixing between the \reduced diagrams of the {\it fictitious} Cweb that have $s \neq 0$, is given by the corresponding $D_{\text{fict}}$, and this corresponds to the mixing between the respective irreducible diagrams of the original Cweb. It is, therefore, useful to define
\begin{itemize}
	\item []\textit{Fused-Webs}: The set of all reducible ($ s\neq0 $) \reduced diagrams that appear in the fictitious Cweb. 
\end{itemize}

\noindent The mixing matrix of a \reducedWeb is given by $D_{\text{fict}}$. However, note that, even though the mixing matrix $D_{\text{fict}}$ satisfies all the properties of a web mixing matrix that are known to date, \reducedWeb is not really a Cweb as not all the diagrams of the fictitious Cweb are part of it.

We need to ensure that the idea of Fused-Webs is consistent with the replica trick algorithm. Towards this end, 
recall that the replica trick algorithm determines the replica ordered colour factors 
$\textbf{R} \big[ C(d) \big| h \big] $
corresponding to each hierarchy $ h $ for all the diagram in a Cweb (see eq.~\eqref{expocolf}).
Given a hierarchy $h$  the replica ordering operator $\textbf{R}$ disentangles the correlators of entangled piece if the replica number associated with the correlators of entangled piece  are different. Thus, an entangled piece remains entangled only when the correlators are associated with same replica number. Further, the operator $\bf R$ can never entangle two correlators that were not entangled to begin with. Therefore, in \reduced diagrams, it is legitimate to replace these replica variables by a single replica variable.

We can distil the above discussion into an algorithm to determine the diagonal blocks of the matrix $ A $ using Fused-Webs. The steps of the algorithm will be explained in detail afterwards with the help of an explicit example.

\subsection*{Algorithm}

\begin{enumerate}
	\item  Identify the completely entangled diagrams of the Cweb. \reduced diagram of each of these forms a \reducedWeb which has only one diagram. The number of these completely entangled diagrams in a Cweb is the order of identity matrix appearing in the Normal ordered mixing matrix of the Cweb.
	\item  {Identify the \textit{distinct} entangled pieces appearing in partially entangled diagrams, and obtain \reduced diagram for each of these. }
	\item  Shuffle the attachments on each Wilson line of a \reduced diagram to generate the corresponding fictitious Cweb, the reducible diagrams of which will form the associated Fused-Web. 
	\item  Obtain the mixing matrix of the Fused-Web. For this we first find the weight vectors and then use the Uniqueness theorem. This 
	 will correspond to the mixing between the associated partially entangled diagrams. 
		\item Order the diagrams of the Cweb associated with a Fused-Web, such that they appear next to each other. 
	\item Repeat steps 3, 4 and 5 for all distinct entangled pieces in a Cweb in order to compute the diagonal blocks of the matrix $ A $.    
\end{enumerate}

In the next subsection, we show an explicit example of a Cweb, where we apply the above steps to calculate the diagonal blocks of the matrix $ A $. 

\subsection{Application of \reducedWeb}

\begin{figure}[H]
	\captionsetup[subfloat]{labelformat=empty}
	\centering
	\subfloat[][]{\includegraphics[height=4cm,width=4cm]{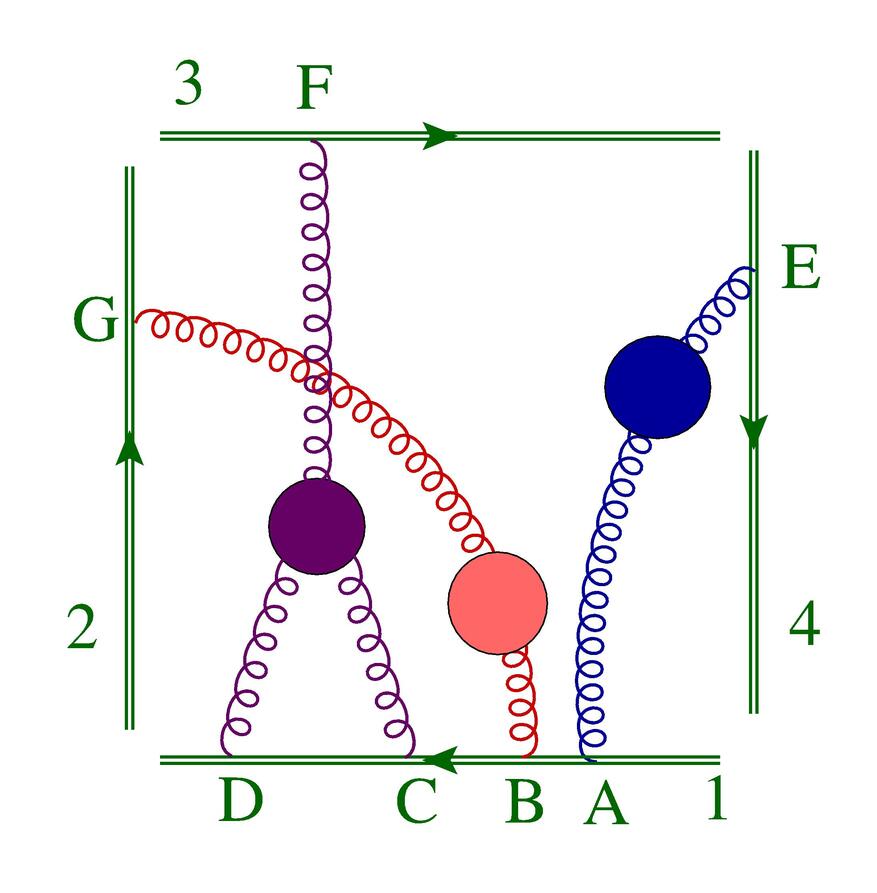} }	
	\caption{A diagram of Cweb $ \text{W}^{(2,1)}_{4} (1,1,1,4) $ }
	\label{fig:EX-boomerang-Avatar-Web}
\end{figure}

Let us study the Cweb $ \text{W}^{(2,1)}_{4} (1,1,1,4) $, shown in fig.~(\ref{fig:EX-boomerang-Avatar-Web}) in the light of the algorithm given above. It has twelve diagrams, out of which six are reducible, two are completely entangled, and four are partially entangled. 
The Normal ordered diagrams and their corresponding shuffles are shown in table (\ref{tab:EX-boomerang-Avatar-Web}).

\begin{table}[b]
	\begin{minipage}[c]{0.5\textwidth}
		\begin{table}[H]
			\begin{center}
				\begin{tabular}{|c|c|c|}
					\hline 
					\textbf{Diagrams}  & \textbf{Sequences}  & \textbf{s-factors}  \\ 
					\hline
					$d_{1}$  & $\{CABD \} $  & 0 \\ \hline
					$d_{2}$  & $\{CBAD \} $  & 0 \\ \hline
					$d_{3}$  & $\{ACBD \} $  & 0 \\ \hline 
					$d_{4}$  & $\{CADB \} $  & 0 \\ \hline
					$d_{5}$  & $\{BCAD \} $  & 0 \\ \hline
					$d_{6}$  & $\{CBDA \} $  & 0 \\ \hline
				\end{tabular}
			\end{center}
		\end{table}
	\end{minipage}
	\hspace{1cm}
	\begin{minipage}[c]{0.5\textwidth}
		\begin{table}[H]
			\begin{center}
				\begin{tabular}{|c|c|c|}
					\hline 
					\textbf{Diagrams}  & \textbf{Sequences}  & \textbf{s-factors}  \\ 
					\hline
					$d_{7}$  & $\{ABCD\} $  & 1 \\ \hline
					$d_{8}$  & $\{ACDB \} $  & 1 \\ \hline
					$d_{9}$  & $ \{BACD \}$  & 1 \\ \hline 
					$d_{10}$  & $\{BCDA \} $  & 1\\ \hline
					$d_{11}$  & $\{CDAB \} $  & 1 \\ \hline
					$d_{12}$  & $\{CDBA \} $  & 1 \\ \hline
				\end{tabular}
			\end{center}
		\end{table}
	\end{minipage}
	\caption{Normal ordered diagrams of Cweb $ W^{(2,1)}_{4} (1,1,1,4) $}
	\label{tab:EX-boomerang-Avatar-Web}
\end{table}

We begin by identifying all possible distinct entangled pieces of the Cweb. We observe that there are three distinct entangled pieces: 

\begin{figure}[tp]
	\captionsetup[subfloat]{labelformat=empty}
	\centering
	\subfloat[][(A)]{\includegraphics[height=4cm,width=4cm]{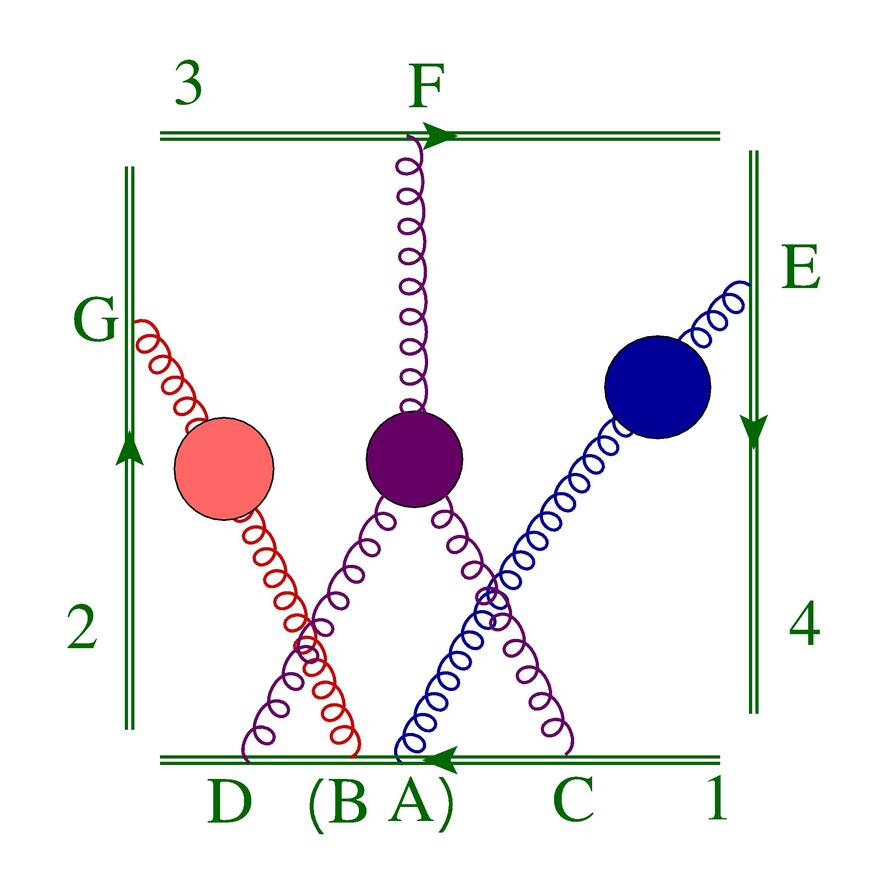} }
	\qquad 
	\subfloat[][(B)]{\includegraphics[height=4cm,width=4cm]{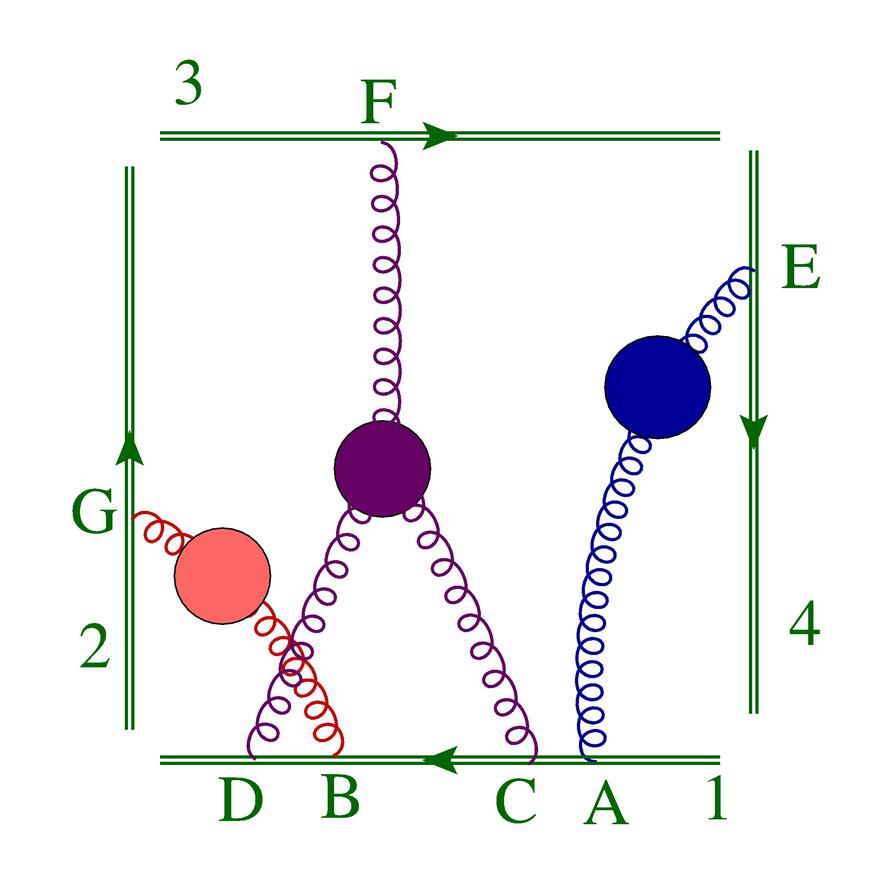} }
	\qquad 
	\subfloat[][(C)]{\includegraphics[height=4cm,width=4cm]{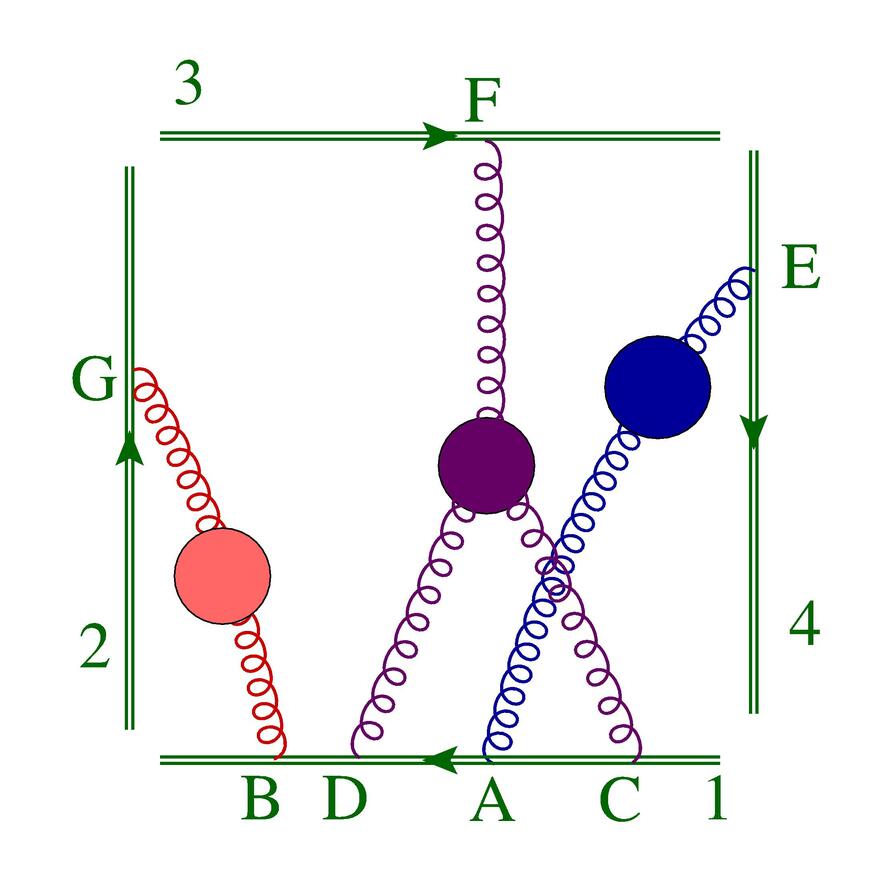} }
	\\
	\subfloat[][(a)]{\includegraphics[height=4cm,width=4cm]{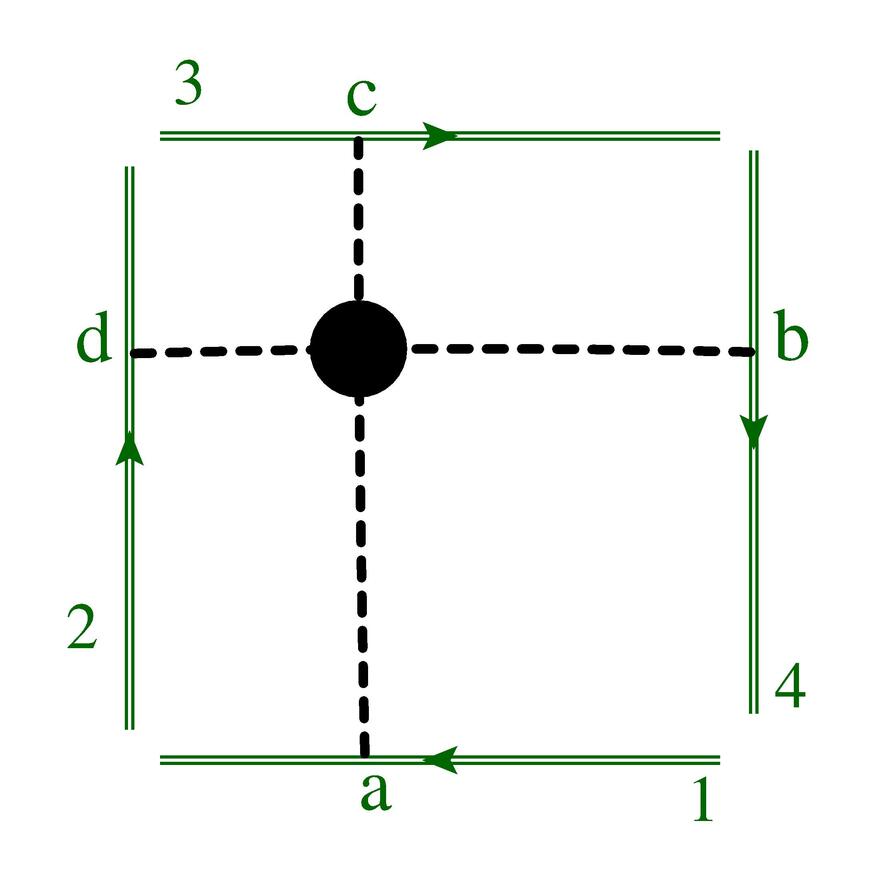} }
	\qquad 
	\subfloat[][(b)]{\includegraphics[height=4cm,width=4cm]{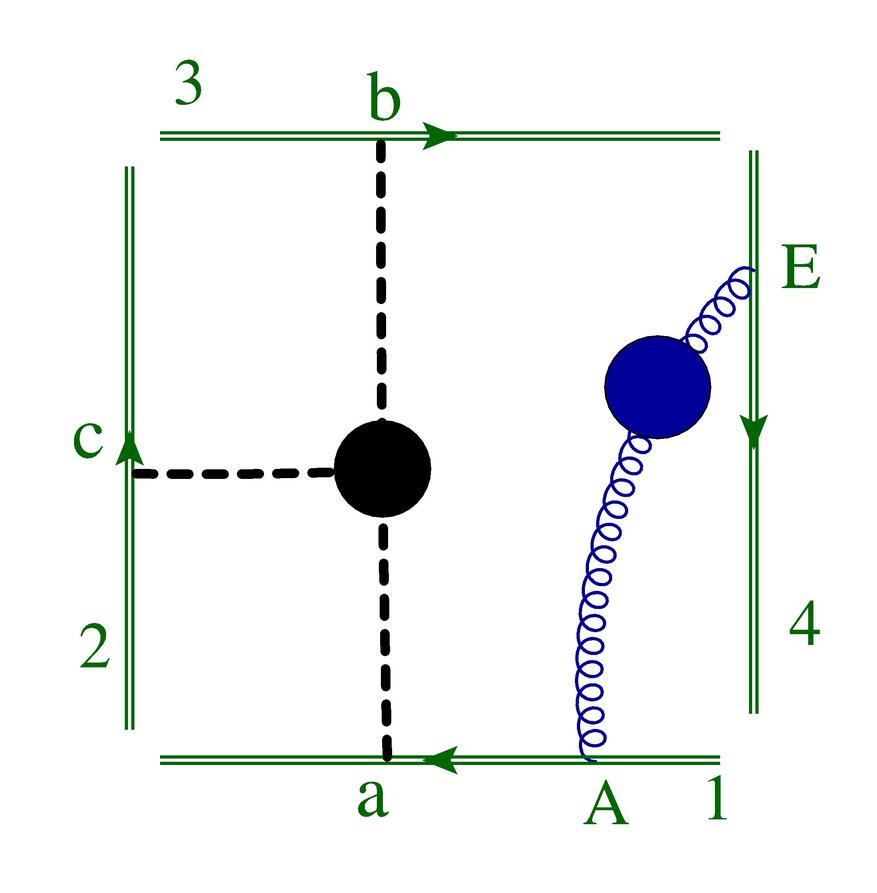} }
	\qquad 
	\subfloat[][(c)]{\includegraphics[height=4cm,width=4cm]{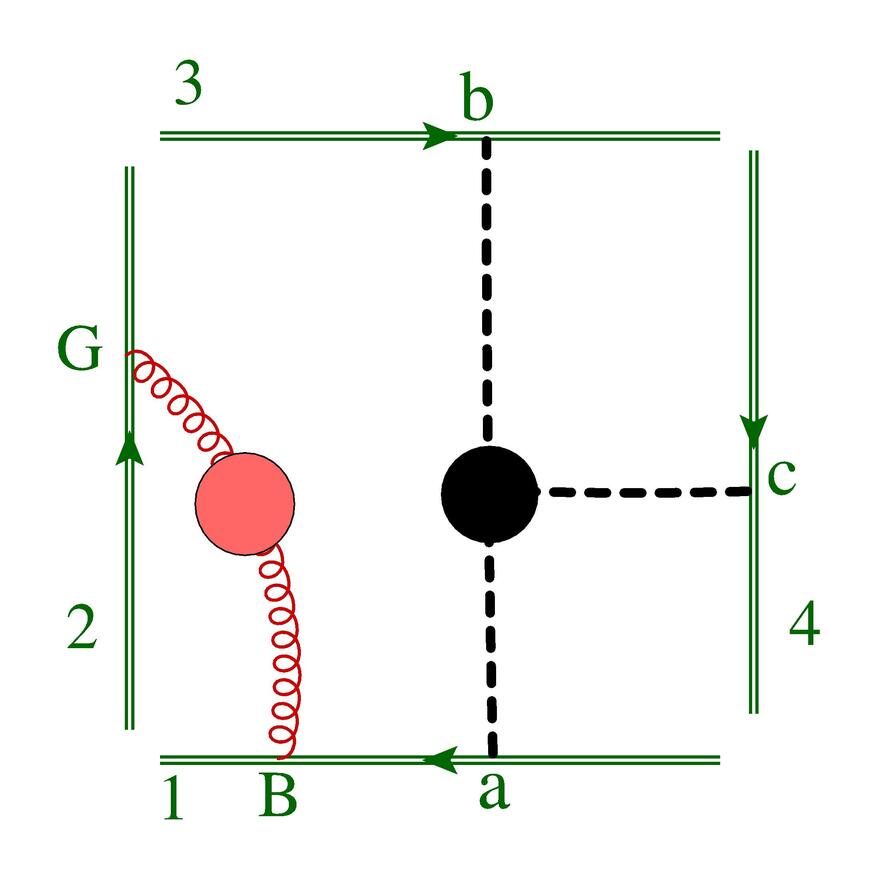} }
	\caption{\reducedWebs for Cweb $ \text{W}^{(2,1)}_{4} (1,1,1,4) $ }
	\label{fig:Avatar-webs-ex}
\end{figure}
\begin{itemize}
	\item  First type is displayed in fig.~(\ref{fig:Avatar-webs-ex}\textcolor{blue}{A}), where both the two point correlators are attached in between the two attachments of the three point correlator on line $ 1 $. Here $ \left(BA\right) $ in the diagram in fig.~(\ref{fig:Avatar-webs-ex}\textcolor{blue}{A}), indicates that both the orderings of attachments, $\{CABD\} $ and $ \{CBAD\} $, belong to this type of entangled piece. These diagrams, $ d_1 $, and $ d_2 $, shown in table (\ref{tab:EX-boomerang-Avatar-Web}), are completely entangled and have the same \reduced diagram, which is  obtained by replacing the entangled piece with a four point \reduced correlator, as shown in fig.~(\ref{fig:Avatar-webs-ex}\textcolor{blue}{a}).

	\item Second type appears in the partially entangled diagram, shown in fig.~(\ref{fig:Avatar-webs-ex}\textcolor{blue}{B}), where the three-point correlator is entangled with the red two-point correlator. This kind of entangled piece appears in two diagrams, with order of attachments on line 1 as: $ \{ACBD\} $, and $ \{CBDA\} $, which are $ d_3 $, and $ d_6 $ in table (\ref{tab:EX-boomerang-Avatar-Web}) respectively. The \reduced diagram corresponding to the diagram $ d_3 $, with order of attachments as $ \{ACBD\} $, is shown in fig.~(\ref{fig:Avatar-webs-ex}\textcolor{blue}{b}). Similarly, \reduced diagram corresponding to $ d_6 $ having the order of attachments $ \{CBDA\} $ can also be obtained.  
	
	\item The third and final kind of entangled piece appears in diagram shown in fig.  (\ref{fig:Avatar-webs-ex}\textcolor{blue}{C}), where the three-point correlator is entangled with the blue two-point correlator, whose \reduced diagram is shown in fig.  (\ref{fig:Avatar-webs-ex}\textcolor{blue}{c}). This kind also belongs to two diagrams having the order of attachments $\{CADB\} $, and $ \{BCAD\} $,  which are $ d_4 $, and $ d_5 $ of table \ref{tab:EX-boomerang-Avatar-Web} respectively.
\end{itemize}

\noindent We discuss below the Fused-Webs generated by the three distinct entangled pieces in order.

 For the \reduced diagram, shown in fig.~(\ref{fig:Avatar-webs-ex}\textcolor{blue}{a}), all the Wilson lines have single attachments, thus the corresponding \reducedWeb contains only one diagram. Hence, the contribution of the two Fused-Webs, corresponding to $ d_1 $ and $ d_2 $, will appear as identity matrix of order two in the mixing matrix given in eq.~(\ref{eq:ex-avatar-web-matrix}). 

 In second \reduced diagram (\ref{fig:Avatar-webs-ex}\textcolor{blue}{b}), generated by second kind of entangled piece, the shuffle on line $ 1 $ generates two diagrams with order of attachments $ \{Aa\} $, and $ \{aA\} $. Following the definition of $ s $-factor of a \reduced diagram, both these diagrams will have $ s=1 $, and together they form a Fused-Web. The mixing matrix of this \reducedWeb  is unique and was computed directly in \cite{Agarwal:2021him}, and is given by, 
\begin{align}
R(1,1)=R\,(1_2)=\frac{1}{2}\left(\begin{array}{cc}
1 & -1 \\
-1 & 1
\end{array}\right)\,.
\label{eq:universalR1_2}
\end{align} 
This matrix generates mixing between the diagrams $ d_3 $, and $ d_6 $  of the Cweb. Choosing the appropriate order of diagrams of the Cweb --- placing $ d_3 $ and $ d_6 $ next to each other, these entries will show up as a two dimensional block $ R(1_2) $. 

The shuffle on line $ 1 $ of third \reduced diagram (\ref{fig:Avatar-webs-ex}\textcolor{blue}{c}), generates two diagrams with order of attachments $ \{aB\} $, and $ \{Ba\} $. Similar to the previous case these two diagrams, having $ s=1 $, also form a \reducedWeb with mixing matrix $ R(1_2) $. This matrix appears as a diagonal block and generates mixing between the diagrams $ d_4 $, and $ d_5 $  of the Cweb.

If Normal ordered diagrams are further ordered such that the partially entangled  diagram $ d_6 $ appears next to $ d_3 $, and $ d_5 $ next to $ d_4 $,
then the mixing matrix for this Cweb will be, 
\begin{align}
R=\left(\begin{array}{c|c}
\begin{array}{c|cc}
\text{I}_{2} & & A_U\\\hline
\text{O}_{4\times 2}& & \begin{array}{cc}
R\,(1_2) & X \\
\text{O}_{2\times 2}& R\,(1_2)
\end{array}	
\end{array} & B \\ 
\hline
\text{O}_{6\times6} & D
\end{array}\right)\,. 
\label{eq:R-LL}
\end{align}
Here $ X $ is null matrix of order two for this case. However, it is not true in general as the action of replica ordering operator on a diagram with one kind distinct entangled piece maps it to that of another kind. From the foregoing discussion it is evident that, the diagonal blocks of any web mixing matrix will always be the matrices of basis Cwebs, except for the class mentioned in the end of the section \ref{sec:Uniqueness-theorem}. 
 
The explicit form of the mixing matrix of the Cweb $ \text{W}^{(2,1)}_{4} (1,1,1,4) $ after ordering the diagrams in the fashion mentioned above is

\begin{align}
R=\frac{1}{6} 
\left(
\begin{array}{cccccccccccc}
6 & 0 & -3 & -3 & -3 & -3 & -1 & 2 & 2 & -1 & 2 & 2 \\
0 & 6 & -3 & -3 & -3 & -3 & 2 & -1 & 2 & 2 & 2 & -1 \\
0 & 0 & 3 & -3 & 0 & 0 & -1 & -1 & -1 & -1 & 2 & 2 \\
0 & 0 & -3 & 3 & 0 & 0 & 2 & -1 & 2 & -1 & -1 & -1 \\
0 & 0 & 0 & 0 & 3 & -3 & -1 & -1 & 2 & 2 & -1 & -1 \\
0 & 0 & 0 & 0 & -3 & 3 & -1 & 2 & -1 & -1 & 2 & -1 \\
0 & 0 & 0 & 0 & 0 & 0 & 2 & -1 & -1 & -1 & -1 & 2 \\
0 & 0 & 0 & 0 & 0 & 0 & -1 & 2 & -1 & 2 & -1 & -1 \\
0 & 0 & 0 & 0 & 0 & 0 & -1 & -1 & 2 & -1 & 2 & -1 \\
0 & 0 & 0 & 0 & 0 & 0 & -1 & 2 & -1 & 2 & -1 & -1 \\
0 & 0 & 0 & 0 & 0 & 0 & -1 & -1 & 2 & -1 & 2 & -1 \\
0 & 0 & 0 & 0 & 0 & 0 & 2 & -1 & -1 & -1 & -1 & 2 \\
\end{array}
\right)\,,
\label{eq:ex-avatar-web-matrix}
\end{align}
which agrees with eq.~(\ref{eq:R-LL}). The $ D $ block of this mixing matrix is $ R(1_6) $ given in appendix \ref{sec:basis}.
The above example shows how the basis matrices appear in a mixing matrix of a Cweb whose column weight vector has one or more zero entries.

\section{Direct construction of two special classes of Cwebs}
\label{sec:explicit}
Attempts have been made to understand the structure of mixing matrices and ideas from Combinatorics such as {\it posets} have proven useful in making all order predictions 
for certain special classes of webs  \cite{Dukes:2013gea,Dukes:2013wa,Dukes:2016ger}.  A systematic approach towards unravelling these structures was initiated in \cite{Agarwal:2021him}, where all the prime dimensional mixing matrices at any order in perturbation theory were obtained using the known properties of the matrices that are listed in section \ref{sec:conje}. Now, armed with the 
learnings of sections \ref{sec:conje} and \ref{sec:\reduced} we push this program further. Our starting point is eq.~\eqref{eq:R-block}, which we rewrite below:
\begin{align}
\tag{\ref{eq:R-block}}
R=\left(\begin{array}{cc}
A & B \\
O & D
\end{array}\right)\,,
\end{align}
in which the block $ D $ is associated with the mixing of only reducible diagrams whereas the block $A$ is associated with the mixing of only irreducible diagrams in a Cweb.
Further in section \ref{sec:D-properties} we showed that the block $ D $ follows all the known properties of web mixing matrices and this fact enables us to write its explicit form using the mixing matrices provided in previous works \cite{Agarwal:2020nyc,Agarwal:2021him,Gardi:2010rn,Gardi:2011wa,Gardi:2011yz,Gardi:2013ita,Gardi:2021gzz}. 
Also we have established the structure of the diagonal blocks of $A$ using the idea of  \reducedWebs in section \ref{sec:\reduced}.
The above form is sufficient to allow us to obtain the rank of the mixing matrix $R$ once we know the diagonal blocks of $A$, and $D$ which are basically the mixing matrices of basis Cwebs. This gives us the total number of independent exponentiated colour factors of that Cweb. To see this we first note that after we have normal ordered the diagrams 
the property of idempotence, $ R^2=R $, tells us that 
\begin{align}
A^2 = A \, , \quad D^2 = D\,, \quad \warning{AB + BD = B\,.}
\label{eq:ref-idea}
\end{align}
Now, the rank of an idempotent matrix is same as its trace since the eigenvalues are only 0 and 1. Therefore the rank of $ R $ is the sum of diagonal entries of $ A $, and $ D $. This gives the rank of $ R $ as the sum of rank $r(A)$ of $ A $  and rank $r(D)$ of  $ D $:

\begin{align}
\label{eq:Rank-formula}
r(R) = r(A) + r(D)\,.
\end{align}
We can utilize this result  to determine the rank without knowing the full form of $R$. 
\warning{In this article, we only utilize the idempotence of $ A $ and $ D $, however, the third relation in eq.~\eqref{eq:ref-idea} provides additional constraints on the elements of mixing matrices which will be explored in the future.\footnote{We thank the anonymous referee for pointing out these additional conditions.}}

At one loop the mixing matrix is identity matrix; at two loops there is only one basis Cweb,  $R(1_{2})$;  at three loops we get two additional basis Cwebs with matrices  $R(1_{6})$ and   $R(1_{2}, 2_{2})$. All these matrices  can appear as $D$ and diagonal blocks of $A$ at four-loops and beyond. 

In section \ref{sec:Direct-cons} we will present all the Cwebs at four loops that contain in their respective $ D $ blocks, the matrices corresponding to the Basis Cwebs upto three loops.
Before that we will first  take up two special classes 
$S=\{0,0,\ldots,0,1_1\} $, and  $ S=\{0,0,\ldots,0,1_2\} $ in turn as we can completely determine their corresponding mixing matrices.

\subsection{Cwebs with $ S=\{0,0,\ldots,0,1_1\} $}\label{sec:DC-class}

Every Cweb belonging to the class for which weight vector $ S=\{0,0,\ldots,0,1_1\} $ has two or more identical correlators that connect the two Wilson lines. 
The Wilson line 1 and line 2  have $k$ and $r$  attachments from each of the correlators, respectively.
The shuffle of the attachments on both the Wilson lines generate only one reducible diagram and all other diagrams are either completely or partially entangled.
\begin{figure}
	\vspace{10mm}
	\captionsetup[subfloat]{labelformat=empty}
	\centering
	\subfloat[][]{\includegraphics[scale=0.15]{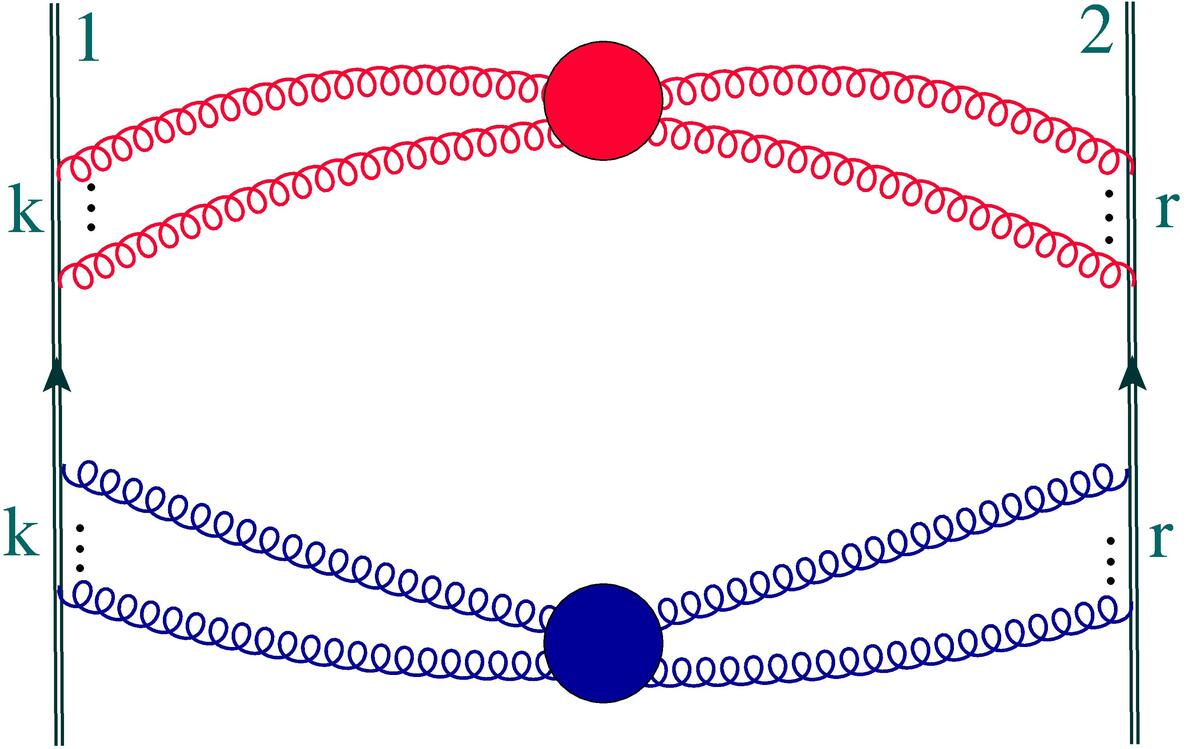} }	
	\caption{General Cweb with two correlators having only one reducible diagram.}
	\label{fig:DoubleCounting-Gen}
\end{figure}
Now we consider a Cweb of the above class which has $l$ diagrams. 
The Normal ordered  matrix has the general form given in eq.~(\ref{eq:R-block}).
The order of the null matrix is $ 1\times(l-1) $, which implies $ D $ is a number, which we will denote by $d$,  and $ B $ is a column vector of dimension $ l-1$.  
Applying zero-row sum rule to the last row, we get
\begin{align}
d\,=\,0\,.
\end{align}
The general form of mixing matrix for this class is, thus
\begin{align}
R=\left(\begin{array}{cc}
A & B \\
O & 0
\end{array}\right)\,. 
\label{eq:R-block-s1-final}
\end{align}
From this we immediately obtain that the rank of mixing matrices for this class of Cwebs is equal to the rank of the matrix $ A $ 
\begin{align}
r (R)\,&=\,r(A)+r(D)\,=\, r(A) \,.
\label{eq:Dblock-DC-zero}
\end{align}

\noindent Though, we can not predict all the matrix elements of $R$  for this class in general, we can completely construct $R$ for a subclass which has only two identical gluon correlators as shown in  fig.~(\ref{fig:DoubleCounting-Gen}). These Cwebs are present only at even loop orders
and their all the $ l-1$ irreducible diagrams are completely entangled.
As we have seen in earlier sections the completely entangled diagrams will give an identity matrix of order $ l-1 $;  the mixing matrix for this subclass, thus, reduces to
\begin{align}
R=\left(
\begin{array}{cccccc}
1 & 0 & 0 & 0 & 0 & b_1 \\
0 & 1 & 0 & 0 & 0 & b_2 \\
\vdots & \vdots & \vdots & \vdots & \vdots & \vdots \\
0 & 0 & 0 & 0 & 1 & b_{l-1} \\
0 & 0 & 0 & 0 & 0 & 0 \\
\end{array}
\right)\,. 
\label{eq:R-s1-DC-gen}
\end{align} 
The remaining elements are fixed to by applying zero-row sum rule:
\begin{align}
b_i=-1 \qquad\qquad\qquad\forall \quad 1 \leq i \leq {l-1}\,.
\end{align}
We now have, for this subclass, the complete mixing matrix  $ R $ and its rank without using the replica trick 
\begin{align}
R=\left(
\begin{array}{cccccc}
1 & 0 & 0 & 0 & 0 & -1 \\
0 & 1 & 0 & 0 & 0 & -1 \\
\vdots & \vdots & \vdots & \vdots & \vdots & \vdots \\
0 & 0 & 0 & 0 & 1 & -1 \\
0 & 0 & 0 & 0 & 0 & 0 \\
\end{array}
\right)\,,\qquad\qquad r(R)= l-1 \,. 
\label{eq:R-s1-DC-gen-final}
\end{align} 
Note that there are two Cwebs $ W_2^{(2)}(2,2) $ and $ W_2^{(0,2)}(2,4) $ of this subclass present at two and four loops respectively. Cweb $ W_2^{(2)}(2,2) $ has two diagrams, thus, it will have two dimensional matrix $R $ of the form above, while $ W_2^{(0,2)}(2,4) $ has six diagrams, and will have the corresponding six dimensional matrix.
These results agree with the explicit matrices for these webs obtained in ~\cite{Agarwal:2021him,Gardi:2010rn}.

The next Cwebs belonging to this subclass will occur at six loops in the perturbation theory.  
We predict that at six loops there will be two Cwebs: $ W_{2}^{(0,0,2)}(4,4) $, and $ W_{2}^{(0,0,2)}(2,6) $, having eighteen and twenty diagrams respectively, 
and their corresponding mixing matrices will be of the form eq.~(\ref{eq:R-s1-DC-gen-final}).
The number of Cwebs of this subclass at order  $\alpha_{s}^{n}$, where $n$ is an even integer is given by 
\begin{align}
\text{Floor}\left[\frac{1}{2} + \frac{n}{4}\right].
\end{align}
The above sequence is known as Non-negative Integers Repeated \cite{NnrSeq}. The elements of the sequence are $1,1,2,2,3,3,4,4,5,5, 6,6,  \cdots$.
\subsection{Octopus-Pair Cwebs with $ S=\{0,0,\ldots,0,1_2\} $ }
\label{sec:singlet-Avatar}
\begin{figure}
	\vspace{10mm}
	\captionsetup[subfloat]{labelformat=empty}
	\centering
	\subfloat[][]{\includegraphics[scale=0.15]{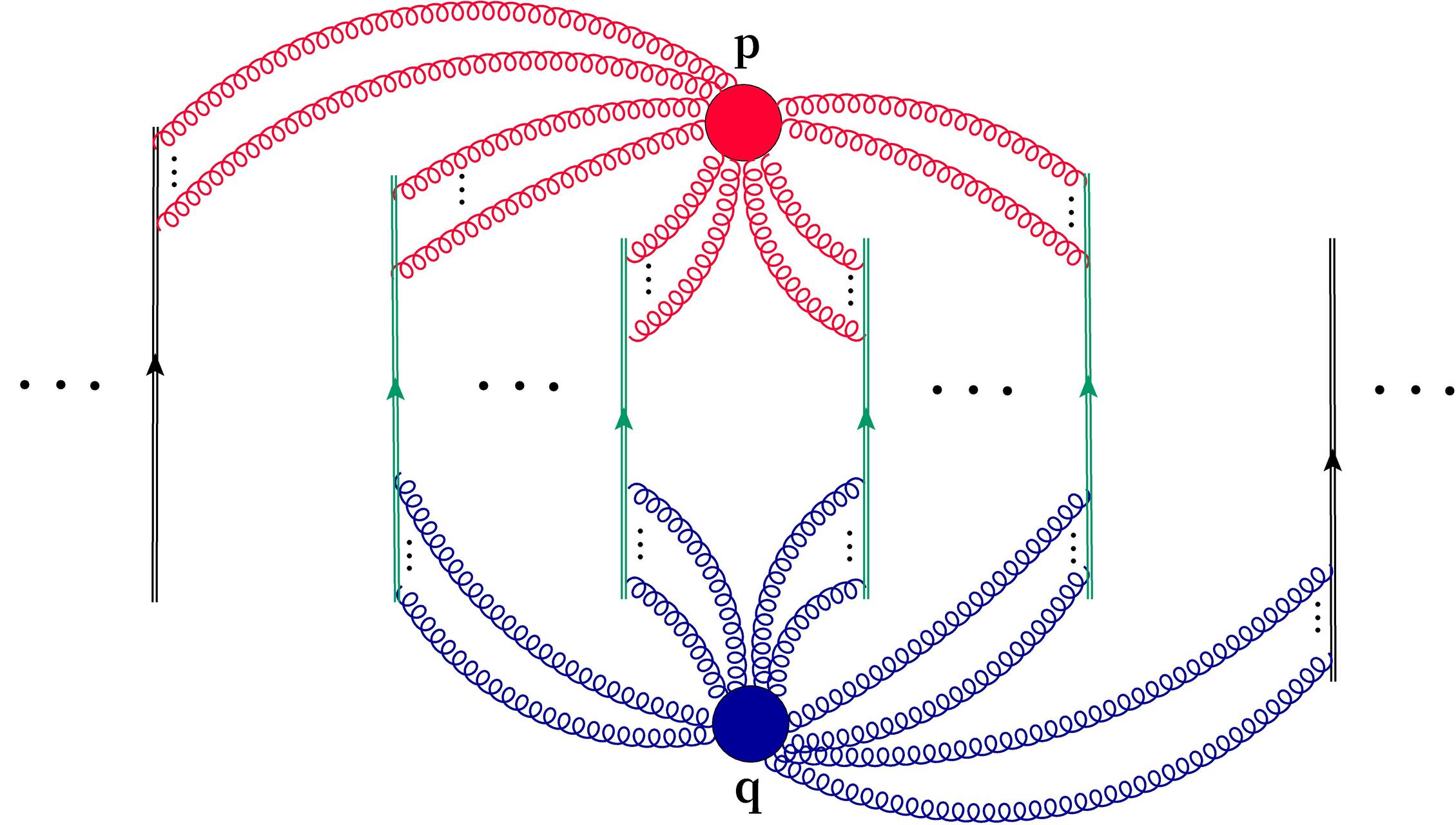} }	
	\caption{One of the two reducible diagrams of the  Cweb belonging to the class for which $ S=\{0,0,\ldots,0,1_2\} $. All other irreducible diagrams are completely entangled.}
	\label{fig:direct-C-singlet}
\end{figure}

Any Cweb which has weight vector $ S=\{0,0,\ldots,0,1_2\} $ is of the configuration displayed in fig.~(\ref{fig:direct-C-singlet}) and it consists 
two distinct gluon correlators. We call these as  {\it Octopus-Pair} Cwebs.
It is easy to see that all the diagrams of this Cweb are completely entangled except two which are reducible. We obtain the two reducible diagrams when all blue coloured gluons are placed either above or below the red gluons on the Wilson lines. Furthermore,
note that for each of these two diagrams there is only one way to sequentially shrink the two correlators to the hard interaction vertex and hence, $ s $-factors for the reducible diagrams are equal to one.  Therefore, after normal ordering the column weight vector for this class of Cwebs is given by, $ S=\{s(d_1), s(d_2), s(d_3), \ldots, s(d_n)\}\,=\,\{0,0,0,\ldots, 0,1,1\} \,$.  The block $D$ of the mixing matrix is fixed to be $ R\,(1_2) $ by the Uniqueness theorem.

Next, we determine the block $A$. As all the irreducible diagrams are completely entangled, the matrix $ A $ is an identity matrix of order $ (n-2) $, that is,  
\begin{align}
A\,=\,\textbf{I}_{(n-2)\times (n-2)}\,.  
\end{align}
Thus the mixing matrix for this type of Cwebs with $ n $ diagrams is constrained to the following form:
\begin{align}
R\,=\,\left(\begin{array}{ccc}
\textbf{I}_{n-2} & & \begin{array}{cc}
c_{1}&  b_{1} \\ 
c_{2} &   b_{2} \\
\vdots & \vdots \\
c_{n-2} &   b_{n-2}
\end{array}
\\ \\
\textbf{O}_{2\times (n-2)}& & \begin{array}{c}
R\,(1_2)
\end{array}	 \\
& & 
\end{array}\right)\,.
\end{align}
The rank of $ R\,(1_2) $ is one and the rank of $ A $ is equal to $n-2$, the rank of the mixing matrix is thus,
\begin{align}
r(0,0,\ldots,1_2)\,=\,r(A)+ 1 = n-1,
\end{align}  
that is, these Cwebs have $n-1$  independent exponentiated colour factors.

The next step is to determine $ c_i $ and $ b_i $ using the properties of the mixing matrices. The row-sum rule imposes
\begin{align}
b_j\,&=-1-c_j, \mkern-18mu & 1 \leq j \leq (n-2), 
\label{eq:row-sum}	
\end{align}
and using idempotence property of mixing matrices we get
\begin{align}
c_j=-\frac{1}{2}\quad 1\leq j \leq (n-2) \,.
\end{align}
With this we have uniquely fixed all the elements of the mixing matrix for this class of  Cwebs, and it  is then given by
\begin{align}
R\,=\,\left(\begin{array}{cccccccc}
1&0& 0 &\ldots & 0 & \, & -1/2 &  -1/2  \\
0&1& 0 &\ldots & 0 & \, &  -1/2 &   -1/2 \\
0&0& 1 &\ldots & 0 & \, &  -1/2 &   -1/2 \\
\vdots &\vdots & \vdots &\vdots & \vdots & \vdots &  \vdots &\vdots \\
0&0& 0 &\ldots & 1 & \, & -1/2 &   -1/2 \\
0&0& 0 &\ldots & 0 & \, & 1/2 &   -1/2 \\
0&0& 0 &\ldots & 0 & \, & -1/2 &   1/2 \\
\end{array}\right)\,.
\label{eq:R-gen-final}
\end{align}

A subclass of the webs shown in fig.~\eqref{fig:direct-C-singlet} was considered in~\cite{Agarwal:2021him}. We observe that eleven non-prime dimensional, and four prime dimensional mixing matrices belong to this class of Cweb for massless Wilson lines at four loops that were calculated in  \cite{Agarwal:2020nyc,Agarwal:2021him}. These are listed in table (\ref{tab:non-prime-singlet2}).
\begin{table}[t]
	\begin{center}
		\begin{tabular}{|c|l|c|c|c|}
			\hline 
			& & &  &\\
			Number of Diagrams   &$ \quad \quad$ Cweb & Perturbative order & Mixing matrix $ R $ & rank r(R) \\
			\hline  
			& &	&								&	\\
			& $ W_{3}^{(1,1)}(1,1,3) $ & $ \mathcal{O}(g^6) $ & &\\
			\cline{2-3}
			&&&&\\
			& $ W_{3}^{(1,1)}(1,2,2) $ & $ \mathcal{O}(g^6) $ & &\\
			\cline{2-3}
			&&&&\\
			& $ W_{4}^{(1,0,1)}(1,1,1,3) $ & $ \mathcal{O}(g^8) $ & &\\
			\cline{2-3}
			3 & 						&				& $ R(0_1,1_2)$ & $2$\\
			& $ W_{4}^{(0,2)}(1,1,1,3) $ & $ \mathcal{O}(g^8) $ &  &\\
			\cline{2-3}
			& 							&		& &	\\
			& $W_{3,\text{I}}^{(0,2)}(1,2,3)$& $ \mathcal{O}(g^8) $ & &\\
			\cline{2-3}
			& 								& & 	&	\\
			& $W_{3}^{(1,0,1)}(1,2,3)$& $ \mathcal{O}(g^8) $ & &\\ 
			\hline
			&     &                    &       &       \\
			& $ W_{4,\text{I}}^{(1,0,1)}(1,1,2,2) $ & $ \mathcal{O}(g^8) $ & &\\
			\cline{2-3}
			4& 									&	&$ R(0_2,1_2)$ & $3$\\
			& $ W_{4,\text{II}}^{(0,2)}(1,1,2,2) $& $ \mathcal{O}(g^8) $ & &\\
			\cline{2-3}
			& 									&	& &\\
			& $W_{3}^{(1,0,1)}(2,2,2)$& $ \mathcal{O}(g^8) $ & &\\
			\cline{2-3}
			& 						&	& &			\\
			& $W_{3}^{(1,0,1)}(1,1,4)$& $ \mathcal{O}(g^8) $ & &\\ 
			\hline
			& 					&		&	&		\\
			& $ W_{3,\text{I}}^{(1,0,1)}(1,2,3) $& $ \mathcal{O}(g^8) $ & &\\
			\cline{2-3}
			6& 							&			&$ R(0_4,1_2)$ & $5$\\
			& $ W_{3,\text{II}}^{(1,0,1)}(1,2,3) $&$ \mathcal{O}(g^8) $ & &\\
			\cline{2-3}
			& 									&	& &\\
			& $ W_{3}^{(0,2)}(1,1,4) $ & $ \mathcal{O}(g^8) $& &\\
			\hline 
			& & & &\\
			8	& $ W_{2}^{(1,0,1)}(2,4) $ & $ \mathcal{O}(g^8) $ &  $ R(0_6,1_2) $& 7 \\
			\hline 	&	& 				&					&	\\
			9		&  $ W_{2}^{(1,0,1)}(3,3) $ & $ \mathcal{O}(g^8) $ &$ R(0_7,1_2)$ & $8$\\
			\cline{2-3}
			& 									& & &	\\
			&	$W_{2}^{(0,2)}(3,3)$ &$ \mathcal{O}(g^8) $ & &\\
			\hline 
		\end{tabular}
	\end{center}
	\caption{Octopus-Pair Cwebs upto four loops that connect massless Wilson lines.}
	\label{tab:non-prime-singlet2}
\end{table}

\section{Determining the number of independent ECFs for Cwebs} 
\label{sec:Direct-cons}
In the previous section we showed how our formalism of \reducedWebs together with the Uniqueness theorem of section \ref{sec:conje} can completely
determine the two classes of Cwebs to all orders in perturbation theory.
Now, in this section we will use  Fused-Web formalism and the Uniqueness theorem to construct the diagonal blocks of mixing matrices of a Cweb at order  $\alpha_{s}^{{n+1}}$ using basis Cwebs that appear upto  order $\alpha_{s}^{n}$. We have seen that  there are three basis Cwebs shown in fig.~(\ref{fig:basisWebs}), and table \ref{tab:conjecture-table} with the mixing matrices $R(1_2)$, $R(1_6)$ and $R(1_2,2_2)$ that appear upto three loops for 
massless Wilson lines. We can use these as the building blocks and construct three sets of Cwebs that contain these mixing matrices as their respective $D$ blocks. These three classes of Cwebs have weight vectors $ S=\{0,0,\ldots,0,1_2\} $,  $ S=\{0,0,\ldots,0,1_6\} $, and  
$ S=\{0,0,\ldots,0,1_{2},2_{2}\} $.

As a demonstration of the concept we will pick a Cweb at four loops for each of these three classes and apply the \reducedWebs formalism to construct the diagonal blocks of $A$,  which will again correspond to the three basis  Cwebs. This will allow us to predict the rank and  thus the total number of independent exponentiated colour factors that these Cwebs have without going through the complete computation by a code based replica trick algorithm.

\subsection{Cwebs with  $ S=\{0,0,\ldots,0,1_2\} $}
We have solved this class completely --- the mixing matrix is given in eq.~(\ref{eq:R-gen-final}) and there are $n-1$ independent exponentiated colour factors if there are $n$ diagrams in the Cweb.

\subsection{Cwebs with $ S=\{0,0,\ldots,0,1_6\} $}
\begin{figure}[H]
	\captionsetup[subfloat]{labelformat=empty}
	\centering
	\subfloat[][]{\includegraphics[height=4cm,width=4cm]{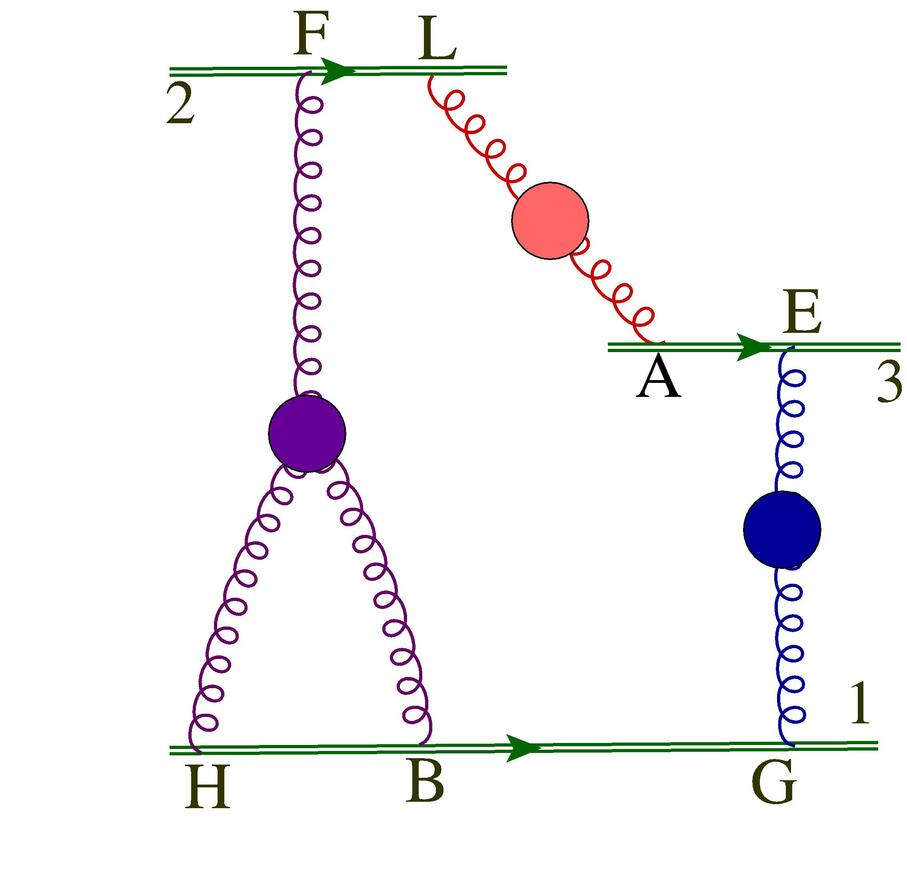} }	
	\caption{A diagram of Cweb $ W_3^{(2,1)}(2,2,3) $ }	\label{fig:EX-SixOne-Avatar-Web}
\end{figure}

In this class of Cwebs, there are six reducible diagrams, each with $ s=1 $. From the Uniqueness theorem given in sec. (\ref{sec:conje}), the general form of $ D $ corresponds to a six dimensional mixing matrix $ R\,(1_6) $, with $ S=\{1_6\} $, shown in appendix \ref{sec:basis}. This matrix appears first at three loops, and was first presented in \cite{Gardi:2013ita}. The rank of this mixing matrix is two, which tells us that the rank of mixing matrix for this class of Cwebs will be, 
\begin{align}
r(R)\,=\,r(A)+2\,.
\label{eq:rank-I-6}
\end{align}  
We show an explicit example of this class using the Cweb $ W_3^{(2,1)}(2,2,3) $ shown in fig.~(\ref{fig:EX-SixOne-Avatar-Web}), present at four loops.
\noindent This Cweb has twelve diagrams out of which six are reducible, each with $ s=1 $. The order of diagrams shown in table (\ref{tab:EX-S-sixone-Avatar-Web}),

\begin{table}[H]
	\begin{minipage}[c]{0.5\textwidth}
		\begin{table}[H]
			\begin{center}
				\begin{tabular}{|c|c|c|}
					\hline 
					\textbf{Diagrams}  & \textbf{Sequences}  & \textbf{s-factors}  \\ 
					\hline
					$d_{1}$  & $\{FL\},\{AE\},\{GHB\}$  & 0 \\ \hline
					$d_{2}$  & $\{LF\},\{EA\},\{HBG\}$  & 0 \\ \hline
					$d_{3}$  & $\{FL\},\{AE\},\{HGB\}$& 0 \\ \hline 
					$d_{4}$  & $\{LF\},\{EA\},\{HGB\}$  & 0 \\ \hline
					$d_{5}$  & $\{FL\},\{EA\},\{HGB\}$  & 0 \\ \hline
					$d_{6}$  & $\{LF\},\{AE\},\{HGB\}$  & 0 \\ \hline
				\end{tabular}
			\end{center}
		\end{table}
	\end{minipage}
	\hspace{1cm}
	\begin{minipage}[c]{0.5\textwidth}
		\begin{table}[H]
			\begin{center}
				\begin{tabular}{|c|c|c|}
					\hline 
					\textbf{Diagrams}  & \textbf{Sequences}  & \textbf{s-factors}  \\ 
					\hline
					$d_{7}$  & $\{FL\},\{AE\},\{HBG\}$  & 1 \\ \hline
					$d_{8}$  & $\{LF\},\{AE\},\{HBG\}$  & 1 \\ \hline
					$d_{9}$  &  $\{FL\},\{EA\},\{HBG\}$  & 1 \\ \hline 
					$d_{10}$  & $\{FL\},\{EA\},\{GHB\}$  & 1\\ \hline
					$d_{11}$  & $\{LF\},\{EA\},\{GHB\}$  & 1 \\ \hline
					$d_{12}$  & $\{LF\},\{AE\},\{GHB\}$  & 1 \\ \hline
				\end{tabular}
			\end{center}
		\end{table}
	\end{minipage}
	\caption{Normal ordered diagrams of Cweb $ W_3^{(2,1)}(2,2,3) $}
	\label{tab:EX-S-sixone-Avatar-Web}
\end{table}

$ D $-block of mixing matrix for this Cwebs is $ R(1_6) $. Now we will use the procedure of \reducedWebs described in section (\ref{sec:avatar-webs}) to find the diagonal blocks of $ A $. The diagrams, $ d_1 $, $ d_2 $, $ d_3 $, and $ d_4 $, given in table (\ref{tab:EX-S-sixone-Avatar-Web}), are completely entangled diagrams. Each of these form a \reducedWeb with single diagram, therefore we get the identity matrix of order four. There is only one kind of entangled piece in the partially entangled diagrams $ d_5 $, and $ d_6 $.
\begin{figure}[H]
	\centering
	\subfloat[][]{\includegraphics[height=4cm,width=4cm]{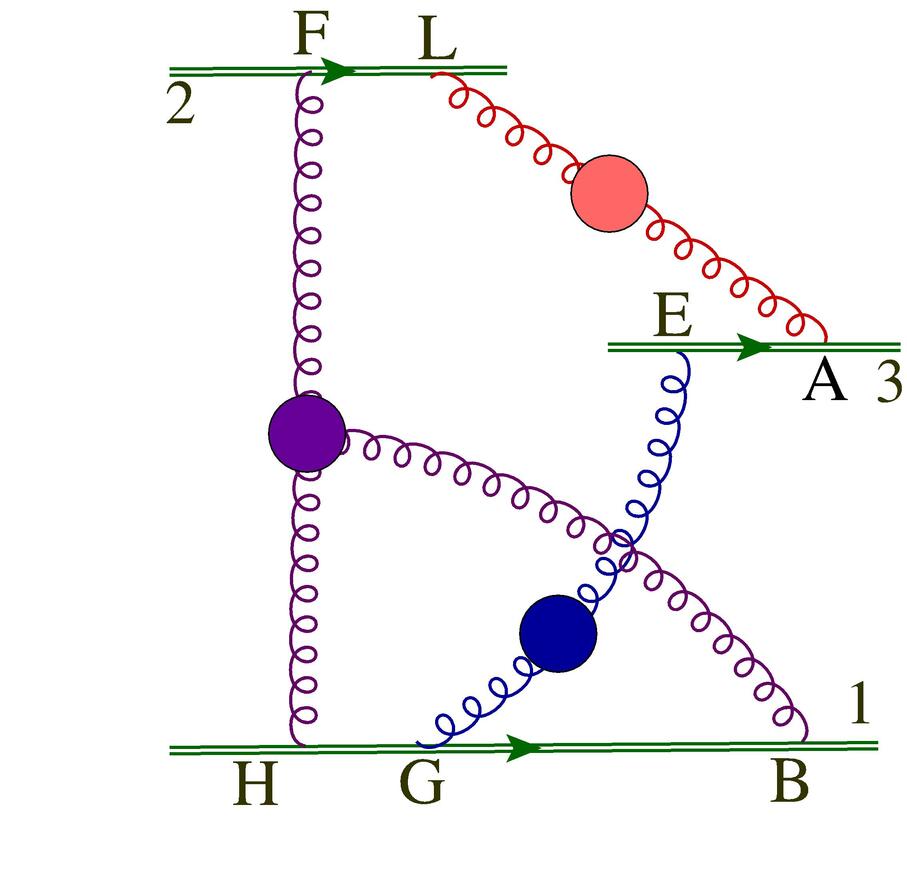} }
	\qquad \quad 
	\subfloat[][]{\includegraphics[height=4cm,width=4cm]{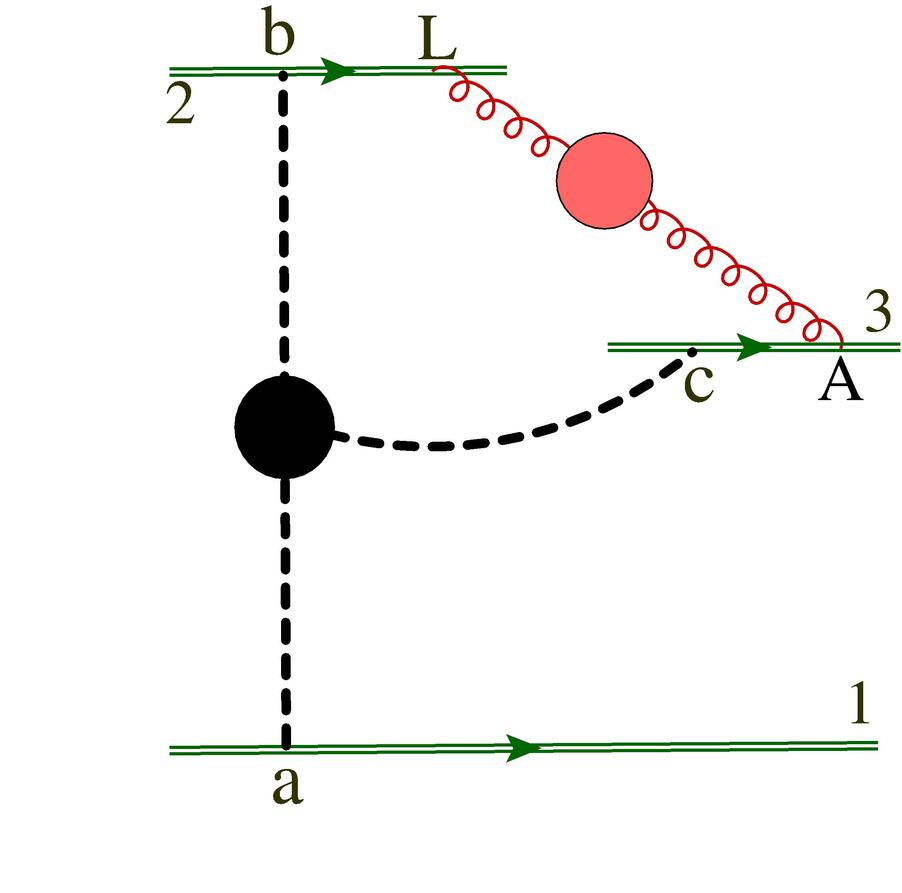} }	
	\caption{\reducedWeb for the Cweb $ W_3^{(2,1)}(2,2,3) $}
	\label{fig:EX-S-sixone-Avatar-Web}
\end{figure}
\noindent The entangled piece is shown in fig.~(\ref{fig:EX-S-sixone-Avatar-Web}\textcolor{blue}{a}), and the corresponding \reduced diagram is shown in fig.~(\ref{fig:EX-S-sixone-Avatar-Web}\textcolor{blue}{b}). The shuffle of attachments of this \reduced diagram generates four diagrams. Out of these four, only two diagrams with order of attachments $ \{\{bL\},\{cA\}\} $, and $ \{\{Lb\},\{Ac\}\} $, are reducible, having $ S=\{1,1\} $ and hence form a \reducedWeb of two diagrams. The mixing matrix associated with this \reducedWeb is $ R(1_2) $.  Hence the diagonal blocks, and rank of $ A $, are given by, 
\begin{align}
A\,=\,\left(\begin{array}{c|cc}
\textbf{I}_{4} & & \cdots\\
\hline 
\vdots & & R(1_2)\\ 	
\end{array}\right)\,,\qquad\qquad \qquad r(A)= 4 + r(R(1_2)) = 5\,.
\end{align}
The diagonal blocks of the mixing matrix for this Cweb is given by, 
\begin{align}
R\,=\,\left(\begin{array}{c|c}
\begin{array}{ccc}
\textbf{I}_{4} & & \cdots\\
{\vdots} & & R(1_2)\\ 	
\end{array}&\ldots \\
\hline
O_{6\times6}&R(1_6)
\end{array}\right)\,.
\end{align}
Therefore, the rank of the mixing matrix for Cweb $ W_3^{(2,1)}(2,2,3) $ is,
\begin{align}
r(R)\;&=\; r(A)\,+\,r(R(1_6))\nonumber\\
&=\;5+2 \;=\;7
\end{align} 
The explicit calculation of the mixing matrix for this Cweb was presented in \cite{Agarwal:2021him}, and is given by,
\begin{align}
R= \dfrac{1}{6}\left(
\begin{array}{cccccccccccc}
6 & 0 & 0 & 0 & -3 & -3 & -4 & 2 & -1 & 2 & -1 & 2 \\
0 & 6 & 0 & 0 & 0 & 0 & -4 & 2 & -4 & 2 & -4 & 2 \\
0 & 0 & 6 & 0 & 0 & 0 & 2 & -4 & 2 & -4 & 2 & -4 \\
0 & 0 & 0 & 6 & -3 & -3 & 2 & -1 & 2 & -1 & 2 & -4 \\
0 & 0 & 0 & 0 & 3 & -3 & -1 & -1 & -1 & 2 & 2 & -1 \\
0 & 0 & 0 & 0 & -3 & 3 & -1 & 2 & 2 & -1 & -1 & -1 \\
0 & 0 & 0 & 0 & 0 & 0 & 2 & -1 & -1 & -1 & -1 & 2 \\
0 & 0 & 0 & 0 & 0 & 0 & -1 & 2 & -1 & -1 & 2 & -1 \\
0 & 0 & 0 & 0 & 0 & 0 & -1 & -1 & 2 & 2 & -1 & -1 \\
0 & 0 & 0 & 0 & 0 & 0 & -1 & -1 & 2 & 2 & -1 & -1 \\
0 & 0 & 0 & 0 & 0 & 0 & -1 & 2 & -1 & -1 & 2 & -1 \\
0 & 0 & 0 & 0 & 0 & 0 & 2 & -1 & -1 & -1 & -1 & 2 \\
\end{array}
\right)\,,\label{eq:matrix-S1-6}
\end{align}
which agrees with the results obtained using the Uniqueness theorem and \reducedWebs. 
The diagonal blocks of other four loop Cwebs of this class are constructed similarly in the appendix \ref{sec:1-6appendix}.

\subsection{Cwebs with $ S=\{0,0,\ldots,0,1_{2},2_{2}\} $}

\begin{figure}[H]
	\centering
	\subfloat[][]{\includegraphics[height=4cm,width=4cm]{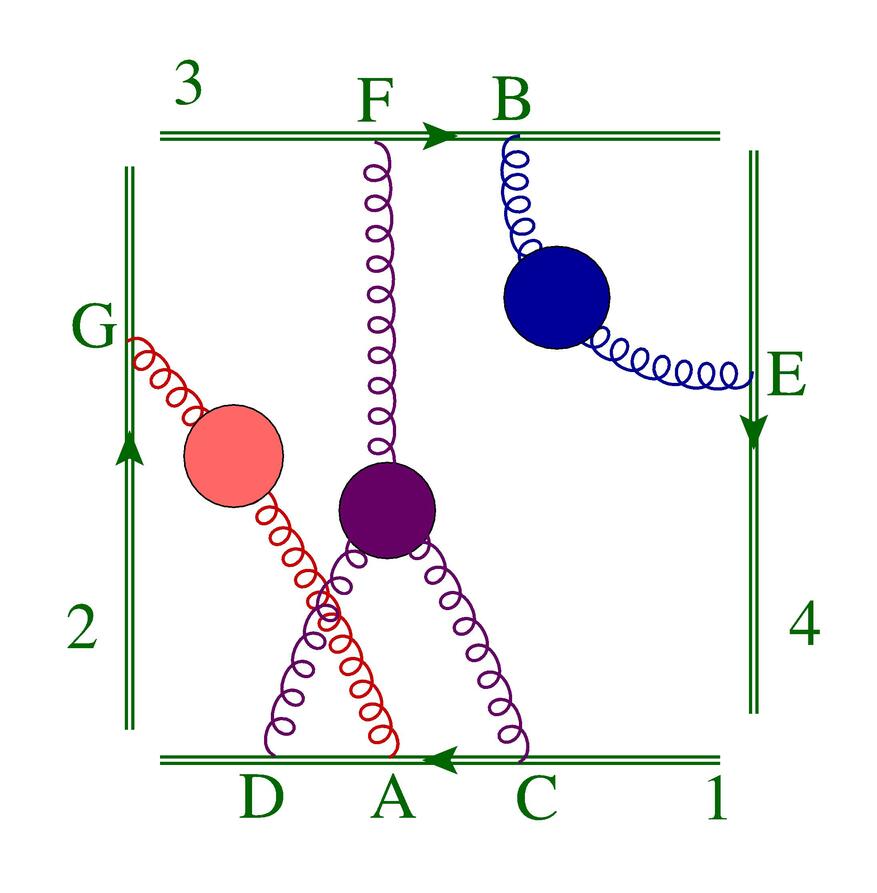} }
	\qquad \quad 
	\subfloat[][]{\includegraphics[height=4cm,width=4cm]{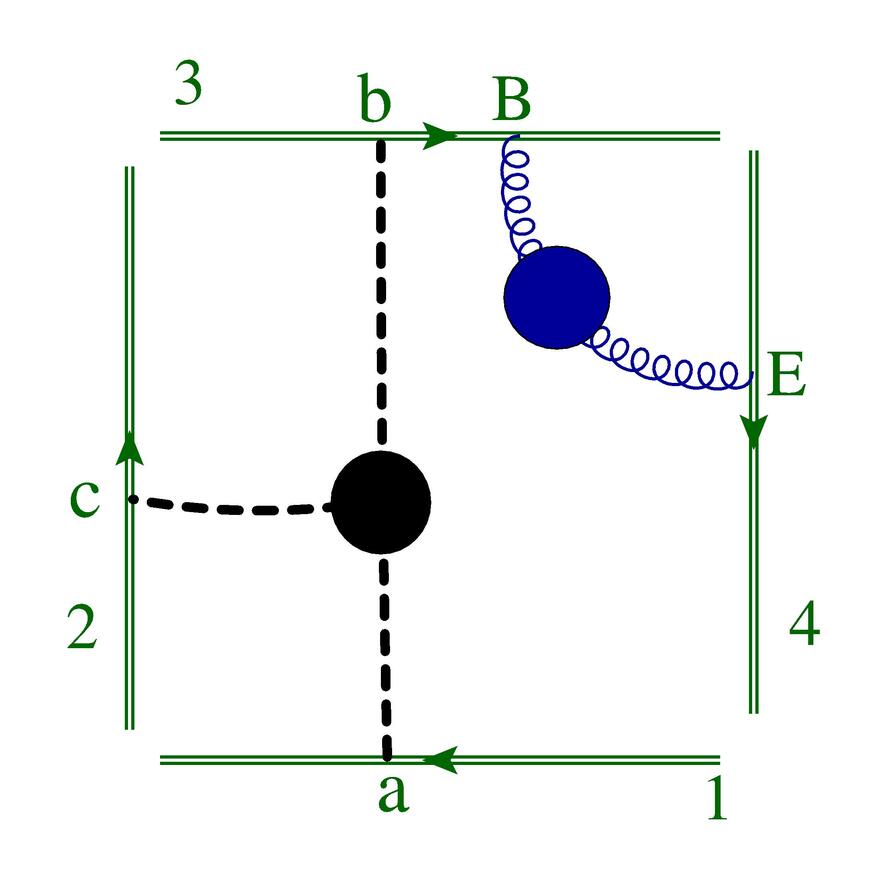} }	
	\caption{(a) Partially entangled diagram  $d_{1}$  and, (b) its corresponding \reduced diagram for the Cweb $ W_4^{(2,1)}(1,1,2,3) $}
	\label{fig:EX-S-1122-Ent-Avatar-Web}
\end{figure}

For this class of Cwebs, the theorem tells that block $ D $ is equal to the  $ 4\times 4 $ mixing matrix $ R\,(1_2,2_2) $ corresponding to $ S=\{1_2,2_2\} $, and has rank one. Following the same procedure as described above, we can write the rank of the mixing matrix for this class as,
\begin{align}
r(R)\,=\,r(A)+1\,.
\label{eq:rank--1122}
\end{align} 
We illustrate the procedure of using \reduceds for this class of Cweb using the $ W_4^{(2,1)}(1,1,2,3) $, present at four loops,  which is shown in figure below. 
\begin{figure}[H]
	\captionsetup[subfloat]{labelformat=empty}
	\centering
	\subfloat[][]{\includegraphics[height=4cm,width=4cm]{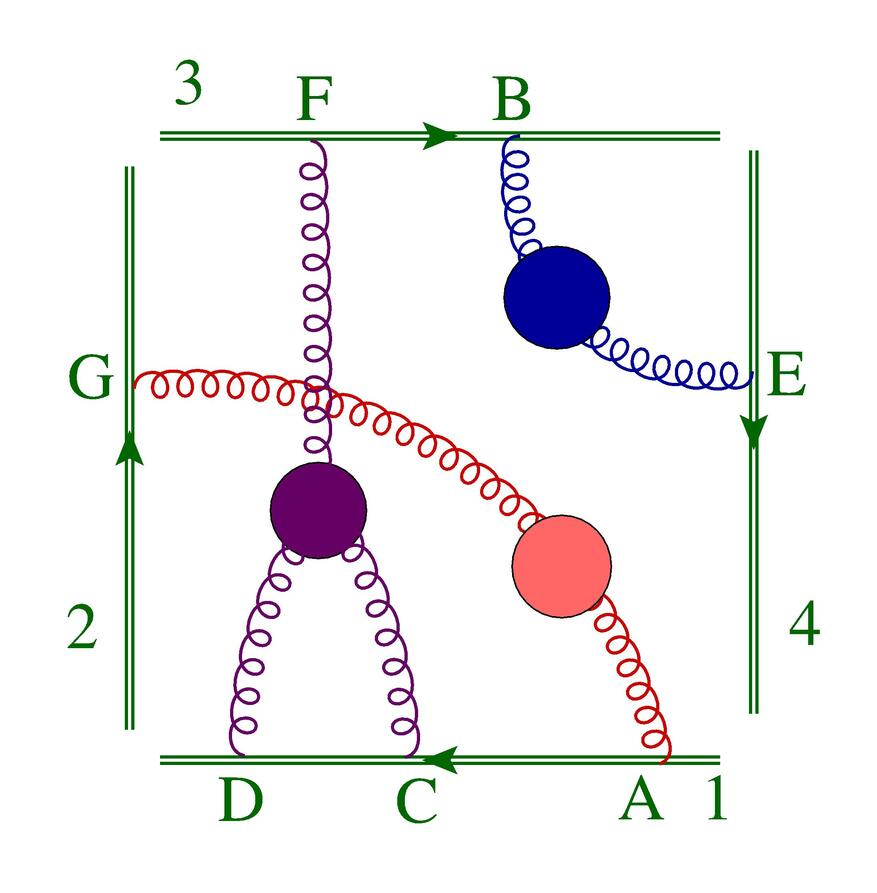} }	
	\caption{A diagram of Cweb $ W_4^{(2,1)}(1,1,2,3) $}	\label{fig:EX-1122-Avatar-Web}
\end{figure}
\noindent This Cweb has six diagrams, whose shuffles are given in the table~\eqref{tab:web1122-avatar}, out of which $ d_3 $, $ d_4 $, $ d_5 $, and $ d_6 $ are reducible with $ S=\{1,1,2,2\} $. 
\begin{table}[H]
	
	\begin{center}
		\begin{tabular}{|c|c|c|}
			\hline 
			\textbf{Diagrams}  & \textbf{Sequences}  & \textbf{s-factors}  \\ 
			\hline
			$d_{1}$  & $\{FB\},\{CAD\}$  & 0 \\ \hline
			$d_{2}$  & $\{BF\},\{CAD\}$& 0 \\ \hline
			$d_{3}$  & $\{FB\},\{ACD\} $  & 1 \\ \hline 
			$d_{4}$  & $\{BF\},\{CDA\}$  & 1 \\ \hline
			$d_{5}$  & $\{BF\},\{ACD\} $  & 2 \\ \hline
			$d_{6}$  & $\{FB\},\{CDA\}$  & 2 \\ \hline
		\end{tabular}
	\end{center}
	\caption{Normal order of diagrams; and $ R $, for the Cweb $ W_4^{(2,1)}(1,1,2,3) $}
	\label{tab:web1122-avatar}
\end{table}
\noindent Thus, block $ D $ corresponds to $ R(1_2,2_2) $. The block diagonals of $ A $, can be determined by identifying the entangled pieces. There is only one kind of entangled piece, which is present in $ d_1 $, and $ d_2 $, which is displayed in fig.~(\ref{fig:EX-S-1122-Ent-Avatar-Web}\textcolor{blue}{a}), which corresponds to a \reduced diagram shown in fig.~(\ref{fig:EX-S-1122-Ent-Avatar-Web}\textcolor{blue}{b}).  
Now, as this entangled piece appears only in diagrams $ d_1 $, and $ d_2 $ of this Cweb, given in table (\ref{tab:web1122-avatar}), thus, both of them are partially entangled diagrams, and hence, there is no completely entangled diagram. Thus, we do not have any identity matrix in block $ A $.   
\noindent Now, the shuffle of attachments of \reduced diagram (\ref{fig:EX-S-1122-Ent-Avatar-Web}\textcolor{blue}{b}), generates two diagrams with order of attachments $ \{bB\} $, and $ \{Bb\} $ on line $ 3 $, both having $ s=1 $. The mixing matrix corresponding to this \reducedWeb is $ R(1_2) $. Therefore, block $ A $ is equal $ R(1_2) $, and its rank is $ 1 $.
\begin{align}
A = R(1_2)
\end{align}
\noindent The diagonal blocks of the mixing matrix for this Cweb are given by, 
\begin{align}
R\,=\,\left(\begin{array}{c|c}
R(1_2)&\ldots \\\hline
O_{6\times2}&R(1_2,2_2)
\end{array}\right)\,.
\label{eq:diagonal}
\end{align}
Hence, the rank of mixing matrix for this Cweb is,
\begin{align}
r(R) \;&=\; r(A)\,+\,r(R(1_2,2_2))\;=\;2
\end{align}
The explicit form of mixing matrix was calculated using replica trick in \cite{Agarwal:2021him}, given by, 
\begin{align}
R=\frac{1}{6}\left(
\begin{array}{cccccc}
3 & -3 & 2 & -1 & -2 & 1 \\
-3 & 3 & -1 & 2 & 1 & -2 \\
0 & 0 & 2 & 2 & -2 & -2 \\
0 & 0 & 2 & 2 & -2 & -2 \\
0 & 0 & -1 & -1 & 1 & 1 \\
0 & 0 & -1 & -1 & 1 & 1 \\
\end{array}
\right)\,. \nonumber
\label{eq:R--1122}
\end{align} 
The diagonal blocks of the mixing matrix predicted from our procedure using \reducedWebs and the Uniqueness theorem  agrees with the explicit form calculated in \cite{Agarwal:2021him} using the replica trick. Results for other four-loop Cwebs of this class are presented in appendix \ref{sec:1-22-2-appendix}.

\section{Summary and outlook}
\label{sec:conclu}
The logarithm of the Soft function can be expressed as a sum over Cwebs that can be written down in terms of Feynman diagrams. This exponentiation of the Cwebs allows us to make predictions of the IR structures in the multiparton scattering amplitudes to all orders in the perturbation theory. The diagrams of a Cweb mix via a mixing matrix such that they select only colour factors that correspond to fully connected diagrams. The mixing matrices have been studied extensively in the literature.

In this article we have developed a new formalism that allows us to predict the number of exponentiated colour factors of several classes of Cwebs at order $\alpha_{s}^{n+1}$ and higher,
if we know the basis Cwebs that are present upto order $\alpha_{s}^{n}$. Our formalism further allows us to predict the mixing matrices for two classes of Cwebs to all orders in perturbation theory.

We have introduced several new ideas: (a) {Normal ordering} of the diagrams of a Cweb, (b) \reducedWebs (c) Basis and Family of Cwebs which prove extremely useful in making the structures present in the mixing matrices very transparent. Basis Cwebs are those that connect $ n+1 $ Wilson lines at $ n $ loops involving only two-point gluon correlators, and family is the set of Cwebs that have the same shuffle.

We have proved a  {Uniqueness theorem} which states that,  for a given column weight vector $S=\{s(d_1),s(d_2),\ldots,s(d_n)\}$
 with all $s(d_{i}) \neq 0$, the mixing matrix is unique.
 We have also  introduced the concept of  \reducedWebs which has helped us determine the diagonal blocks of the mixing matrices that correspond to the mixing between the irreducible diagrams
of a Cweb. 
Together these ideas provide us with an ability to predict the rank of the mixing matrices or equivalently the number of 
 independent exponentiated colour factors that are present for a given Cweb.

Using our formalism we can predict, without doing the explicit calculations using  the replica trick algorithm,  the explicit form of mixing matrices of $ 26 $ out of $ 60 $ Cwebs present at four loops connecting massless Wilson lines using the matrices from two and three loops which is $ 43\% $ of total number of Cwebs present at four loops. Using \reducedWebs we can further predict the diagonal blocks of $ 9 $ mixing matrices at four loops. Thus, in total, we can predict the rank of $ 35 $ mixing matrices at four loops which is $ 58\% $ of the total Cwebs without using the replica trick. All the predictions match with the known results presented in \cite{Agarwal:2020nyc,Agarwal:2021him}. 

It would be interesting to see if this framework can be expanded further to provide more understanding of the structures present in the mixing matrices. It would also be interesting to see the implications of this formalism on the kinematic structure of the Cwebs.

\section*{Acknowledgement}

\noindent 
NA, SP and AT would like to thank Lorenzo Magnea for collaboration on earlier projects on Cwebs.
SP would like to thank MoE, Govt. of India, for an SRF 
fellowship, AS would like to thank CSIR, Govt. of India, for a JRF fellowship (09/1001(0075)/2020-EMR-I). 
\appendix 
\section*{Appendix}
\section{Replica trick}
\label{sec:repl}
One of the powerful techniques in the combinatorial problems in physics, which involves exponentiation is the  replica trick \cite{MezaPariVira}. For Wilson line correlators, the replica trick algorithm was developed in \cite{Gardi:2010rn,Laenen:2008gt}. The same replica trick was adopted in \cite{Agarwal:2020nyc,Agarwal:2021him} for the calculation of the mixing matrices for four-loop Cwebs. Here, we briefly discuss the replica trick algorithm, which was used in the calculation of the mixing matrices for Cwebs at four loops. To start with, we consider the path integral of the Wilson line correlators as, 
\begin{align}
\mathcal{S}_n(\gamma_i)=\,\int \mathcal{D}A_\mu^a\,\exp(iS(A_\mu ^a)) \prod _{k=1}^n\phi_k(\gamma_k)=\exp[\mathcal{W}_n(\gamma_i)]\,
\end{align}
where $ S(A_\mu ^a) $ is the classical action of the gauge fields. In order to proceed with the replica trick algorithm, one introduces $ N_r $ non-interacting identical copies of each gluon field $ A_\mu $, which means, we replace each $ A_\mu $ by $ A_\mu ^i $, where,  $ i=1,\ldots, N_r $. Now, for each replica, we associate a copy of each Wilson line, thereby, replacing each Wilson line by a product of $ N_r $ Wilson lines. Thus, in the replicated theory, the path integral of the Wilson line correlator can then be written as,
\begin{align}
{\cal S}_n^{\, {\rm repl.}} \left( \gamma_i \right) \, = \,   \Big[ 
{\cal S}_n \left( \gamma_i \right) \Big]^{N_r} \, = \, \exp \Big[ N_r \,
{\cal W}_n (\gamma_i) \Big] \, =  \, {\bf 1} + N_r \, {\cal W}_n (\gamma_i) 
+ {\cal O} (N_r^2) \, .
\label{exprepl}
\end{align}
Now, using this equation, one can calculate $ \mathcal{W}_n $ by calculating $ \mathcal{O}(N_r) $ terms of the Wilson line correlator in the replicated theory. The method of replicas involves five steps, which are summarized below. 
\begin{enumerate}
	\item [-] Associate a replica number to each connected gluon correlator in a Cweb. 
	\item[-]  Define a replica ordering operator $ \textbf{R} $, which acts on the colour generators on each Wilson line and order them according to their replica numbers. Thus, if $ \textbf{T}_i $ denotes a colour generator for a correlator belonging to replica number $ i $, then action of $ \textbf{R} $ on $ \textbf{T}_i \textbf{T}_j$ preserves the order for $ i\leq j $, and reverses the order for $ i>j $. Thus, replica ordered colour factor for a diagram in a Cweb will always be a diagram of the same Cweb.
	\item [-] The next step in order to calculate the exponentiated colour factors, one needs to find the hierarchies between the replica numbers present in a Cweb. If a Cweb has $ m $ connected pieces, we call hierarchies $ h(m) $. $ h(m) $ are known as Bell number or Fubini number \cite{IntSeq} in the number theory and combinatorics. The first few Fubini numbers are given by $ h(m)=\{1,1,3,13,75,541\} $ for $ m= 0,1,2,3,4,5$. At four loops, the highest number of correlator in a Cweb is $ m_{\text{max}}=4 $, which corresponds to $ h_{\text{max}}=75 $   
	\item [-]  The next object is to calculate $ M_{N_r}(h) $, which counts the number of appearances of a particular hierarchy in the presence of $ N_r $ replicas.  For a given hierarchy $ h $, which contains $ n_r(h) $ distinct replicas, the multiplicity $ M_{N_r}(h) $ is given by, 
	\begin{align}
	M_{N_r}(h) \, = \, \frac{N_r!}{\big( N_r - n_r(h) \big)! \,\, n_r(h)!}  \,
	\end{align}  
	\item [-] The exponentiated colour factor for a diagram $ d $ is then given by, 
	\begin{align}
	C_{N_r}^{\, {\rm repl.}}  (d) \, = \, \sum_h M_{N_r} (h) \, \textbf{R} \big[ C(d) \big| h 
	\big]  \, ,
	\label{expocolf}
	\end{align}
	where $ \textbf{R} \big[ C(d) \big| h $ is the replica ordered colour factor of diagram $ d $, for hierarchy $ h $. Finally, the exponentiated colour factor for diagram $ d $ is computed by extracting the coefficient of $ \mathcal{O}(N_r) $ terms of the above equation.  	
\end{enumerate}

\section{Cwebs with \reducedWebs}\label{sec:table-of-av-webs}
In this appendix, we show the direct construction of the diagonal blocks of web mixing matrices, using Fused-Webs. 
\subsection{\reducedWebs of Cweb having $ S=\{0,0,\cdots,0,1_6\} $}\label{sec:1-6appendix}
In this class of Cwebs, the $ D $-block of mixing matrices is $ R(1_6) $, whose full form is given in appendix \ref{sec:basis}. Thus the rank for this class of matrices will be, 
\begin{align}
r(R)\,&=\,r(A)+r(R(1_6))\nonumber\\
\,&=\,r(A) + 2\,.
\label{eq:rank-I-6-class}
\end{align}
Therefore after determining the diagonals blocks of sub-matrix $ A $ of the mixing matrix, we can determine the rank of $ R $.
\vspace{0.5cm}

\noindent\textbf{1.}\, $ \textbf{W}^{(2,1)}_{3,\text{I}}(2,2,3) $

\begin{figure}[H]
	\vspace*{-0.6cm}
	\centering
	\subfloat[][]{\includegraphics[height=4cm,width=4cm]{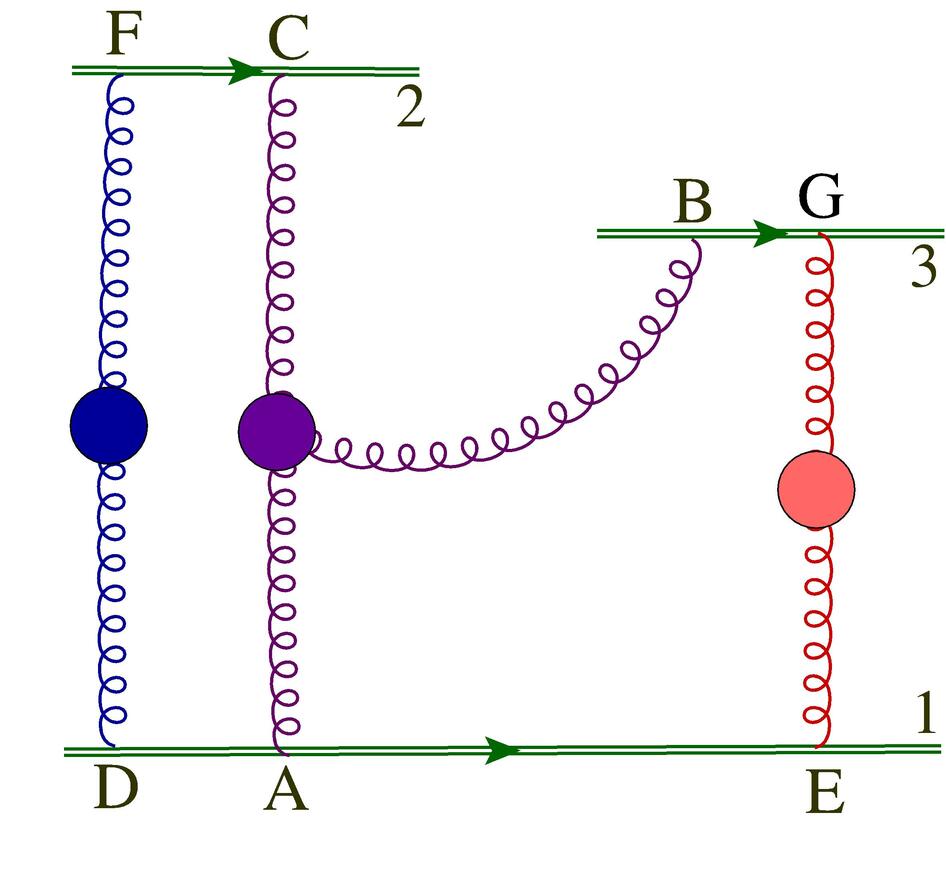} }
	\caption{Cweb $ W^{(2,1)}_{3,\text{I}}(2,2,3) $}
	\label{fig:six-one-web7-av}
\end{figure}

\vspace{0.2cm}
\noindent This Cweb, shown in fig.~(\ref{fig:six-one-web7-av}) has twenty-four diagrams, out of which six are reducible, ten are completely entangled, and remaining eight are partially   entangled The Normal ordered diagrams and their $ s $-factors are given in table \ref{tab:six-one-web7-av}.

\begin{table}[H]
	\vspace*{-6mm}
	\begin{minipage}[c]{0.5\textwidth}
		\begin{table}[H]
			\begin{center}
				\resizebox{0.9\textwidth}{!}{%
				\begin{tabular}{|c|c|c|}
					\hline 
					\textbf{Diagrams}  & \textbf{Sequences}  & \textbf{s-factors}  \\ 
					\hline
					$d_1$ & $\lbrace\lbrace EDA\rbrace,\lbrace BG \rbrace, \lbrace FC \rbrace\rbrace$ & 0 \\ \hline
					$d_2$ & $\lbrace\lbrace EDA\rbrace,\lbrace BG \rbrace, \lbrace CF \rbrace\rbrace$ & 0 \\ \hline
					$d_3$ & $\lbrace\lbrace DEA\rbrace,\lbrace GB \rbrace, \lbrace CF \rbrace\rbrace$ & 0 \\ \hline
					$d_4$ & $\lbrace\lbrace DEA\rbrace,\lbrace BG \rbrace, \lbrace CF \rbrace\rbrace$ & 0 \\ \hline
					$d_{5}$ & $\lbrace\lbrace EAD\rbrace,\lbrace BG \rbrace, \lbrace FC \rbrace\rbrace$ & 0 \\ \hline
					$d_{6}$ & $\lbrace\lbrace AED\rbrace,\lbrace GB \rbrace, \lbrace FC \rbrace\rbrace$ & 0 \\ \hline
					$d_{7}$ & $\lbrace\lbrace AED \rbrace,\lbrace BG \rbrace, \lbrace FC \rbrace\rbrace$ & 0 \\ \hline
					$d_{8}$ & $\lbrace\lbrace DAE\rbrace,\lbrace GB \rbrace, \lbrace CF \rbrace\rbrace$ & 0 \\ \hline
					$d_{9}$ & $\lbrace\lbrace ADE\rbrace,\lbrace GB \rbrace, \lbrace FC \rbrace\rbrace$ & 0 \\ \hline
					$d_{10}$ & $\lbrace\lbrace ADE\rbrace,\lbrace GB \rbrace, \lbrace CF \rbrace\rbrace$ & 0 \\ \hline
					$d_{11}$ & $\lbrace\lbrace EDA\rbrace,\lbrace GB \rbrace, \lbrace CF \rbrace\rbrace$ & 0 \\ \hline
					$d_{12}$ & $\lbrace\lbrace DAE \rbrace,\lbrace BG \rbrace, \lbrace CF \rbrace\rbrace$ & 0 \\ \hline
				\end{tabular}}
			\end{center}
		\end{table}
	\end{minipage}
	\hspace{0.08cm}
	\begin{minipage}[c]{0.5\textwidth}
		\begin{table}[H]
			\begin{center}
				\resizebox{0.9\textwidth}{!}{%
				\begin{tabular}{|c|c|c|}
					\hline 
					\textbf{Diagrams}  & \textbf{Sequences}  & \textbf{s-factors}  \\ 
					\hline
					$d_{13}$ & $\lbrace\lbrace DEA\rbrace,\lbrace BG \rbrace, \lbrace FC \rbrace\rbrace$ & 0 \\ \hline
					$d_{14}$ & $\lbrace\lbrace EAD\rbrace,\lbrace BG \rbrace, \lbrace CF \rbrace\rbrace$ & 0 \\ \hline
					$d_{15}$ & $\lbrace\lbrace EAD\rbrace,\lbrace GB \rbrace, \lbrace FC \rbrace\rbrace$ & 0 \\ \hline
					$d_{16}$ & $\lbrace\lbrace ADE \rbrace,\lbrace BG \rbrace, \lbrace FC \rbrace\rbrace$ & 0 \\ \hline 
					$d_{17}$ & $\lbrace\lbrace AED\rbrace,\lbrace GB \rbrace, \lbrace CF \rbrace\rbrace$ & 0 \\ \hline
					$d_{18}$ & $\lbrace\lbrace DAE\rbrace,\lbrace GB \rbrace, \lbrace FC \rbrace\rbrace$ & 0 \\ \hline
					$d_{19}$ & $\lbrace\lbrace EDA\rbrace,\lbrace GB \rbrace, \lbrace FC \rbrace\rbrace$ & 1 \\ \hline
					$d_{20}$ & $\lbrace\lbrace DEA\rbrace,\lbrace GB \rbrace, \lbrace FC \rbrace\rbrace$ & 1 \\ \hline
					$d_{21}$ & $\lbrace\lbrace EAD\rbrace,\lbrace GB \rbrace, \lbrace CF \rbrace\rbrace$ & 1 \\ \hline
					$d_{22}$ & $\lbrace\lbrace AED \rbrace,\lbrace BG \rbrace, \lbrace CF \rbrace\rbrace$ & 1 \\ \hline
					$d_{23}$ & $\lbrace\lbrace DAE \rbrace,\lbrace BG \rbrace, \lbrace FC \rbrace\rbrace$ & 1 \\ \hline
					$d_{24}$ & $\lbrace\lbrace ADE \rbrace,\lbrace BG \rbrace, \lbrace CF \rbrace\rbrace$ & 1 \\ \hline
				\end{tabular}}
			\end{center}
		\end{table}
	\end{minipage}
	\caption{Normal ordered diagrams of Cweb $ W^{(2,1)}_{3,\text{I}}(2,2,3) $}
	\label{tab:six-one-web7-av}
\end{table}
\begin{figure}[H]
	\captionsetup[subfloat]{labelformat=empty}
	\centering
	\subfloat[][(a)]{\includegraphics[height=4cm,width=4cm]{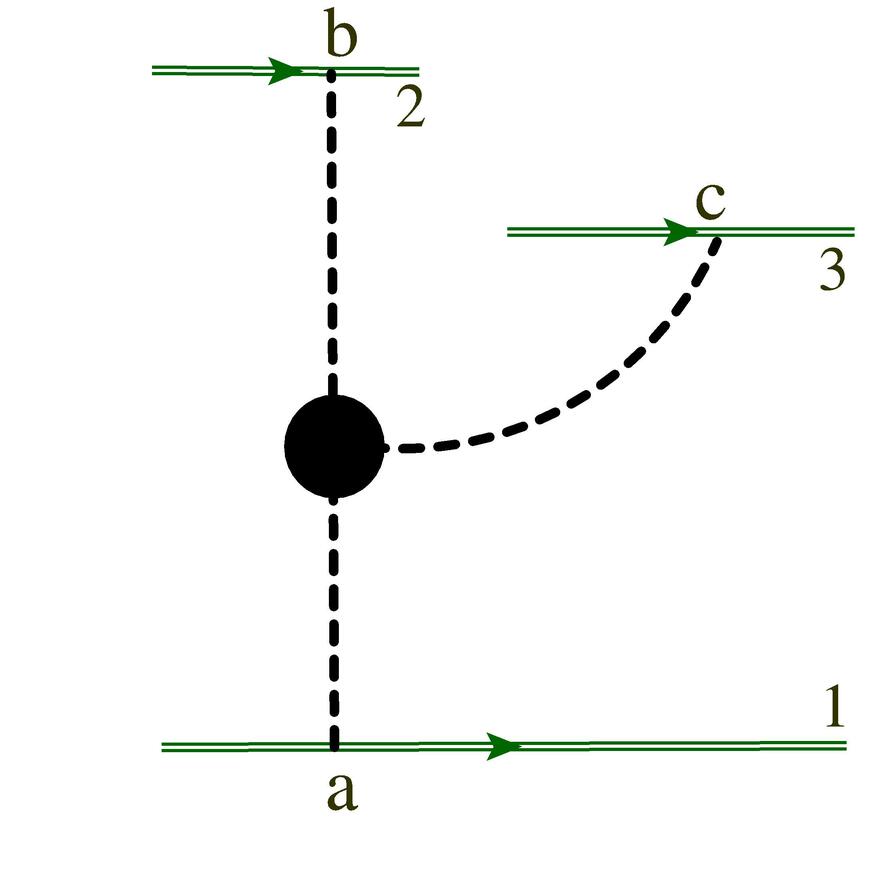} }
	\qquad 
	\subfloat[][(b)]{\includegraphics[height=4cm,width=4cm]{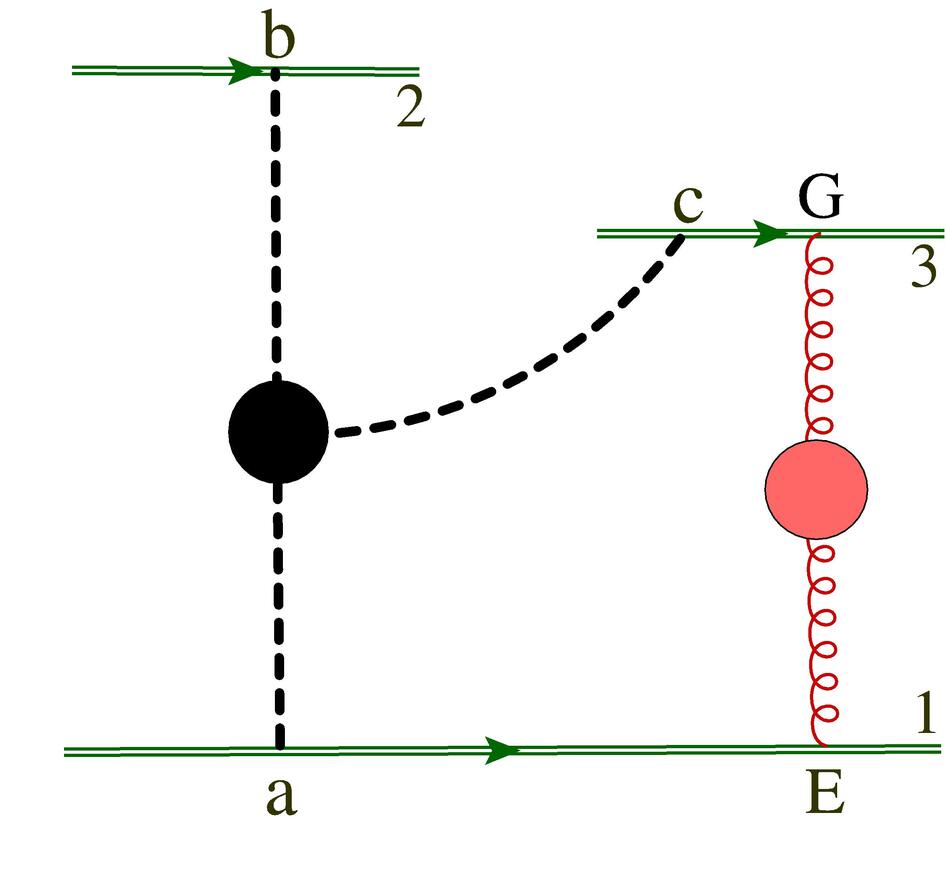} }
	\qquad 
	\subfloat[][(c)]{\includegraphics[height=4cm,width=4cm]{Red3LW7Ent1} }
	\qquad 
	\subfloat[][(d)]{\includegraphics[height=4cm,width=4cm]{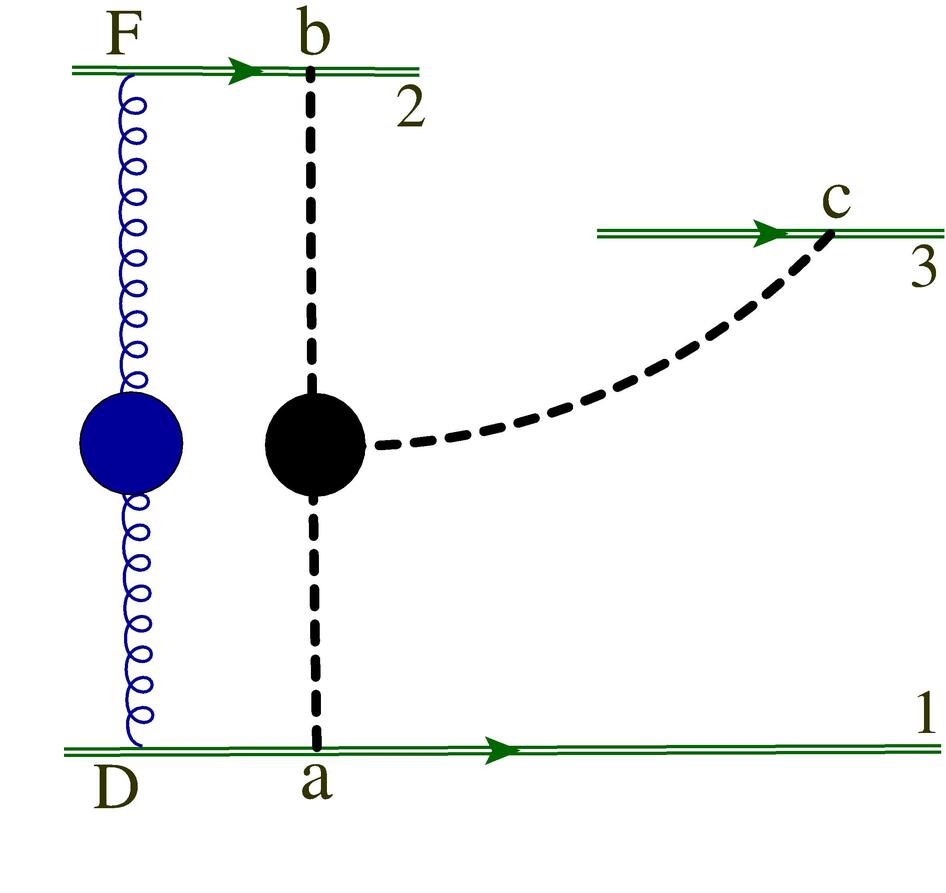} }
	\qquad 
	\subfloat[][(e)]{\includegraphics[height=4cm,width=4cm]{Red3LW7Ent3} }
	\caption{\reducedWebs for Cweb $ W^{(2,1)}_{3,\text{I}}(2,2,3) $ }
	\label{fig:six-one-web7-allAVAtar-WEBS}
\end{figure}
\noindent \noindent The procedure developed in section \ref{sec:avatar-webs}, is applied to this Cweb, which results in table \ref{tab:six-one-web7-av-Ent}. It classifies the irreducible diagrams of the Cweb according to the entangled pieces. This table provides the \reducedWebs with the associated mixing matrices for the Cweb. For example, in this Cweb, there are ten completely entangled diagrams, whose order of attachments can be read of from table \ref{tab:six-one-web7-av}. This then form the \reducedWebs with single diagram, whose mixing matrix is identity matrix of order ten. Further, there are four distinct entangled pieces appearing in partially entangled diagrams. Each entangled piece is associated with two diagram of Cweb, whose \reduced diagrams form a Fused-Web. Thus there are four \reducedWebs with their corresponding mixing matrices.   
\begin{table}
	\begin{center}
		\begin{tabular}{|c|c|c|c|c|c|}
			\hline
			Entanglement & Diagrams of  & \reducedWeb & Diagrams in  & $ s $-factors   & $ R $ \\ 
			& Cweb &  & \reducedWeb &    &  \\
			\hline & & & & &\\
			Complete entangled  & $ d_1 $, $ d_2 $, $ d_3 $, $ d_4 $, $ d_5 $ & \ref{fig:six-one-web7-allAVAtar-WEBS}\textcolor{blue}{a} & - & 1 & $ I_{10} $ \\ 
			& $ d_6 $, $ d_7 $, $ d_8 $, $ d_9 $, $ d_{10} $ & & & &\\ \hline
			First Partial Entangled  & $ d_{11} $, $ d_{12} $ & \ref{fig:six-one-web7-allAVAtar-WEBS}\textcolor{blue}{b} & $ \{c\,G\},\,\{a\,E\} $ & 1 & $ R(1_2) $ \\ 
			&  & & $ \{G\,c\},\,\{E\,a\} $ & 1 & \\ \hline
			Second Partial Entangled  & $ d_{13} $, $ d_{14} $ & \ref{fig:six-one-web7-allAVAtar-WEBS}\textcolor{blue}{c} & $ \{c\,G\},\,\{a\,E\} $ & 1 & $ R(1_2) $ \\ 
			&  & & $ \{G\,c\},\,\{E\,a\} $ & 1 & \\ \hline
			Third Partial Entangled  & $ d_{15} $, $ d_{16} $ & \ref{fig:six-one-web7-allAVAtar-WEBS}\textcolor{blue}{d} & $ \{F\,b\},\,\{D\,a\} $ & 1 & $ R(1_2) $ \\ 
			&  & & $ \{b\,F\},\,\{a\,D\} $ & 1 & \\ \hline
			Fourth Partial Entangled  & $ d_{17} $, $ d_{18} $ & \ref{fig:six-one-web7-allAVAtar-WEBS}\textcolor{blue}{e} & $ \{F\,b\},\,\{D\,a\} $ & 1 & $ R(1_2) $ \\ 
			&  & & $ \{b\,F\},\,\{a\,D\} $ & 1 & \\ \hline
		\end{tabular}	
	\end{center}
	\caption{\reducedWebs and their mixing matrices for Cweb $ W^{(2,1)}_{3,\text{I}}(2,2,3) $}
	\label{tab:six-one-web7-av-Ent}
\end{table}

\noindent The order of diagrams in the Cweb given in table \ref{tab:six-one-web7-av}, is chosen such that diagrams with same kind of entangled piece appear together. Therefore, mixing matrices of the \reducedWebs for this Cweb, present on the diagonal blocks of $ A $, are given as, 
\begin{align}
A\,=\,\left(\begin{array}{c|cc}
\textbf{I}_{10} & & \cdots\\
\hline 
\textbf{O}_{8\times 10}& & \begin{array}{cccc}
R\,(1_2) & && \\
& R\,(1_2)&&\\
& & R\,(1_2)&\\
&&&R\,(1_2)\\
\end{array}	
\end{array}\right)\,,\qquad 
\end{align}
and rank of $ A $ is,
\begin{align}
r(A) &= r(I_{10})  +  4\,r(R(1_2))\nonumber\\
&=  14\,.
\end{align}

\noindent The number of exponentiated colour factors is the rank of the mixing matrix. Thus rank of $ R $, using eq.~(\ref{eq:rank-I-6-class}), is given as

\begin{align}
r(R) = r(A)  + 2  = 16\,.
\end{align}

\vspace{0.5cm}
\noindent\textbf{2.}\, $ \textbf{W}^{(2,1)}_{3,\text{I}}(1,3,3) $

\begin{figure}[H]
	\captionsetup[subfloat]{labelformat=empty}
	\centering
	\subfloat[][]{\includegraphics[height=4cm,width=4cm]{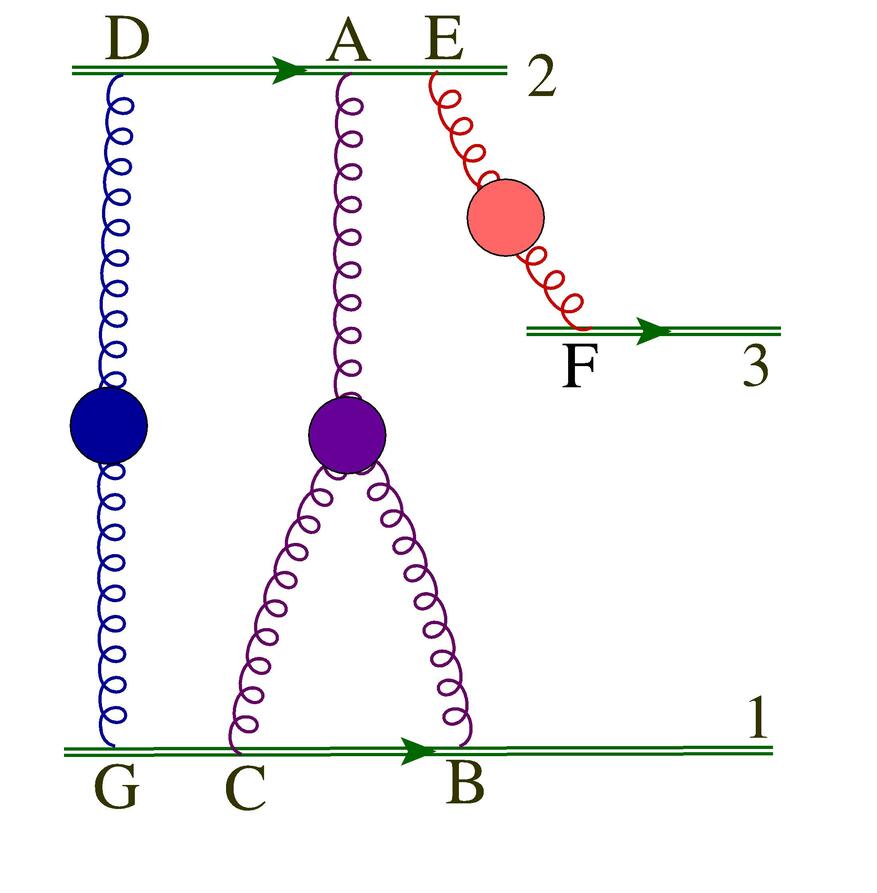} }
	\caption{Cweb $ W^{(2,1)}_{3,\text{I}}(1,3,3) $}
	\label{fig:six-one-web14-av}
\end{figure}

\vspace{0.2cm}
\noindent This Cweb, shown in fig.~(\ref{fig:six-one-web14-av}) has eighteen diagrams, out of which six are reducible, four are completely entangled, and remaining eight are partially entangled. The Normal ordered diagrams and their $ s $-factors are shown in table \ref{tab:six-one-web14-av}. 

\begin{table}[H]
	\begin{minipage}[c]{0.5\textwidth}
		\begin{table}[H]
			\begin{center}
					\resizebox{0.9\textwidth}{!}{%
				\begin{tabular}{|c|c|c|}
					\hline 
					\textbf{Diagrams}  & \textbf{Sequences}  & \textbf{s-factors}  \\ 
					\hline
					$d_1$ & $\lbrace\lbrace DEA\rbrace,\lbrace CGB \rbrace \rbrace$ & 0 \\ \hline
					$d_2$ & $\lbrace\lbrace DEA\rbrace,\lbrace CBG \rbrace \rbrace$ & 0 \\ \hline
					$d_{3}$ & $\lbrace\lbrace AED\rbrace,\lbrace GCB \rbrace \rbrace$ & 0 \\ \hline
					$d_{4}$ & $\lbrace\lbrace AED\rbrace,\lbrace CGB \rbrace \rbrace$ & 0 \\ \hline
					$d_5$ & $\lbrace\lbrace EDA\rbrace,\lbrace CGB \rbrace \rbrace$ & 0 \\ \hline
					$d_{6}$ & $\lbrace\lbrace DAE\rbrace,\lbrace CGB \rbrace \rbrace$ & 0 \\ \hline
					$d_7$ & $\lbrace\lbrace EDA\rbrace,\lbrace CBG \rbrace \rbrace$ & 0 \\ \hline
					$d_{8}$ & $\lbrace\lbrace DAE\rbrace,\lbrace CBG \rbrace \rbrace$ & 0 \\ \hline
					$d_9$ & $\lbrace\lbrace EAD\rbrace,\lbrace GCB \rbrace \rbrace$ & 0 \\ \hline
				\end{tabular}}
			\end{center}
		\end{table}
	\end{minipage}
	\hspace{0.2cm}
	\begin{minipage}[c]{0.5\textwidth}
		\begin{table}[H]
			\begin{center}
					\resizebox{0.9\textwidth}{!}{%
				\begin{tabular}{|c|c|c|}
					\hline 
					\textbf{Diagrams}  & \textbf{Sequences}  & \textbf{s-factors}  \\ 
					\hline
					$d_{10}$ & $\lbrace\lbrace ADE\rbrace,\lbrace GCB \rbrace \rbrace$ & 0 \\ \hline
					$d_{11}$ & $\lbrace\lbrace EAD\rbrace,\lbrace CGB \rbrace \rbrace$ & 0 \\ \hline
					$d_{12}$ & $\lbrace\lbrace ADE\rbrace,\lbrace CGB \rbrace \rbrace$ & 0 \\ \hline
					$d_{13}$ & $\lbrace\lbrace EDA\rbrace,\lbrace GCB \rbrace \rbrace$ & 1 \\ \hline
					$d_{14}$ & $\lbrace\lbrace DEA\rbrace,\lbrace GCB \rbrace \rbrace$ & 1 \\ \hline
					$d_{15}$ & $\lbrace\lbrace EAD\rbrace,\lbrace CBG \rbrace \rbrace$ & 1 \\ \hline
					$d_{16}$ & $\lbrace\lbrace AED\rbrace,\lbrace CBG \rbrace \rbrace$ & 1 \\ \hline
					$d_{17}$ & $\lbrace\lbrace DAE\rbrace,\lbrace GCB \rbrace \rbrace$ & 1 \\ \hline
					$d_{18}$ & $\lbrace\lbrace ADE\rbrace,\lbrace CBG \rbrace \rbrace$ & 1 \\ \hline
				\end{tabular}}
			\end{center}
		\end{table}
	\end{minipage}
	\caption{Normal ordered diagrams of Cweb $ W^{(2,1)}_{3,\text{I}}(1,3,3) $}
	\label{tab:six-one-web14-av}
\end{table}

\begin{figure}[H]
	\captionsetup[subfloat]{labelformat=empty}
	\centering
	\subfloat[][(a)]{\includegraphics[height=4cm,width=4cm]{Red3LWanyEnt0} }
	\qquad 
	\subfloat[][(b)]{\includegraphics[height=4cm,width=4cm]{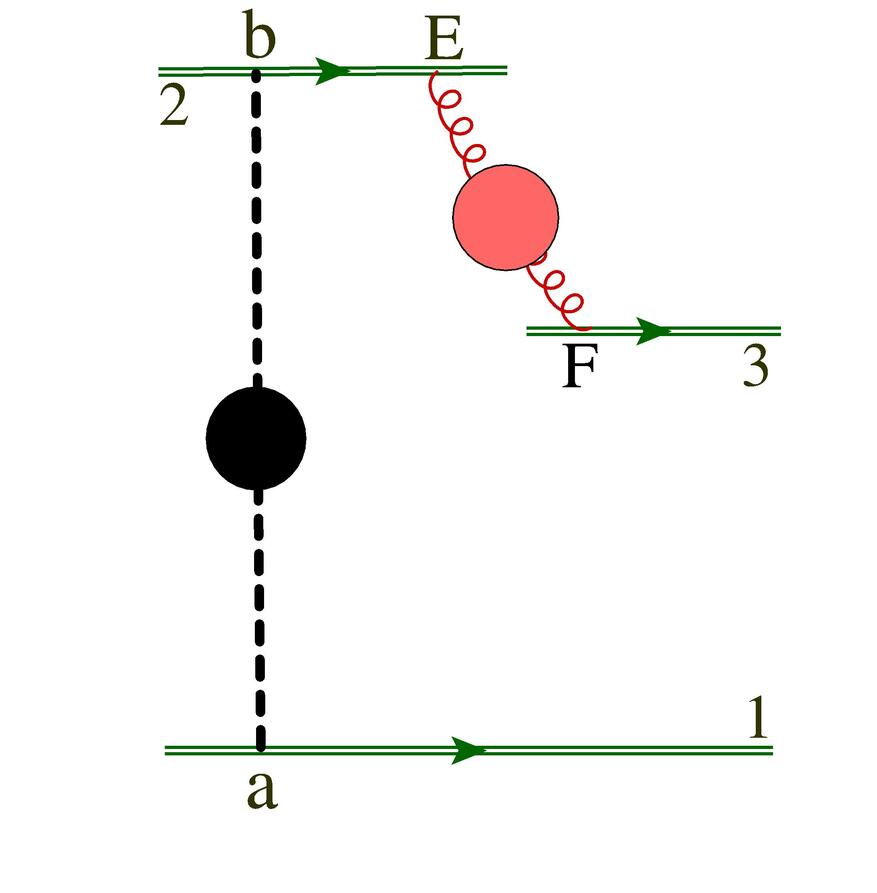} }
	\qquad 
	\subfloat[][(c)]{\includegraphics[height=4cm,width=4cm]{Red3LW14Ent2} }
	\qquad 
	\subfloat[][(d)]{\includegraphics[height=4cm,width=4cm]{Red3LW14Ent2} }
	\qquad 
	\subfloat[][(e)]{\includegraphics[height=4cm,width=4cm]{Red3LW14Ent2} }
	\caption{\reducedWebs for Cweb $ W^{(2,1)}_{3,\text{I}}(1,3,3) $ }
	\label{fig:six-one-web14-allAVAtar-WEBS}
\end{figure}
\noindent The procedure Fused-Webs is applied to this Cweb, which results in table \ref{tab:six-one-web14-av-Ent}. This table classifies the irreducible diagrams of the Cweb according to the entangled pieces and provides the \reducedWebs with the associated mixing matrices for the Cweb.

\begin{table}[H]
	\begin{center}
		\begin{tabular}{|c|c|c|c|c|c|}
			\hline
			Entanglement & Diagrams of  & \reducedWeb & Diagrams in  & $ s $-factors   & $ R $ \\ 
			& Cweb &  & \reducedWeb &    &  \\
			\hline & & & & &\\
			Complete entangled  & $ d_1 $, $ d_2 $,  & \ref{fig:six-one-web14-allAVAtar-WEBS}\textcolor{blue}{a} & - & 1 & $ I_{4} $ \\ 
			& $ d_3 $, $ d_4 $& & & &\\
			\hline
			First Partial Entangled  & $ d_{5} $, $ d_{6} $ & \ref{fig:six-one-web14-allAVAtar-WEBS}\textcolor{blue}{b} & $ \{b,\,E\} $ & 1 & $ R(1_2) $ \\ 
			&  & & $ \{E,\,b\} $ & 1 & \\ \hline
			Second Partial Entangled  & $ d_{7} $, $ d_{8} $ & \ref{fig:six-one-web14-allAVAtar-WEBS}\textcolor{blue}{c} & $ \{b,\,E\} $ & 1 & $ R(1_2) $ \\ 
			&  & & $ \{E,\,b\} $ & 1 & \\ \hline
			Third Partial Entangled  & $ d_{9} $, $ d_{10} $ & \ref{fig:six-one-web14-allAVAtar-WEBS}\textcolor{blue}{d} & $ \{b,\,E\} $ & 1 & $ R(1_2) $ \\ 
			&  & & $ \{E,\,b\} $ & 1 & \\ \hline
			Fourth Partial Entangled  & $ d_{11} $, $ d_{12} $ & \ref{fig:six-one-web14-allAVAtar-WEBS}\textcolor{blue}{e} & $ \{b,\,E\} $ & 1 & $ R(1_2) $ \\ 
			&  & & $ \{E,\,b\} $ & 1 & \\ \hline
		\end{tabular}	
	\end{center}
	\caption{\reducedWebs and their mixing matrices for Cweb $ W^{(2,1)}_{3,\text{I}}(1,3,3) $}
	\label{tab:six-one-web14-av-Ent}
\end{table}

The order of diagrams in the Cweb given in table \ref{tab:six-one-web14-av}, is chosen such that diagrams with same kind of entangled piece appear together. Therefore, mixing matrices of the \reducedWebs for this Cweb, present on the diagonal blocks of $ A $, are given as,

\begin{align}
A\,=\,\left(\begin{array}{c|cc}
\textbf{I}_{4} & & \cdots\\
\hline 
\textbf{O}_{8\times 4}& & \begin{array}{cccc}
R\,(1_2) & && \\
& R\,(1_2)&&\\
& & R\,(1_2)&\\
&&&R\,(1_2)\\
\end{array}	
\end{array}\right)\,,\qquad 
\end{align}
the rank of $ A $ is,

\begin{align}
r(A) &= r(I_{4})  +  4\,r(R(1_2))\nonumber\\
&=  8\,.
\end{align}

\noindent The number of exponentiated colour factors is the rank of the mixing matrix. Thus using eq.~(\ref{eq:rank-I-6-class}), the rank of $ R $  is,

\begin{align}
r(R) = r(A)  + 2  = 10\,.
\end{align}
\noindent\textbf{3.}\, $ \textbf{W}^{(2,1)}_{3}(1,2,4) $

\vspace{0.2cm}

\begin{figure}[H]
	\captionsetup[subfloat]{labelformat=empty}
	\centering
	\subfloat[][]{\includegraphics[height=4cm,width=4cm]{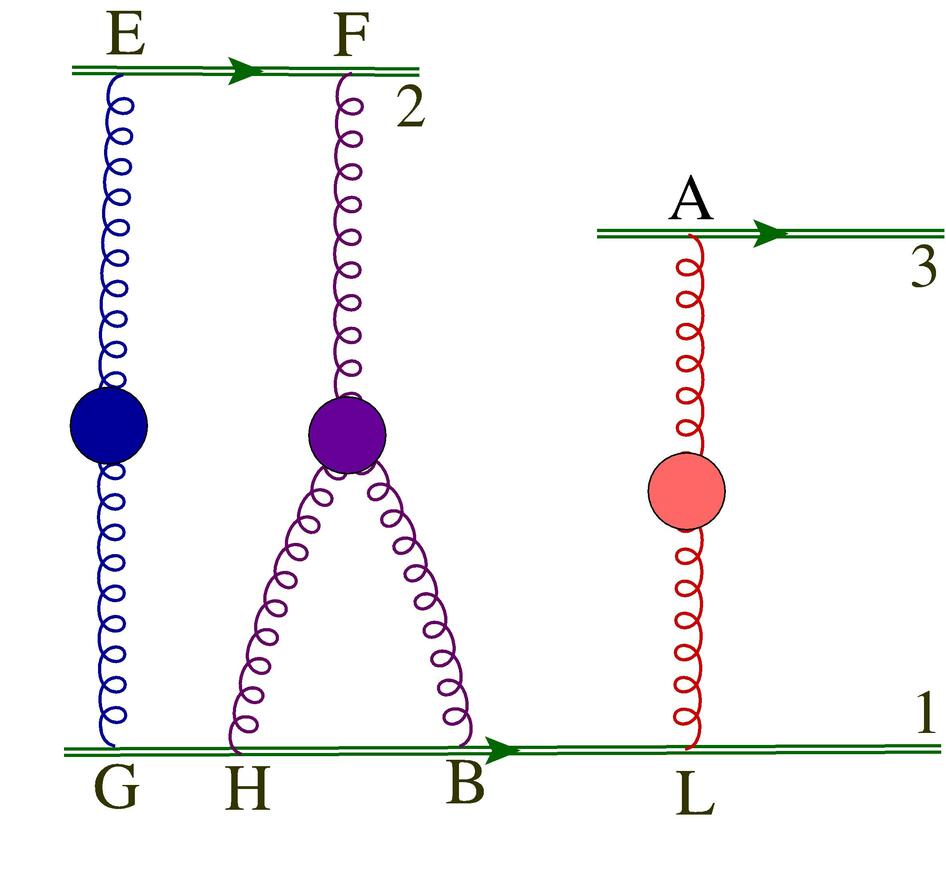} }
	\vspace*{-0.5cm}
	\caption{Cweb $ W^{(2,1)}_{3}(1,2,4) $}
	\label{fig:six-one-web18-av}
\end{figure}

\noindent This Cweb, shown in fig.~(\ref{fig:six-one-web18-av}) has twenty-four diagrams, out of which six are reducible, eight are completely entangled, and remaining ten are partially entangled The Normal ordered diagrams and their $ s $-factors are shown in table \ref{tab:six-one-web18-av}. 

\begin{table}[H]
	\vspace*{-2.2cm}
	\begin{minipage}[c]{0.5\textwidth}
		\begin{table}[H]
			\begin{center}
					\resizebox{0.9\textwidth}{!}{%
				\begin{tabular}{|c|c|c|}
					\hline 
					\textbf{Diagrams}  & \textbf{Sequences}  & \textbf{s-factors}  \\ 
					\hline
					$d_1$ & $\lbrace\lbrace GLHB \rbrace, \lbrace FE\rbrace \rbrace$ & 0 \\ \hline
					$d_2$ & $\lbrace\lbrace HLGB \rbrace, \lbrace EF\rbrace \rbrace$ & 0 \\ \hline
					$d_3$ & $\lbrace\lbrace HLGB \rbrace, \lbrace FE\rbrace \rbrace$ & 0 \\ \hline
					$d_{4}$ & $\lbrace\lbrace GHLB \rbrace, \lbrace FE\rbrace \rbrace$ & 0 \\ \hline
					$d_{5}$ & $\lbrace\lbrace HGLB \rbrace, \lbrace EF\rbrace \rbrace$ & 0 \\ \hline
					$d_{6}$ & $\lbrace\lbrace HGLB \rbrace, \lbrace FE\rbrace \rbrace$ & 0 \\ \hline
					$d_{7}$ & $\lbrace\lbrace HLBG \rbrace, \lbrace EF\rbrace \rbrace$ & 0 \\ \hline
					$d_{8}$ & $\lbrace\lbrace HBLG \rbrace, \lbrace EF\rbrace \rbrace$ & 0 \\ \hline
					$d_9$ & $\lbrace\lbrace LGHB \rbrace, \lbrace FE\rbrace \rbrace$ & 0 \\ \hline
					$d_{10}$ & $\lbrace\lbrace GHBL \rbrace, \lbrace FE\rbrace \rbrace$ & 0 \\ \hline
					$d_{11}$ & $\lbrace\lbrace LHGB \rbrace, \lbrace EF\rbrace \rbrace$ & 0 \\ \hline
					$d_{12}$ & $\lbrace\lbrace HGBL \rbrace, \lbrace EF\rbrace \rbrace$ & 0 \\ \hline				
				\end{tabular}}
			\end{center}
		\end{table}
	\end{minipage}
	\begin{minipage}[c]{0.5\textwidth}
		\begin{table}[H]
			\begin{center}
					\resizebox{0.9\textwidth}{!}{%
				\begin{tabular}{|c|c|c|}
					\hline 
					\textbf{Diagrams}  & \textbf{Sequences}  & \textbf{s-factors}  \\ 
					\hline
					$d_{13}$ & $\lbrace\lbrace LHGB \rbrace, \lbrace FE\rbrace \rbrace$ & 0 \\ \hline
					$d_{14}$ & $\lbrace\lbrace HGBL \rbrace, \lbrace FE\rbrace \rbrace$ & 0 \\ \hline
					$d_{15}$ & $\lbrace\lbrace GHLB \rbrace, \lbrace EF\rbrace \rbrace$ & 0 \\ \hline
					$d_{16}$ & $\lbrace\lbrace HLBG \rbrace, \lbrace FE\rbrace \rbrace$ & 0 \\ \hline		
					$d_{17}$ & $\lbrace\lbrace LHBG \rbrace, \lbrace EF\rbrace \rbrace$ & 0\\ \hline
					$d_{18}$ & $\lbrace\lbrace HBGL \rbrace, \lbrace EF\rbrace \rbrace$ & 0 \\ \hline
					$d_{19}$ & $\lbrace\lbrace LGHB \rbrace, \lbrace EF\rbrace \rbrace$ & 1 \\ \hline
					$d_{20}$ & $\lbrace\lbrace GLHB\rbrace, \lbrace EF\rbrace \rbrace$ & 1 \\ \hline
					$d_{21}$ & $\lbrace\lbrace LHBG \rbrace, \lbrace FE\rbrace \rbrace$ & 1 \\ \hline
					$d_{22}$ & $\lbrace\lbrace HBLG \rbrace, \lbrace FE\rbrace \rbrace$ & 1 \\ \hline
					$d_{23}$ & $\lbrace\lbrace GHBL \rbrace, \lbrace EF\rbrace \rbrace$ & 1 \\ \hline
					$d_{24}$ & $\lbrace\lbrace HBGL \rbrace, \lbrace FE\rbrace \rbrace$ & 1 \\ \hline
				\end{tabular}}
			\end{center}
		\end{table}
	\end{minipage}
	\caption{Normal ordered diagrams of Cweb $ W^{(2,1)}_{3}(1,2,4) $}
	\label{tab:six-one-web18-av}
\end{table}

\begin{figure}[H]
	\vspace*{-0.5cm}
	\captionsetup[subfloat]{labelformat=empty}
	\centering
	\subfloat[][(a)]{\includegraphics[height=4cm,width=4cm]{Red3LWanyEnt0} }
	\qquad 
	\subfloat[][(b)]{\includegraphics[height=4cm,width=4cm]{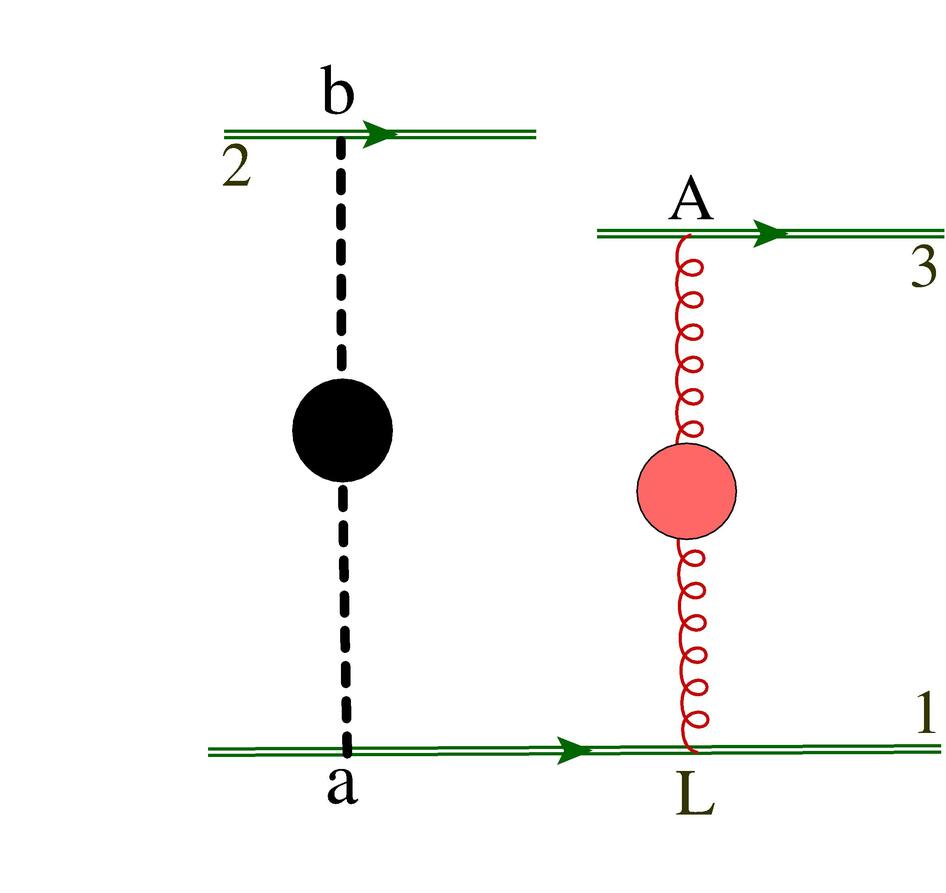} }
	\qquad 
	\subfloat[][(c)]{\includegraphics[height=4cm,width=4cm]{Red3LW18Ent1} }
	\qquad 
	\subfloat[][(d)]{\includegraphics[height=4cm,width=4cm]{Red3LW18Ent1} }
	\qquad 
	\subfloat[][(e)]{\includegraphics[height=4cm,width=4cm]{Red3LW18Ent1} }
	\qquad 
	\subfloat[][(f)]{\includegraphics[height=4cm,width=4cm]{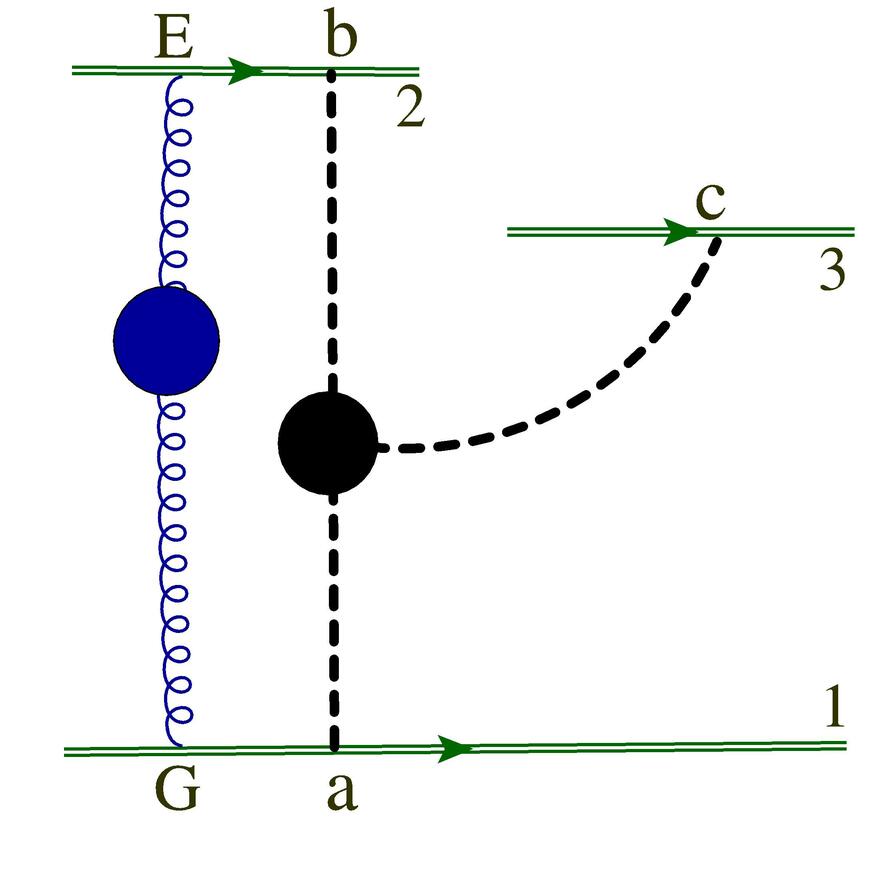} }
	\caption{\reducedWebs for Cweb $ W^{(2,1)}_{3}(1,2,4) $ }
	\label{fig:six-one-web18-allAVAtar-WEBS}
\end{figure}
\noindent Fused-Webs algorithm is applied to this Cweb, which results in table \ref{tab:six-one-web18-av-Ent}. It classifies the irreducible diagrams of the Cweb according to the entangled pieces. This table provides the \reducedWebs with the associated mixing matrices for the Cweb.

\begin{table}[H]
	\begin{center}
		\begin{tabular}{|c|c|c|c|c|c|}
			\hline
			Entanglement & Diagrams of  & \reducedWeb & Diagrams in  & $ s $-factors   & $ R $ \\ 
			& Cweb &  & \reducedWeb &    &  \\
			\hline & & & & &\\
			Complete entangled  & $ d_1 $, $ d_2 $, $ d_3 $, $ d_4 $  & \ref{fig:six-one-web18-allAVAtar-WEBS}\textcolor{blue}{a} & - & 1 & $ I_{8} $ \\ 
			& $ d_5 $, $ d_6 $, $ d_7 $, $ d_8 $ & & & &\\
			\hline
			First Partial Entangled  & $ d_{9} $, $ d_{10} $ & \ref{fig:six-one-web18-allAVAtar-WEBS}\textcolor{blue}{b} & $ \{a\,L\} $ & 1 & $ R(1_2) $ \\ 
			&  & & $\{L\,a\} $ & 1 & \\ \hline
			Second Partial Entangled  & $ d_{11} $, $ d_{12} $ & \ref{fig:six-one-web18-allAVAtar-WEBS}\textcolor{blue}{c} & $\,\{a\,L\} $ & 1 & $ R(1_2) $ \\ 
			&  & & $ \{L\,a\} $ & 1 & \\ \hline
			Third Partial Entangled  & $ d_{13} $, $ d_{14} $ & \ref{fig:six-one-web18-allAVAtar-WEBS}\textcolor{blue}{d} & $\{a\,L\}  $ & 1 & $ R(1_2) $ \\ 
			&  & & $\{L\,a\} $ & 1 & \\ \hline
			Fourth Partial Entangled  & $ d_{15} $, $ d_{16} $ & \ref{fig:six-one-web18-allAVAtar-WEBS}\textcolor{blue}{e} & $ \{a\,L\}  $ & 1 & $ R(1_2) $ \\ 
			&  & & $ \{L\,a\} $ & 1 & \\ \hline
			Fifth Partial Entangled  & $ d_{17} $, $ d_{18} $ & \ref{fig:six-one-web18-allAVAtar-WEBS}\textcolor{blue}{f} & $ \{G\,a\},\{E\,b\}  $ & 1 & $ R(1_2) $ \\ 
			&  & & $ \{a\,G\},\{b\,E\} $ & 1 & \\ \hline  
		\end{tabular}	
	\end{center}
	\caption{\reducedWebs and their mixing matrices for Cweb $ W^{(2,1)}_{3}(1,2,4) $ }
	\label{tab:six-one-web18-av-Ent}
\end{table}

\noindent The order of diagrams in the Cweb given in table \ref{tab:six-one-web18-av}, is chosen such that diagrams with same kind of entangled piece appear together. Therefore, mixing matrices of the \reducedWebs for this Cweb, present on the diagonal blocks of $ A $, are given as,

\begin{align}
A\,=\,\left(\begin{array}{c|cc}
\textbf{I}_{8} & & \cdots\\
\hline 
\textbf{O}_{10\times 8}& & \begin{array}{ccccc}
R\,(1_2) & && &\\
& R\,(1_2)&&&\\
& & R\,(1_2)&&\\
&&&R\,(1_2)&\\
&&&&R(1_2)\\
\end{array}	
\end{array}\right)\,,\qquad 
\end{align}
the rank of $ A $ is,

\begin{align}
r(A) &= r(I_{8})  +  5\,r(R(1_2))\nonumber\\
&=  13\,. 
\end{align}

\noindent The number of exponentiated colour factors is the rank of the mixing matrix. The rank is, using eq.~(\ref{eq:rank-I-6-class}), given as,

\begin{align}
r(R) = r(A)  + 2  = 15\,.
\end{align}
\noindent \textbf{4.}\, $ \textbf{W}^{(2,1)}_{4}(1,1,1,4) $\\

\begin{figure}[H]
	\captionsetup[subfloat]{labelformat=empty}
	\centering
	\subfloat[][]{\includegraphics[height=4cm,width=4cm]{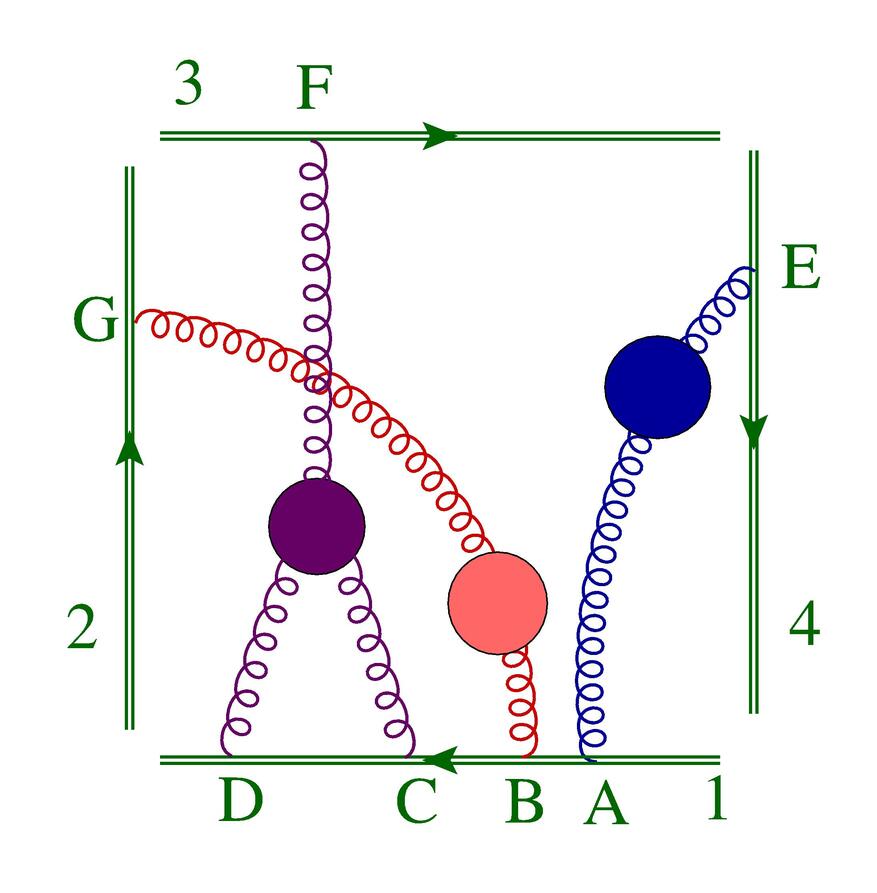} }
	\vspace{-0.5cm}
	\caption{Cweb $ W^{(2,1)}_{4}(1,1,1,4) $}
	\label{fig:six-one-web4-6-av}
\end{figure}

\noindent This Cweb, shown in fig.~(\ref{fig:six-one-web4-6-av}) has twelve diagrams, out of which six are reducible, two are completely entangled, and remaining four are partially entangled. The Normal ordered diagrams and their $ s $-factors are shown in table \ref{tab:six-one-web4-6-av}. 

\begin{table}[H]
	\begin{minipage}[c]{0.5\textwidth}
		\begin{table}[H]
			\begin{center}
					\resizebox{0.9\textwidth}{!}{%
				\begin{tabular}{|c|c|c|}
					\hline 
					\textbf{Diagrams}  & \textbf{Sequences}  & \textbf{s-factors}  \\ 
					\hline
					$d_1$ & $\lbrace\lbrace CABD\rbrace\rbrace$ & 0\\ \hline
					$d_2$ & $\lbrace\lbrace CBAD\rbrace\rbrace$ & 0\\ \hline
					$d_3$ & $\lbrace\lbrace ACBD\rbrace\rbrace$ & 0\\ \hline
					$d_{4}$ & $\lbrace\lbrace CBDA\rbrace\rbrace$ & 0\\ \hline
					$d_5$ & $\lbrace\lbrace BCAD\rbrace\rbrace$ & 0\\ \hline
					$d_6$ & $\lbrace\lbrace CADB\rbrace\rbrace$ & 0\\ \hline
				\end{tabular}}
			\end{center}
		\end{table}
	\end{minipage}
	\begin{minipage}[c]{0.5\textwidth}
		\begin{table}[H]
			\begin{center}
				\resizebox{0.9\textwidth}{!}{%
				\begin{tabular}{|c|c|c|}
					\hline 
					\textbf{Diagrams}  & \textbf{Sequences}  & \textbf{s-factors}  \\ 
					\hline
					$d_7$ & $\lbrace\lbrace ABCD\rbrace\rbrace$ & 1 \\ \hline
					$d_8$ & $\lbrace\lbrace BACD\rbrace\rbrace$ & 1\\ \hline
					$d_9$ & $\lbrace\lbrace ACDB\rbrace\rbrace$ & 1\\ \hline
					$d_{10}$ & $\lbrace\lbrace CDAB\rbrace\rbrace$ & 1\\ \hline
					$d_{11}$ & $\lbrace\lbrace BCDA\rbrace\rbrace$ & 1\\ \hline
					$d_{12}$ & $\lbrace\lbrace CDBA\rbrace\rbrace$ & 1\\ \hline
				\end{tabular}}
			\end{center}
		\end{table}
	\end{minipage}
	\caption{Normal ordered diagrams of Cweb $ W^{(2,1)}_{4}(1,1,1,4) $}
	\label{tab:six-one-web4-6-av}
\end{table}

\begin{figure}[H]
	\vspace*{-1cm}
	\captionsetup[subfloat]{labelformat=empty}
	\centering
	\subfloat[][(a)]{\includegraphics[height=4cm,width=4cm]{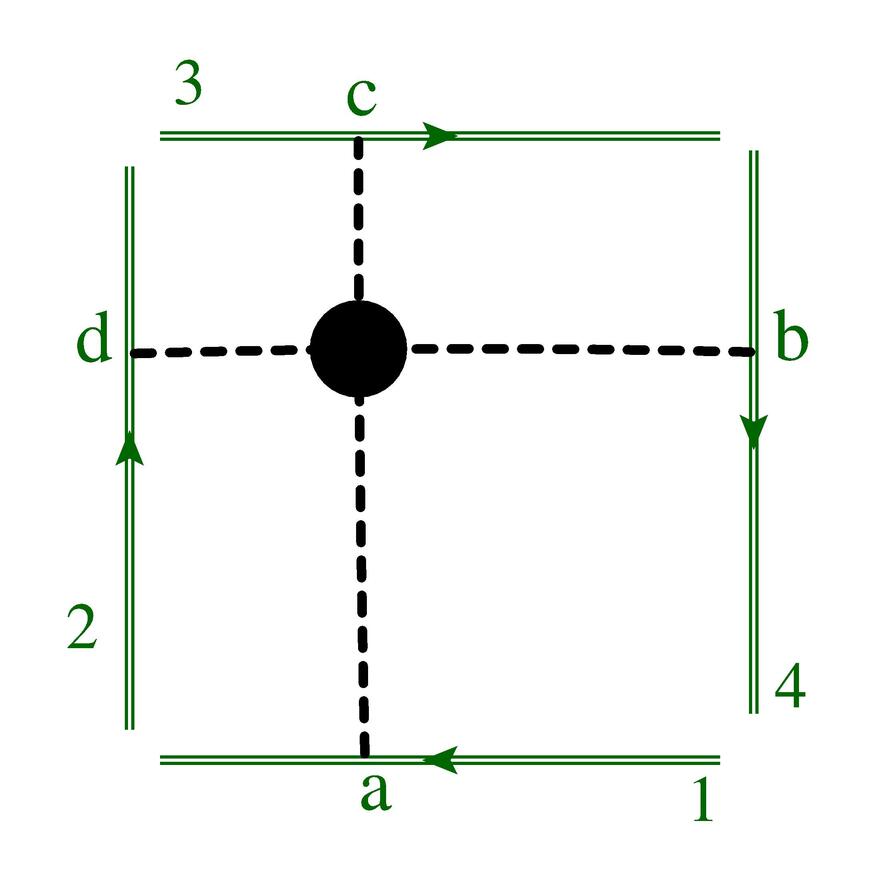} }
	\qquad 
	\subfloat[][(b)]{\includegraphics[height=4cm,width=4cm]{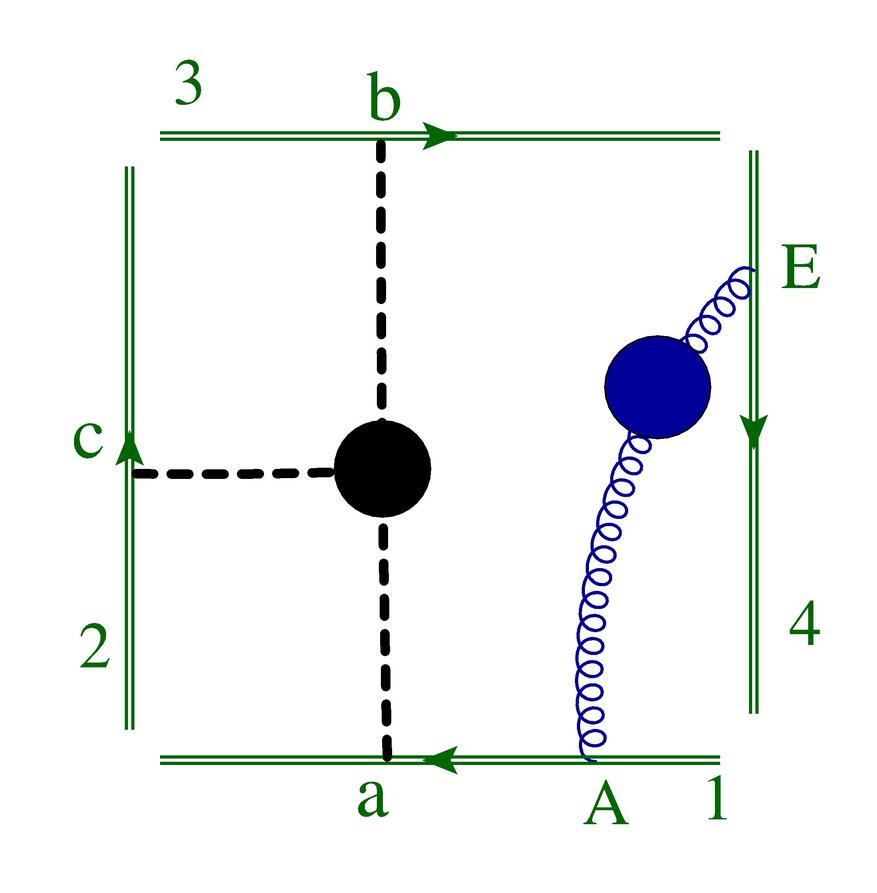} }
	\qquad 
	\subfloat[][(c)]{\includegraphics[height=4cm,width=4cm]{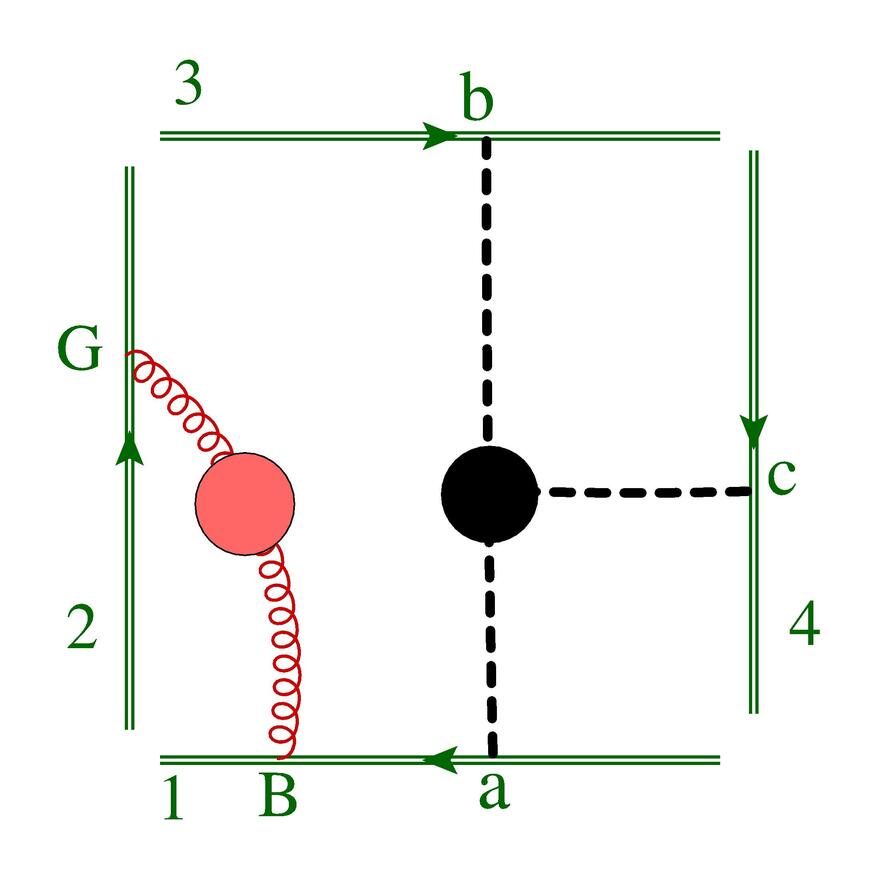} }
	\caption{\reducedWebs for Cweb $ W^{(2,1)}_{4}(1,1,1,4) $ }
	\label{fig:six-one-web4-6-allAVAtar-WEBS}
\end{figure}
\noindent The procedure developed in section \ref{sec:avatar-webs}, is applied to this Cweb, which results in table \ref{tab:six-one-web4-6-av-Ent}. It classifies the irreducible diagrams of the Cweb according to the entangled pieces. This table provides the \reducedWebs with the associated mixing matrices for the Cweb.

\begin{table}[H]
	\begin{center}
		\begin{tabular}{|c|c|c|c|c|c|}
			\hline
			Entanglement & Diagrams of  & \reducedWeb & Diagrams in  & $ s $-factors   & $ R $ \\ 
			& Cweb &  & \reducedWeb &    &  \\
			\hline
			Complete entangled  & $ d_1 $, $ d_2 $  & \ref{fig:six-one-web4-6-allAVAtar-WEBS}\textcolor{blue}{a} & - & 1 & $ I_{2} $ \\ 
			&&&& & \\
			\hline
			First Partial Entangled  & $ d_{3} $, $ d_{4} $ & \ref{fig:six-one-web4-6-allAVAtar-WEBS}\textcolor{blue}{b} & $ \{A,\,a\} $ & 1 & $ R(1_2) $ \\ 
			&  & & $ \{a,\,A\} $ & 1 & \\ \hline
			Second Partial Entangled  & $ d_{5} $, $ d_{6} $ & \ref{fig:six-one-web4-6-allAVAtar-WEBS}\textcolor{blue}{c} & $ \{a,\,B\} $ & 1 & $ R(1_2) $ \\ 
			&  & & $ \{B,\,a\} $ & 1 & \\ \hline
			
		\end{tabular}	
	\end{center}
	\caption{\reducedWebs and their mixing matrices for Cweb $ W^{(2,1)}_{4}(1,1,1,4) $}
	\label{tab:six-one-web4-6-av-Ent}
\end{table}

\noindent The order of diagrams in the Cweb given in table \ref{tab:six-one-web4-6-av}, is chosen such that diagrams with same kind of entangled piece appear together. Therefore, mixing matrices of the \reducedWebs for this Cweb, present on the diagonal blocks of $ A $, are given as,

\begin{align}
A\,=\,\left(\begin{array}{c|cc}
\textbf{I}_{2} & & \cdots\\
\hline 
\textbf{O}_{4\times 2}& & \begin{array}{cc}
R\,(1_2) & \\
& R\,(1_2)
\end{array}	
\end{array}\right)\,,\qquad 
\end{align}
the rank of $ A $ is,

\begin{align}
r(A) &= r(I_{2})  +  2\,r(R(1_2))\nonumber\\
&=  4\,.
\end{align}

\noindent The number of exponentiated colour factors is the rank of the mixing matrix, which is given as

\begin{align}
r(R) = r(A)  + 2  = 6\,.
\end{align}

\noindent \textbf{5.}\, $ \textbf{W}^{(2,1)}_{4,\text{III}}(1,2,2,2) $

\vspace{0.2cm}

\begin{figure}
	\centering
	\subfloat[][]{\includegraphics[height=4cm,width=4cm]{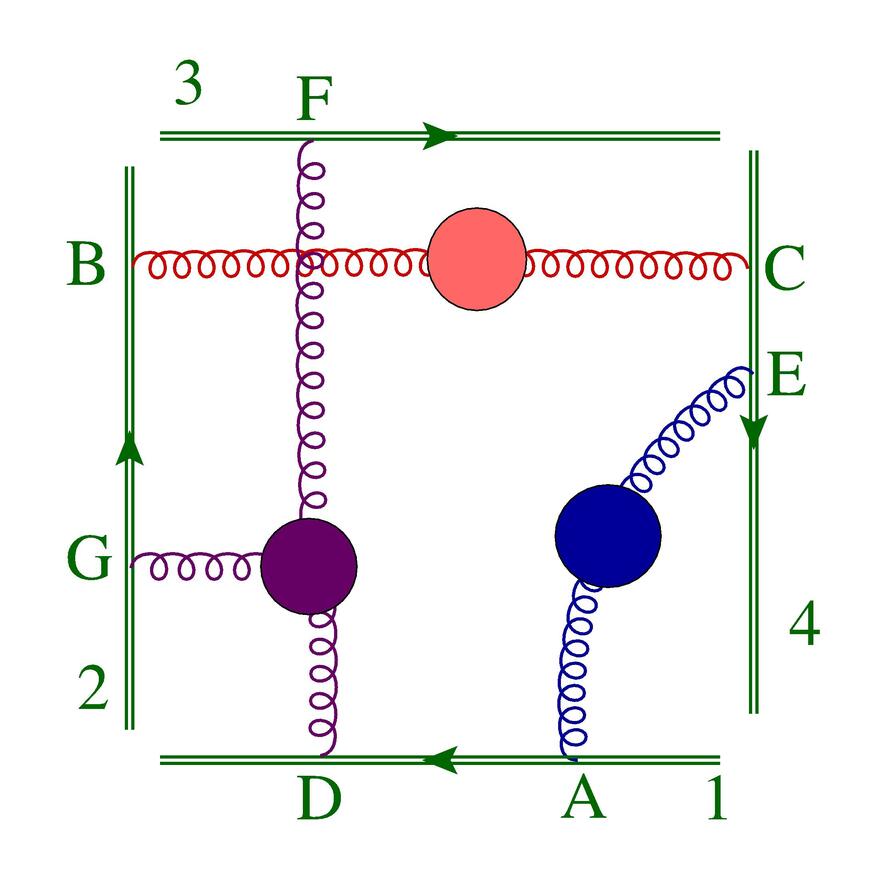} }
	\caption{Cweb $ W^{(2,1)}_{4,\text{III}}(1,2,2,2) $}
	\label{fig:six-one-web4-14-av}
\end{figure}

\noindent This Cweb, shown in fig.~(\ref{fig:six-one-web4-14-av}) has eight diagrams, out of which six are reducible, two are completely entangled, and there are no partially entangled diagram. The Normal ordered diagrams and their $ s $-factors are shown in table \ref{tab:six-one-web4-14-av}. 

\begin{table}[H]
	\begin{minipage}[c]{0.5\textwidth}
		\begin{table}[H]
			\begin{center}
					\resizebox{0.9\textwidth}{!}{%
				\begin{tabular}{|c|c|c|}
					\hline 
					\textbf{Diagrams}  & \textbf{Sequences}  & \textbf{s-factors}  \\ 
					\hline
					$d_1$ & $\lbrace \lbrace AD\rbrace,\lbrace GB \rbrace, \lbrace CE\rbrace \rbrace$ & 0 \\ 
					\hline
					$d_2$ & $\lbrace \lbrace DA\rbrace,\lbrace BG \rbrace, \lbrace EC\rbrace \rbrace$ & 0 \\ 
					\hline
					$d_3$ & $\lbrace \lbrace AD\rbrace,\lbrace BG \rbrace, \lbrace CE\rbrace \rbrace$ & 1 \\ 
					\hline
					$d_4$ & $\lbrace \lbrace AD\rbrace,\lbrace BG \rbrace, \lbrace EC\rbrace \rbrace$ & 1 \\ 
					\hline
				\end{tabular}}
			\end{center}
		\end{table}
	\end{minipage}
	\begin{minipage}[c]{0.5\textwidth}
		\begin{table}[H]
			\begin{center}
					\resizebox{0.9\textwidth}{!}{%
				\begin{tabular}{|c|c|c|}
					\hline 
					\textbf{Diagrams}  & \textbf{Sequences}  & \textbf{s-factors}  \\ 
					\hline
					$d_5$ & $\lbrace \lbrace AD\rbrace,\lbrace GB \rbrace, \lbrace EC\rbrace \rbrace$ & 1 \\ 
					\hline
					$d_6$ & $\lbrace \lbrace DA\rbrace,\lbrace BG \rbrace, \lbrace CE\rbrace \rbrace$ & 1 \\ 
					\hline
					$d_7$ & $\lbrace \lbrace DA\rbrace,\lbrace GB \rbrace, \lbrace CE\rbrace \rbrace$ & 1 \\ 
					\hline
					$d_8$ & $\lbrace \lbrace DA\rbrace,\lbrace GB \rbrace, \lbrace EC\rbrace \rbrace$ & 1 \\ 
					\hline
				\end{tabular}}
			\end{center}
		\end{table}
	\end{minipage}
	\caption{Normal ordered diagrams of Cweb $ W^{(2,1)}_{4,\text{III}}(1,2,2,2) $}
	\label{tab:six-one-web4-14-av}
\end{table}

\begin{figure}[H]
	\captionsetup[subfloat]{labelformat=empty}
	\centering
	\subfloat[][(a)]{\includegraphics[height=4cm,width=4cm]{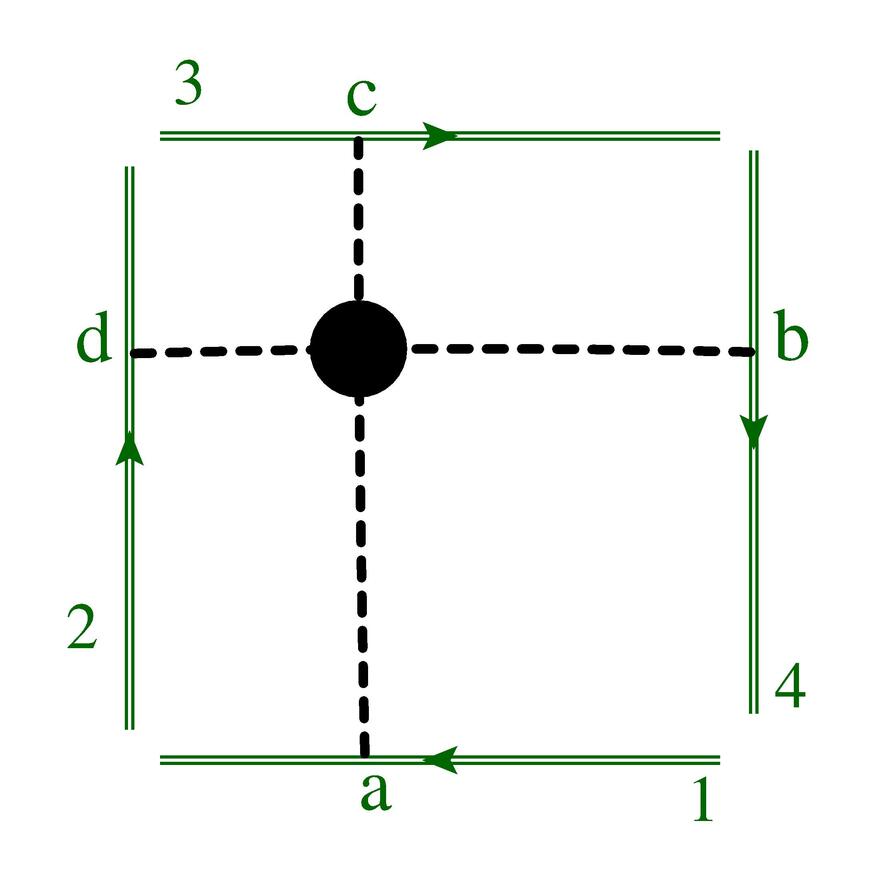} }
	\caption{\reducedWeb for Cweb $ W^{(2,1)}_{4,\text{III}}(1,2,2,2) $ }
	\label{fig:six-one-web4-14-allAVAtar-WEBS}
\end{figure}

\begin{table}[H]
	\begin{center}
		\begin{tabular}{|c|c|c|c|c|c|}
			\hline
			Entanglement & Diagrams of  & \reducedWeb & Diagrams in  & $ s $-factors   & $ R $ \\ 
			& Cweb &  & \reducedWeb &    &  \\
			\hline & & & & &\\
			Complete entangled  & $ d_1 $, $ d_2 $  & \ref{fig:six-one-web4-14-allAVAtar-WEBS}\textcolor{blue}{a} & - & 1 & $ I_{2} $ \\ 
			& & & & &\\
			\hline
		\end{tabular}	
	\end{center}
	\caption{\reducedWebs and their mixing matrices for Cweb $ W^{(2,1)}_{4,\text{III}}(1,2,2,2) $}
	\label{tab:six-one-web4-14-av-Ent}
\end{table}

\noindent The application of \reducedWebs results in table \ref{tab:six-one-web4-14-av-Ent}. It classifies the irreducible diagrams of the Cweb according to the entangled pieces. This table provides the \reducedWebs with the associated mixing matrices for the Cweb. The order of diagrams in the Cweb given in table \ref{tab:six-one-web4-14-av}, is chosen such that diagrams with same kind of entangled piece appear together. Therefore the sub-matrix $ A $, for this Cweb is,

\begin{align}
A  =  \textbf{I}_2\,.
\end{align}
Thus the rank of $ A $ is,

\begin{align}
r(A) &= r(I_{2}) \nonumber\\
&=  2 
\end{align}

\noindent The number of exponentiated colour factors for this Cweb is,
\begin{align}
r(R) = r(A)  + 2  = 4\,.
\end{align}
\noindent\textbf{6.}\, $ \textbf{W}^{(2,1)}_{3,\text{III}}(2,2,3) $

\begin{figure}[H]
	\captionsetup[subfloat]{labelformat=empty}
	\centering
	\subfloat[][]{\includegraphics[height=4cm,width=4cm]{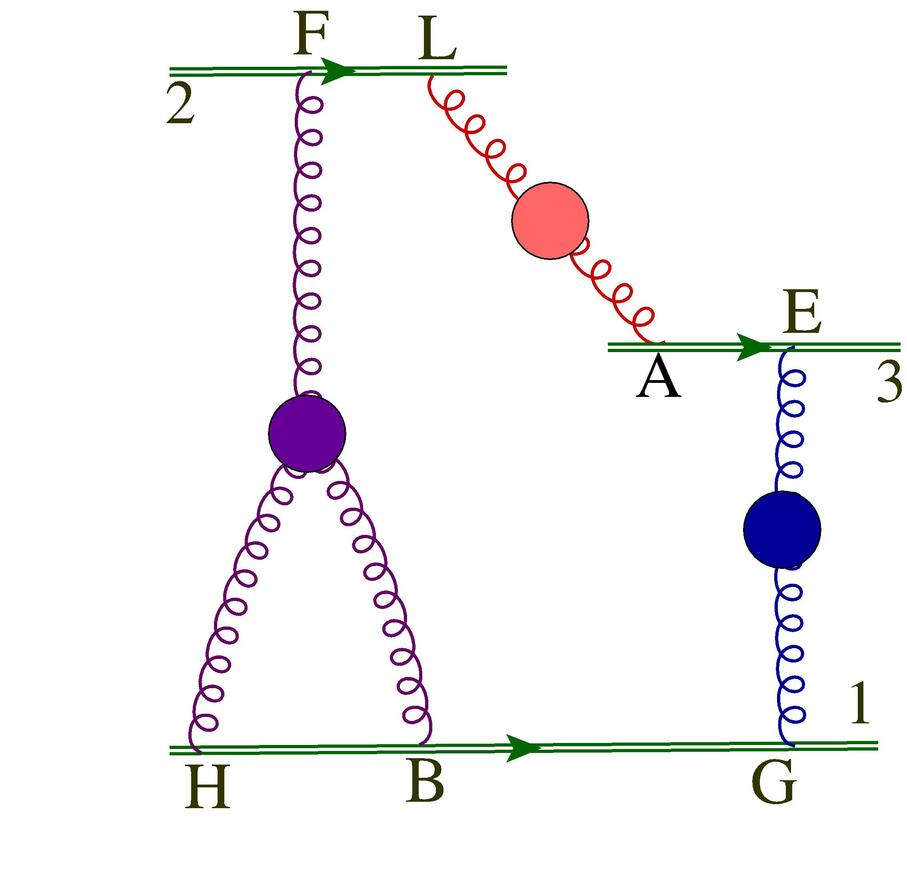} }
	\caption{Cweb $ W^{(2,1)}_{3,\text{III}}(2,2,3) $}
	\label{fig:six-one-web3-9-av}
\end{figure}

\vspace{0.2cm}
\noindent This Cweb, shown in fig.~(\ref{fig:six-one-web3-9-av}) has twelve diagrams, out of which six are reducible, four are completely entangled, and remaining two are partially entangled. The Normal ordered diagrams and their $ s $-factors are shown in table \ref{tab:six-one-web3-9-av}.

\begin{table}[H]
	\begin{minipage}[c]{0.5\textwidth}
		\begin{table}[H]
			\begin{center}
					\resizebox{0.9\textwidth}{!}{%
				\begin{tabular}{|c|c|c|}
					\hline 
					\textbf{Diagrams}  & \textbf{Sequences}  & \textbf{s-factors}  \\ 
					\hline
					$d_{1}$  & $\{FL\},\{AE\},\{GHB\}$  & 0 \\ \hline
					$d_{2}$  & $\{LF\},\{EA\},\{HBG\}$  & 0 \\ \hline
					$d_{3}$  & $\{FL\},\{AE\},\{HGB\}$& 0 \\ \hline 
					$d_{4}$  & $\{LF\},\{EA\},\{HGB\}$  & 0 \\ \hline
					$d_{5}$  & $\{FL\},\{EA\},\{HGB\}$  & 0 \\ \hline
					$d_{6}$  & $\{LF\},\{AE\},\{HGB\}$  & 0 \\ \hline
				\end{tabular}}
			\end{center}
		\end{table}
	\end{minipage}
	\begin{minipage}[c]{0.5\textwidth}
		\begin{table}[H]
			\begin{center}
					\resizebox{0.9\textwidth}{!}{%
				\begin{tabular}{|c|c|c|}
					\hline 
					\textbf{Diagrams}  & \textbf{Sequences}  & \textbf{s-factors}  \\ 
					\hline
					$d_{7}$  & $\{FL\},\{AE\},\{HBG\}$  & 1 \\ \hline
					$d_{8}$  & $\{LF\},\{AE\},\{HBG\}$  & 1 \\ \hline
					$d_{9}$  &  $\{FL\},\{EA\},\{HBG\}$  & 1 \\ \hline 
					$d_{10}$  & $\{FL\},\{EA\},\{GHB\}$  & 1\\ \hline
					$d_{11}$  & $\{LF\},\{EA\},\{GHB\}$  & 1 \\ \hline
					$d_{12}$  & $\{LF\},\{AE\},\{GHB\}$  & 1 \\ \hline
				\end{tabular}}
			\end{center}
		\end{table}
	\end{minipage}
	\caption{Normal ordered diagrams of Cweb $ {W}^{(2,1)}_{3,\text{III}}(2,2,3) $ }
	\label{tab:six-one-web3-9-av}
\end{table}

\begin{figure}[H]
	\captionsetup[subfloat]{labelformat=empty}
	\centering
	\subfloat[][(a)]{\includegraphics[height=4cm,width=4cm]{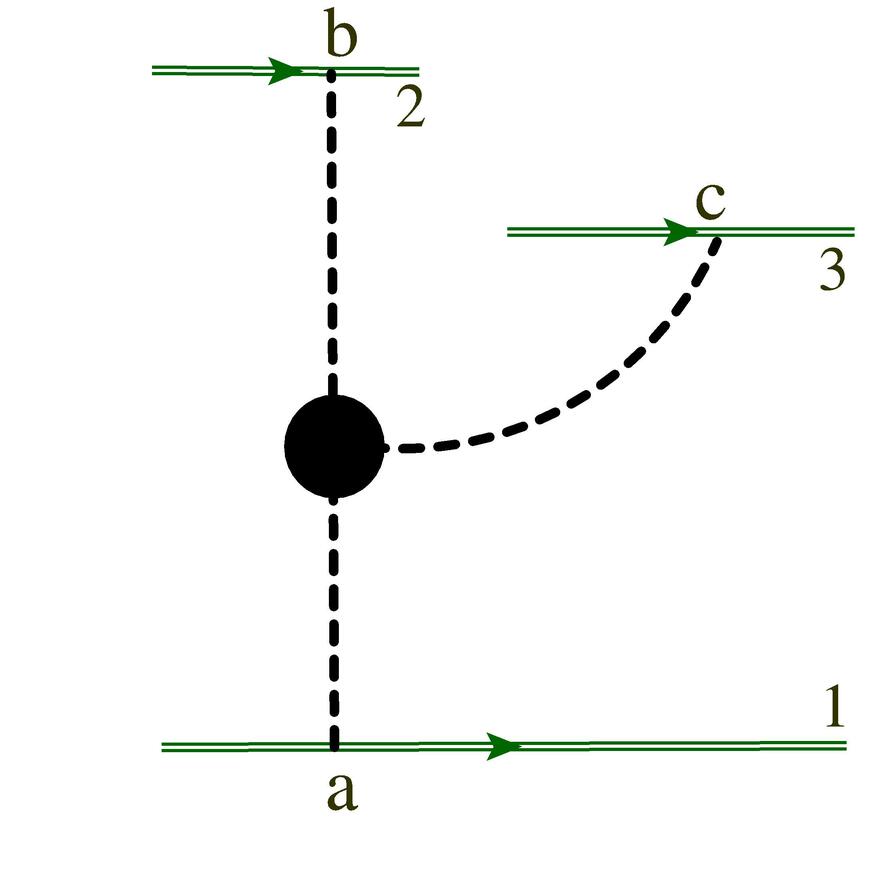} }
	\qquad 
	\subfloat[][(b)]{\includegraphics[height=4cm,width=4cm]{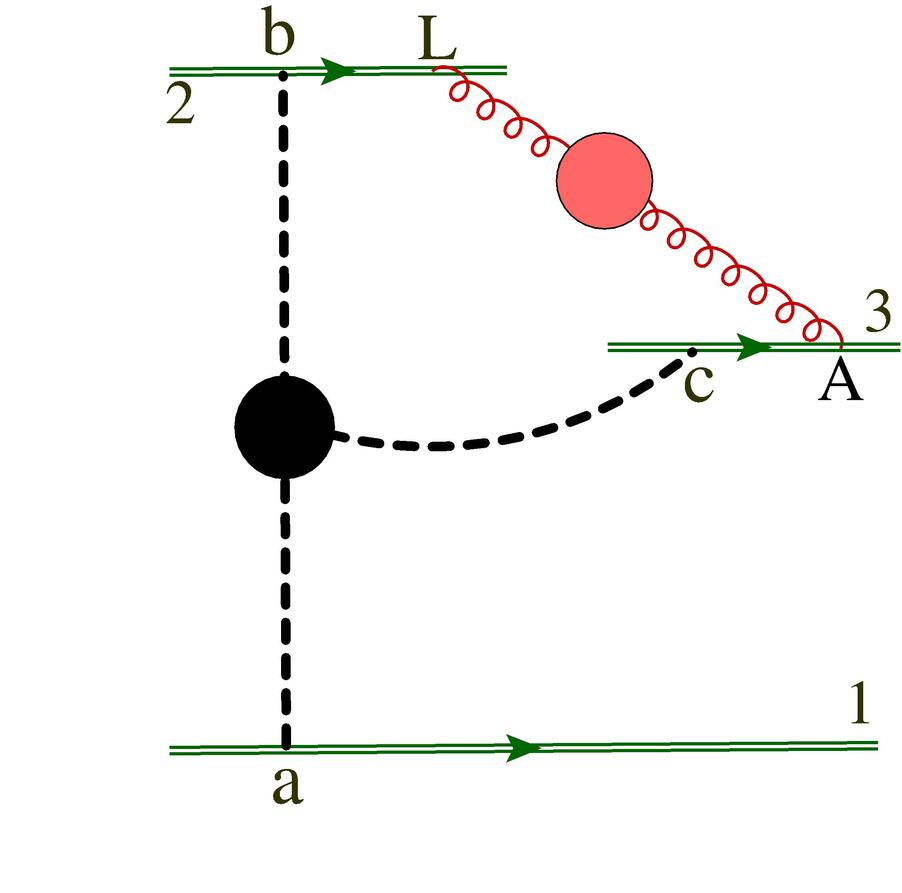} }
	\caption{\reducedWebs for Cweb $ W^{(2,1)}_{3,\text{III}}(2,2,3) $ }
	\label{fig:six-one-web3-9-allAVAtar-WEBS}
\end{figure}
\noindent The procedure of Fused-Webs is applied to this Cweb, which results in table \ref{tab:six-one-web3-9-av-Ent}. It classifies the irreducible diagrams of the Cweb according to the entangled pieces. This table provides the \reducedWebs with the associated mixing matrices for the Cweb.

\begin{table}[H]
	\begin{center}
		\begin{tabular}{|c|c|c|c|c|c|}
			\hline
			Entanglement & Diagrams of  & \reducedWeb & Diagrams in  & $ s $-factors   & $ R $ \\ 
			& Cweb &  & \reducedWeb &    &  \\
			\hline
			Complete entangled  & $ d_1 $, $ d_2 $,  & \ref{fig:six-one-web3-9-allAVAtar-WEBS}\textcolor{blue}{a} & - & 1 & $ I_{4} $ \\ 
			& $ d_3 $, $ d_4 $& & & &\\
			\hline
			First Partial Entangled  & $ d_{5} $, $ d_{6} $ & \ref{fig:six-one-web3-9-allAVAtar-WEBS}\textcolor{blue}{b} & $ \{c\,A\},\,\{b\,L\} $ & 1 & $ R(1_2) $ \\ 
			&  & & $ \{A\,c\},\,\{L,\,b\} $ & 1 & \\ \hline
		\end{tabular}	
	\end{center}
	\caption{\reducedWebs and their mixing matrices for Cweb $ W_{3,\text{III}}^{(2,1)}(2,2,3) $ }
	\label{tab:six-one-web3-9-av-Ent}
\end{table}

\noindent The order of diagrams in the Cweb given in table \ref{tab:six-one-web3-9-av}, is chosen such that diagrams with same kind of entangled piece appear together. Therefore, mixing matrices of the \reducedWebs for this Cweb, present on the diagonal blocks of $ A $, are given as,

\begin{align}
A\,=\,\left(\begin{array}{c|cc}
\textbf{I}_{4} & & \cdots\\
\hline 
\textbf{O}_{2\times 4}& & R(1_2)
\end{array}\right)\,,\qquad 
\end{align}
The rank of $ A $ is then,

\begin{align}
r(A) &= r(I_{4})  +  \,r(R(1_2))\nonumber\\
&=  5\,.
\end{align}

\noindent The number of exponentiated colour factors is the rank of the mixing matrix which is given as
\begin{align}
r(R) = r(A)  + 2  = 7\,.
\end{align}
\subsection{\reducedWebs of Cweb having $ S=\{0,0,\cdots,0,1_2,2_2\} $}\label{sec:1-22-2-appendix}

In this class of Cwebs, the $ D $-block of mixing matrices is $ R(1_2,2_2) $, whose full form is given in appendix \ref{sec:basis}. Thus the rank for this class of matrices will be, 

\begin{align}
r(R)\,&=\,r(A)+r(R(1_2,2_2))\nonumber \\
& =r(A) + 1 .
\label{eq:rank--1122-class}
\end{align} 

\noindent Hence, the diagonals blocks of sub-matrix $ A $ determines the rank for this class of Cwebs. 
\vspace{0.5cm}

\noindent\textbf{1.}\, $ \textbf{W}^{(2,1)}_{4,\text{I}}(1,1,2,3) $

\vspace{0.2cm}

\begin{figure}[H]
	\captionsetup[subfloat]{labelformat=empty}
	\centering
	\subfloat[][]{\includegraphics[height=4cm,width=4cm]{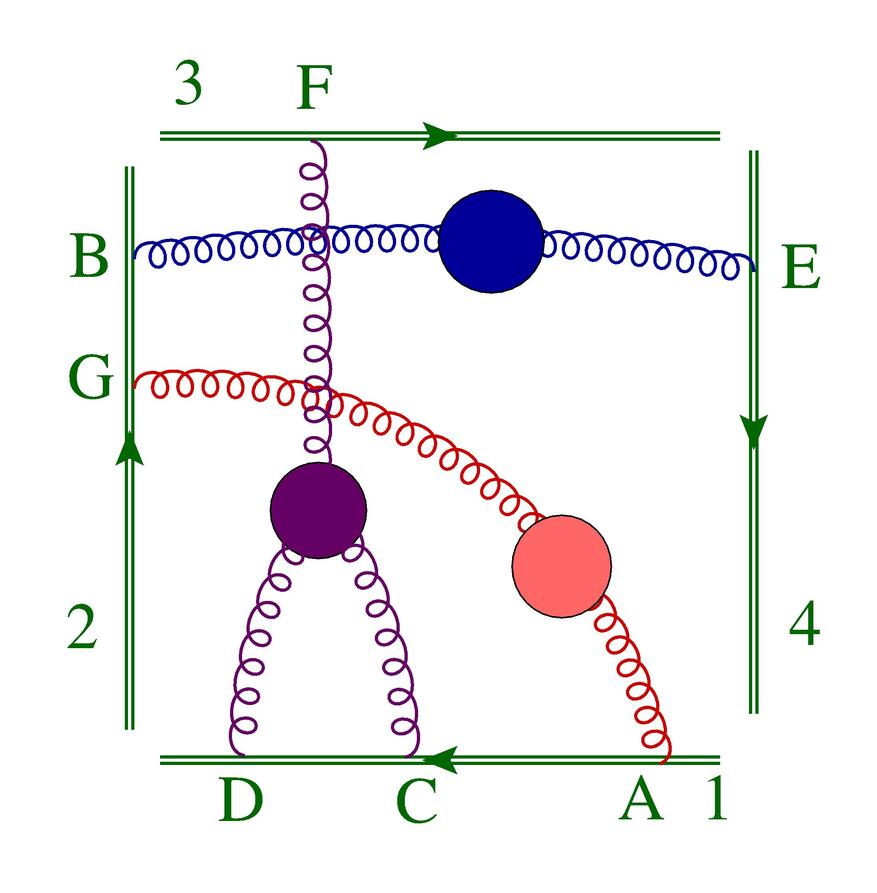} }
	\vspace{-0.5cm}
	\caption{Cweb $ W^{(2,1)}_{4,\text{I}}(1,1,2,3) $}
	\label{fig:six-one-web4-7-av}
\end{figure}

\noindent This Cweb, shown in fig.~(\ref{fig:six-one-web4-7-av}) has six diagrams, out of which four are reducible, and remaining two are partially entangled. The Normal ordered diagrams and their $ s $-factors are shown in table \ref{tab:six-one-web4-7-av}.

\begin{table}[H]
	\begin{minipage}[c]{0.5\textwidth}
		\begin{table}[H]
			\begin{center}
					\resizebox{0.9\textwidth}{!}{%
				\begin{tabular}{|c|c|c|}
					\hline 
					\textbf{Diagrams}  & \textbf{Sequences}  & \textbf{s-factors}  \\ 
					\hline
					$d_1$ & $\lbrace \lbrace BG\rbrace,\lbrace CAD\rbrace\rbrace$ & 0 \\ \hline
					$d_2$ & $\lbrace \lbrace GB\rbrace,\lbrace CAD\rbrace\rbrace$ & 0 \\ \hline
					$d_3$ & $\lbrace \lbrace BG\rbrace,\lbrace ACD\rbrace\rbrace$ & 1 \\ \hline
				\end{tabular}}
			\end{center}
		\end{table}
	\end{minipage}
	\begin{minipage}[c]{0.5\textwidth}
		\begin{table}[H]
			\begin{center}
					\resizebox{0.9\textwidth}{!}{%
				\begin{tabular}{|c|c|c|}
					\hline 
					\textbf{Diagrams}  & \textbf{Sequences}  & \textbf{s-factors}  \\ 
					\hline
					$d_4$ & $\lbrace \lbrace GB\rbrace,\lbrace CDA\rbrace\rbrace$ & 1 \\ \hline
					$d_5$ & $\lbrace \lbrace GB\rbrace,\lbrace ACD\rbrace\rbrace$ & 2 \\ \hline
					$d_6$ & $\lbrace \lbrace BG\rbrace,\lbrace CDA\rbrace\rbrace$ & 2 \\ \hline
				\end{tabular}}
			\end{center}
		\end{table}
	\end{minipage}
	\caption{Normal ordered diagrams of Cweb $ W^{(2,1)}_{4,\text{I}}(1,1,2,3) $}
	\label{tab:six-one-web4-7-av}
\end{table}

\begin{figure}[H]
	\vspace*{-0.9cm}
	\captionsetup[subfloat]{labelformat=empty}
	\centering
	\subfloat[][(a)]{\includegraphics[height=4cm,width=4cm]{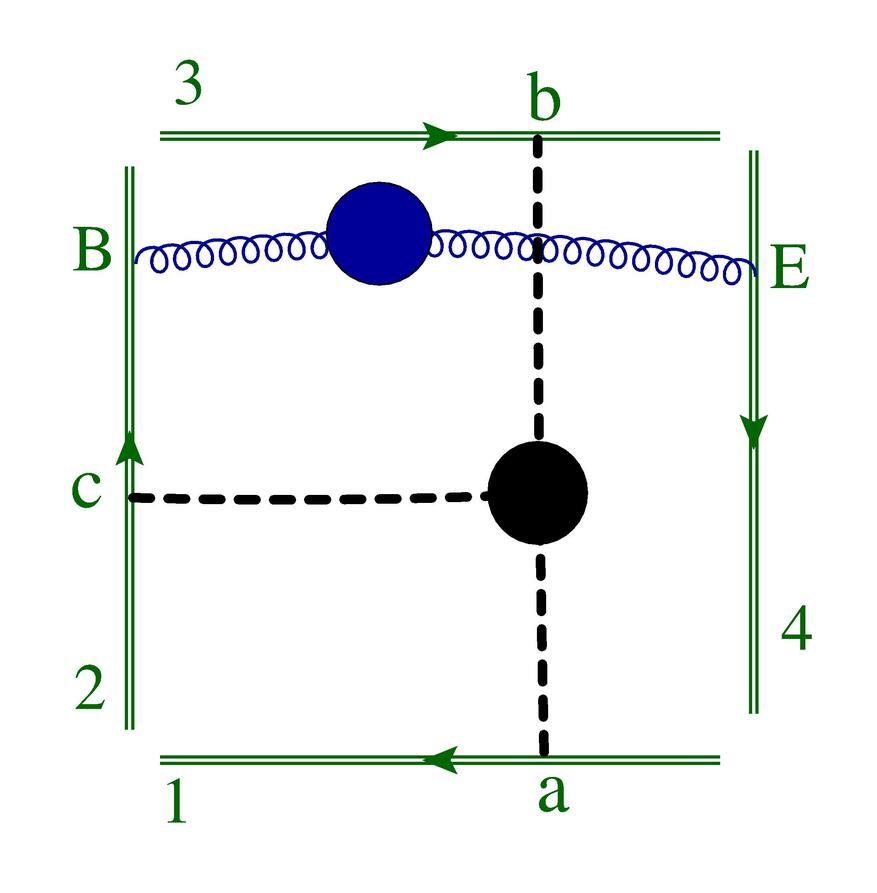} }
	\caption{\reducedWebs for Cweb $ W^{(2,1)}_{4,\text{I}}(1,1,2,3) $}
	\label{fig:six-one-web4-7-allAVAtar-WEBS}
\end{figure}

\begin{table}[H]
	\begin{center}
		\begin{tabular}{|c|c|c|c|c|c|}
			\hline
			Entanglement & Diagrams of  & \reducedWeb & Diagrams in  & $ s $-factors   & $ R $ \\ 
			& Cweb &  & \reducedWeb &    &  \\
			\hline &&&&&\\
			First Partial Entangled  & $ d_{1} $, $ d_{2} $ & \ref{fig:six-one-web4-7-allAVAtar-WEBS}\textcolor{blue}{a} & $ \{B,\,c\} $ & 1 & $ R(1_2) $ \\ 
			&  & & $ \{c,\,B\} $ & 1 & \\ \hline
		\end{tabular}	
	\end{center}
	\caption{\reducedWebs and their mixing matrices for Cweb $ W^{(2,1)}_{4,\text{I}}(1,1,2,3) $}
	\label{tab:six-one-web4-7-av-Ent}
\end{table}
\noindent The algorithm of Fused-Webs is applied to this Cweb, which results in table \ref{tab:six-one-web4-7-av-Ent}. It classifies the irreducible diagrams of the Cweb according to the entangled pieces. This table provides the \reducedWebs with the associated mixing matrices for the Cweb. The order of diagrams in the Cweb given in table \ref{tab:six-one-web4-7-av}, is chosen such that diagrams with same kind of entangled piece appear together. Therefore, mixing matrices of the \reducedWebs for this Cweb, present on the diagonal blocks of $ A $, is given as, 
\begin{align}
A\,=\,R(1_2)\,,
\end{align}
and the rank of $ A $ is
\begin{align}
r(A) &= r(R(1_2))\nonumber\\
&=  1\,.
\end{align}

\noindent The number of exponentiated colour factors is the rank of the mixing matrix; which is given as
\begin{align}
r(R) = r(A)  + 1  = 2\,.
\end{align}

\noindent \textbf{2.}\, $ \textbf{W}^{(2,1)}_{4,\text{II}}(1,2,2,2) $

\vspace{0.2cm}

\begin{figure}[H]
	\centering
	\subfloat[][]{\includegraphics[height=4cm,width=4cm]{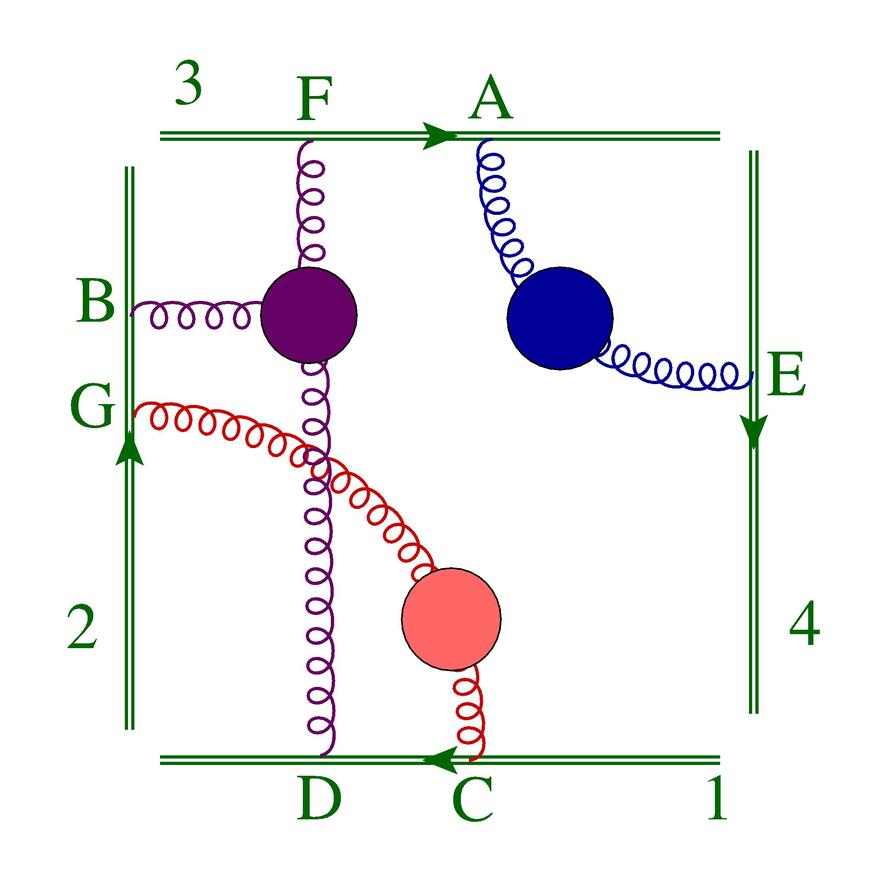} }
	\caption{Cweb $ W^{(2,1)}_{4,\text{II}}(1,2,2,2) $}
	\label{fig:six-one-web4-13-av}
\end{figure}

\noindent This Cweb, shown in fig.~(\ref{fig:six-one-web4-13-av}) has eight diagrams, out of which four are reducible, and remaining four are partially entangled. The Normal ordered diagrams and their $ s $-factors are shown in table \ref{tab:six-one-web4-13-av}.

\begin{table}[H]
	\begin{minipage}[c]{0.5\textwidth}
		\begin{table}[H]
			\begin{center}
					\resizebox{0.9\textwidth}{!}{%
				\begin{tabular}{|c|c|c|}
					\hline 
					\textbf{Diagrams}  & \textbf{Sequences}  & \textbf{s-factors}  \\ 
					\hline
					$d_1$ & $\lbrace \lbrace CD\rbrace,\lbrace BG \rbrace, \lbrace AF\rbrace \rbrace$ & 0 \\ 
					\hline
					$d_2$ & $\lbrace \lbrace CD\rbrace,\lbrace BG \rbrace, \lbrace FA\rbrace \rbrace$ & 0 \\ 
					\hline
					$d_3$ & $\lbrace \lbrace DC\rbrace,\lbrace GB \rbrace, \lbrace AF\rbrace \rbrace$ & 0 \\ 
					\hline
					$d_4$ & $\lbrace \lbrace DC\rbrace,\lbrace GB \rbrace, \lbrace FA\rbrace \rbrace$ & 0 \\ 
					\hline
				\end{tabular}}
			\end{center}
		\end{table}
	\end{minipage}
	\begin{minipage}[c]{0.5\textwidth}
		\begin{table}[H]
			\begin{center}
					\resizebox{0.9\textwidth}{!}{%
				\begin{tabular}{|c|c|c|}
					\hline 
					\textbf{Diagrams}  & \textbf{Sequences}  & \textbf{s-factors}  \\ 
					\hline
					$d_5$ & $\lbrace \lbrace CD\rbrace,\lbrace GB \rbrace, \lbrace FA\rbrace \rbrace$ & 1 \\ 
					\hline
					$d_6$ & $\lbrace \lbrace DC\rbrace,\lbrace BG \rbrace, \lbrace AF\rbrace \rbrace$ & 1 \\ 
					\hline
					$d_7$ & $\lbrace \lbrace CD\rbrace,\lbrace GB \rbrace, \lbrace AF\rbrace \rbrace$ & 2 \\ 
					\hline
					$d_8$ & $\lbrace \lbrace DC\rbrace,\lbrace BG \rbrace, \lbrace FA\rbrace \rbrace$ & 2 \\ 
					\hline
				\end{tabular}}
			\end{center}
		\end{table}
	\end{minipage}
	\caption{Normal ordered diagrams of Cweb $ W^{(2,1)}_{4,\text{II}}(1,2,2,2) $}
	\label{tab:six-one-web4-13-av}
\end{table}

\begin{figure}[H]
	\captionsetup[subfloat]{labelformat=empty}
	\centering
	\subfloat[][(a)]{\includegraphics[height=4cm,width=4cm]{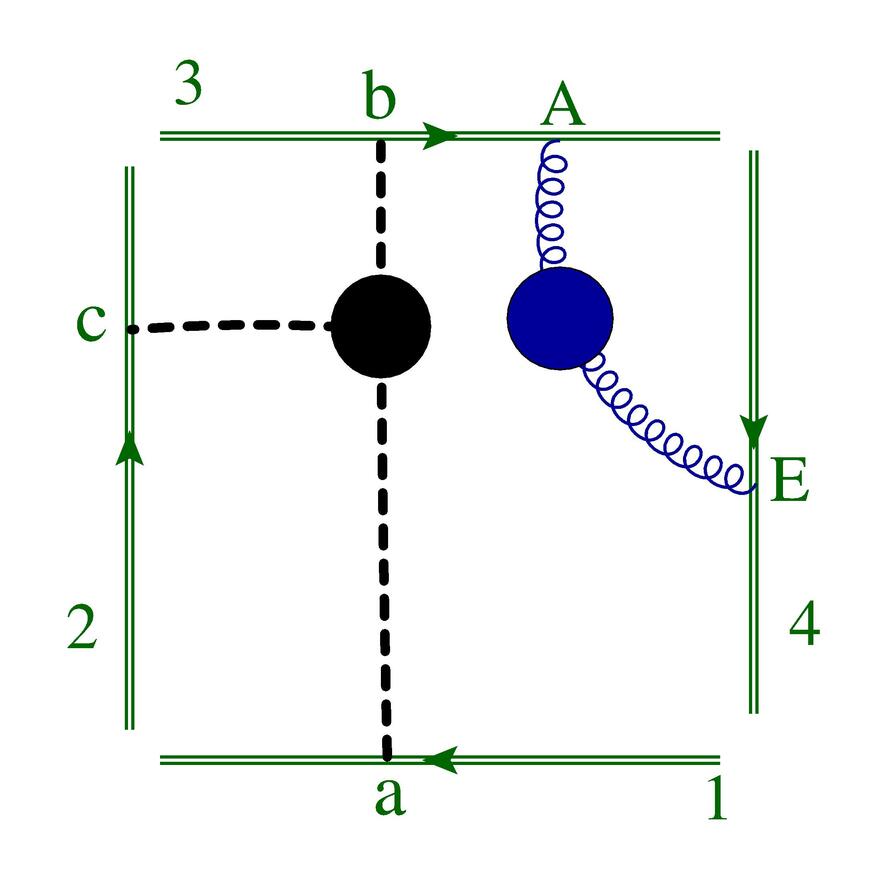} }
	\qquad 
	\subfloat[][(b)]{\includegraphics[height=4cm,width=4cm]{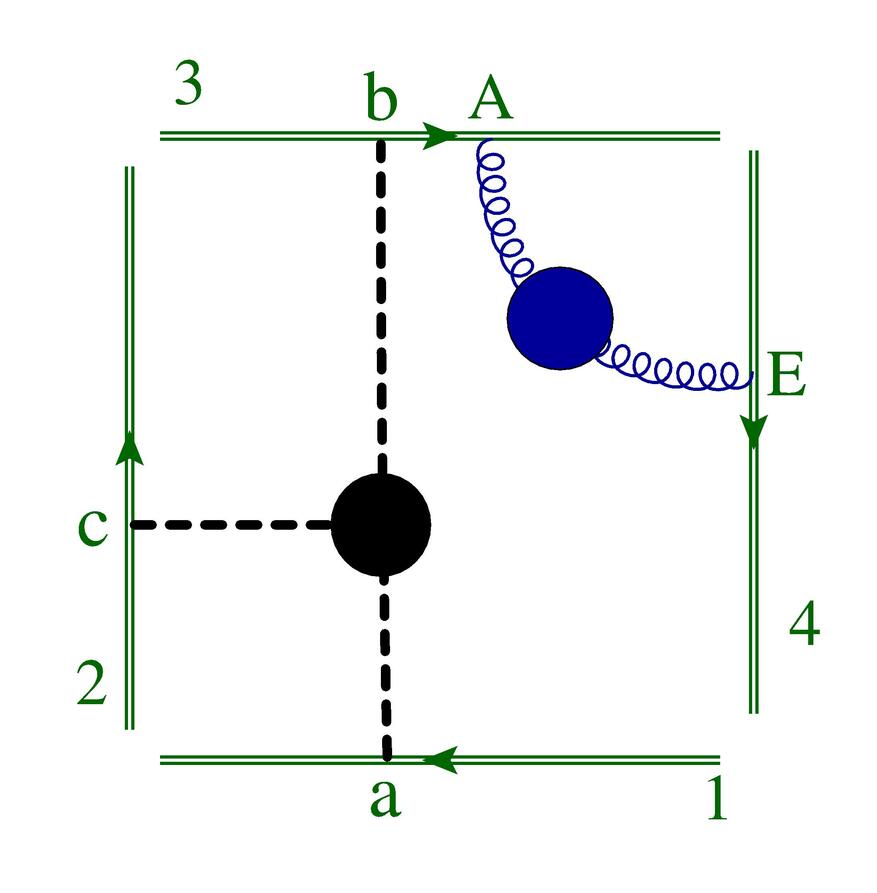} }
	\caption{\reducedWebs for Cweb $ W^{(2,1)}_{4,\text{II}}(1,2,2,2) $ }
	\label{fig:six-one-web4-13-allAVAtar-WEBS}
\end{figure}
\noindent The procedure of Fused-Webs is applied to this Cweb, which results in table \ref{tab:six-one-web4-13-av-Ent}. It classifies the irreducible diagrams of the Cweb according to the entangled pieces. This table provides the \reducedWebs with the associated mixing matrices for the Cweb.

\begin{table}[H]
	\begin{center}
		\begin{tabular}{|c|c|c|c|c|c|}
			\hline
			Entanglement & Diagrams of  & \reducedWeb & Diagrams in  & $ s $-factors   & $ R $ \\ 
			& Cweb &  & \reducedWeb &    &  \\
			\hline
			First Partial Entangled  & $ d_{1} $, $ d_{2} $ & \ref{fig:six-one-web4-13-allAVAtar-WEBS}\textcolor{blue}{a} & $ \{b,\,A\} $ & 1 & $ R(1_2) $ \\ 
			&  & & $ \{A,\,b\} $ & 1 & \\ \hline
			Second Partial Entangled  & $ d_{3} $, $ d_{4} $ & \ref{fig:six-one-web4-13-allAVAtar-WEBS}\textcolor{blue}{b} & $ \{b,\,A\} $ & 1 & $ R(1_2) $ \\ 
			&  & & $ \{A,\,b\} $ & 1 & \\ \hline
		\end{tabular}	
	\end{center}
	\caption{\reducedWebs and their mixing matrices for Cweb $ W^{(2,1)}_{4,\text{II}}(1,2,2,2) $}
	\label{tab:six-one-web4-13-av-Ent}
\end{table}

\noindent The order of diagrams in the Cweb given in table \ref{tab:six-one-web4-13-av}, is chosen such that diagrams with same kind of entangled piece appear together. Therefore, mixing matrices of the \reducedWebs for this Cweb, present on the diagonal blocks of $ A $, is given as,

\begin{align}
A\,=\,\left(\begin{array}{c|cc}
R(1_2) & & \cdots\\
\hline 
\cdots & & R(1_2)
\end{array}\right)\,,
\end{align}

\noindent  and the rank of $ A $ is,
\begin{align}
r(A) &= 2\,r(R(1_2))\nonumber\\
&=  2\,.
\end{align}
\noindent The rank of the mixing matrix, using eq.~(\ref{eq:rank--1122-class}) for this Cweb is given as,
\begin{align}
r(R) = r(A)  + 1  = 3\,.
\end{align}


\noindent \textbf{3.}\, $ \textbf{W}_4^{(2,1)}(1,1,2,3) $

\vspace{0.2cm}

\begin{figure}[H]
	\captionsetup[subfloat]{labelformat=empty}
	\centering
	\subfloat[][]{\includegraphics[height=4cm,width=4cm]{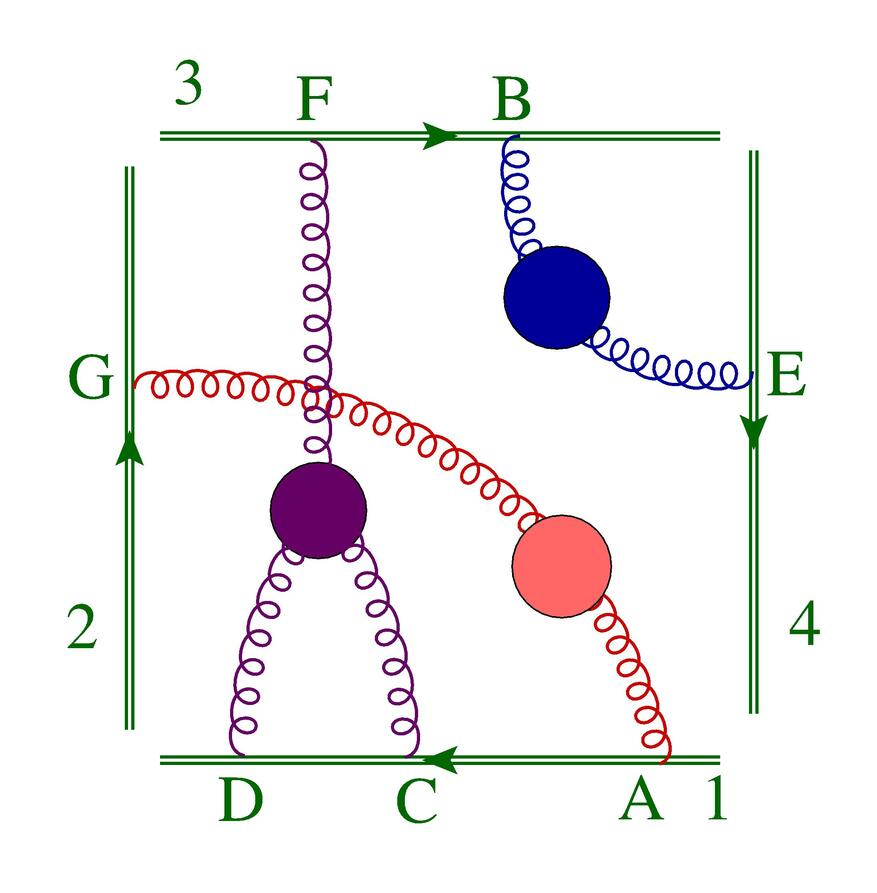} }
	\vspace{-0.5cm}
	\caption{Cweb  $ {W}_4^{(2,1)}(1,1,2,3) $}
	\label{fig:six-one-web4-8-av}
\end{figure}

\noindent This Cweb, shown in fig.~(\ref{fig:six-one-web4-8-av}) has six diagrams, out of which four are reducible, and remaining two are partially entangled. The Normal ordered diagrams and their $ s $-factors are shown in table \ref{tab:six-one-web4-8-av}.

\begin{table}[H]
	\begin{minipage}[c]{0.5\textwidth}
		\begin{table}[H]
			\begin{center}
					\resizebox{0.9\textwidth}{!}{%
				\begin{tabular}{|c|c|c|}
					\hline 
					\textbf{Diagrams}  & \textbf{Sequences}  & \textbf{s-factors}  \\ 
					\hline
					$d_{1}$  & $\{FB\},\{CAD\}$  & 0 \\ \hline
					$d_{2}$  & $\{BF\},\{CAD\}$& 0 \\ \hline
					$d_{3}$  & $\{FB\},\{ACD\} $  & 1 \\ \hline 
				\end{tabular}}
			\end{center}
		\end{table}
	\end{minipage}
	\begin{minipage}[c]{0.5\textwidth}
		\begin{table}[H]
			\begin{center}
					\resizebox{0.9\textwidth}{!}{%
				\begin{tabular}{|c|c|c|}
					\hline 
					\textbf{Diagrams}  & \textbf{Sequences}  & \textbf{s-factors}  \\ 
					\hline
					$d_{4}$  & $\{BF\},\{CDA\}$  & 1 \\ \hline
					$d_{5}$  & $\{BF\},\{ACD\} $  & 2 \\ \hline
					$d_{6}$  & $\{FB\},\{CDA\}$  & 2 \\ \hline
				\end{tabular}}
			\end{center}
		\end{table}
	\end{minipage}
	\caption{Normal ordered diagrams of Cweb  $ {W}_4^{(2,1)}(1,1,2,3) $}
	\label{tab:six-one-web4-8-av}
\end{table}

\begin{figure}[H]
	\captionsetup[subfloat]{labelformat=empty}
	\centering
	\subfloat[][(a)]{\includegraphics[height=4cm,width=4cm]{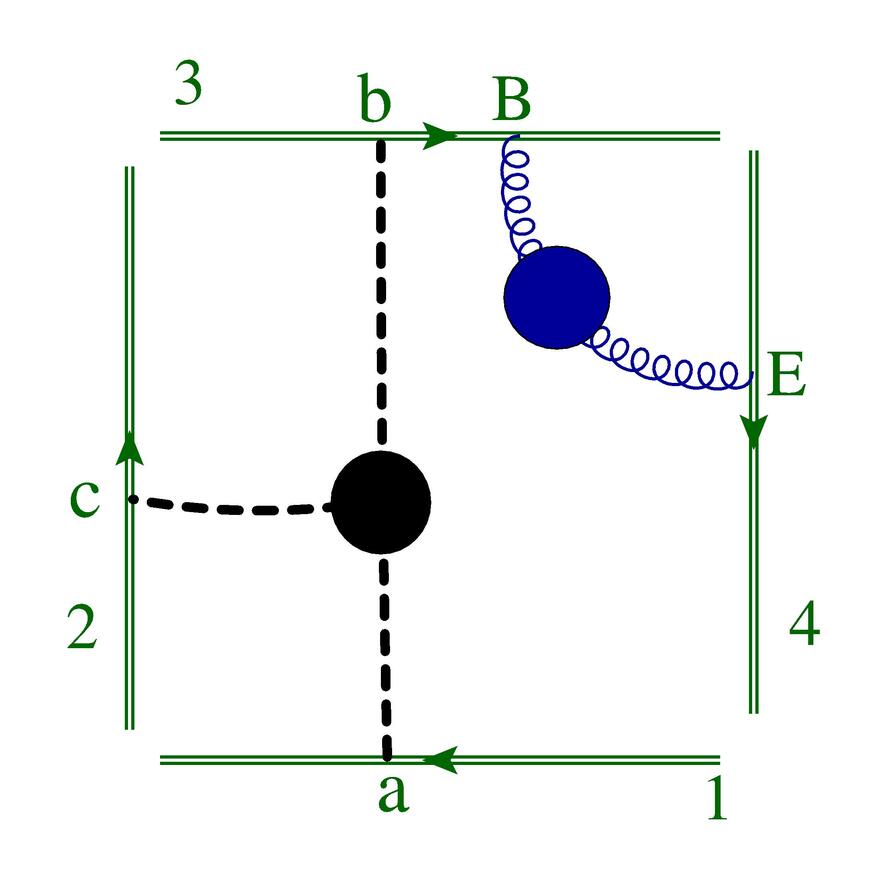} }
	\caption{\reducedWebs for Cweb  $ {W}_4^{(2,1)}(1,1,2,3) $}
	\label{fig:six-one-web4-8-allAVAtar-WEBS}
\end{figure}
\noindent The procedure developed in section \ref{sec:avatar-webs}, is applied to this Cweb, which results in table \ref{tab:six-one-web4-8-av-Ent}. It classifies the irreducible diagrams of the Cweb according to the entangled pieces. This table provides the \reducedWebs with the associated mixing matrices for the Cweb.

\begin{table}[H]
	\begin{center}
		\begin{tabular}{|c|c|c|c|c|c|}
			\hline
			Entanglement & Diagrams of  & \reducedWeb & Diagrams in  & $ s $-factors   & $ R $ \\ 
			& Cweb &  & \reducedWeb &    &  \\
			\hline &&&&&\\
			First Partial Entangled  & $ d_{1} $, $ d_{2} $ & \ref{fig:six-one-web4-8-allAVAtar-WEBS}\textcolor{blue}{a} & $ \{b,\,B\} $ & 1 & $ R(1_2) $ \\ 
			&  & & $ \{B,\,b\} $ & 1 & \\ \hline
		\end{tabular}	
	\end{center}
	\caption{\reducedWebs and their mixing matrices for Cweb  $ {W}_4^{(2,1)}(1,1,2,3) $}
	\label{tab:six-one-web4-8-av-Ent}
\end{table}

\noindent The order of diagrams in the Cweb given in table \ref{tab:six-one-web4-8-av}, is chosen such that diagrams with same kind of entangled piece appear together. Therefore, mixing matrices of the \reducedWebs for this Cweb, present on the diagonal blocks of $ A $, is given as, 
\begin{align}
A\,=\,R(1_2)\,,
\end{align}
and the rank of $ A $ is
\begin{align}
r(A) &= \,r(R(1_2))\nonumber\\
&=  1\,.
\end{align}
The number of exponentiated colour factors for this Cweb is the rank of $ R $, thus using eq.~(\ref{eq:rank--1122-class}) we get,
\begin{align}
r(R) = r(A)  + 1  = 2\,.
\end{align}

\noindent  The number of exponentiated colour factors predicted in sections \ref{sec:1-6appendix} and \ref{sec:1-22-2-appendix}, are in agreement with the results of Cwebs present at four loop in \cite{Agarwal:2021him,Agarwal:2020nyc}.

\section{Mixing matrices for basis Cwebs}
\label{sec:basis}
In this appendix, we present the mixing matrices used as a basis to write down $ D $, and the diagonal blocks of $ A $. 
\begin{itemize}
	\item [\bf{1}.] Mixing matrix for $ S=\{1_2\} $

The unique mixing matrix for this type of Cweb first appears at two loops, and has the form, 
	\begin{align}
	R\,(1_2)=\frac{1}{2}\left(\begin{array}{cc}
	1 & -1 \\
	-1 & 1
	\end{array}\right)\,.
	\end{align}
This matrix was constructed directly using the known properties in \cite{Agarwal:2021him}. 	
	\item[\bf{2}.] Mixing matrix for $ S=\{1_6\} $
	
	The mixing matrix for this type of Cweb first appears at three loops, and has the form, 
\begin{align}
R(1_6) =\frac{1}{6} \left(
\begin{array}{cccccc}
      2 & -1 & -1 & -1 & -1 & 2 \\
      -1 & 2 & -1 & 2 & -1 & -1 \\
      -1 & -1 & 2 & -1 & 2 & -1 \\
      -1 & 2 & -1 & 2 & -1 & -1 \\
      -1 & -1 & 2 & -1 & 2 & -1 \\
      2 & -1 & -1 & -1 & -1 & 2 \\
\end{array}
\right)\, .
\end{align}

The explicit form of this matrix was computed directly in \cite{Dukes:2013gea}, using the idea of posets. 
	\item[\bf{3}.] Mixing matrix for $ S=\{1_{24}\} $

	The mixing matrix for this type of Cweb first appears at four loops, and has the form, 
	\resizebox{.9\linewidth}{!}{
	\begin{minipage}{\linewidth}
	\begin{align}
		R(1_{24})=\frac{1}{12} \left(
	\begin{array}{cccccccccccccccccccccccc}
	3 & -1 & -1 & -1 & -1 & 1 & -1 & 1 & -1 & -1 & -1 & 1 & -1 & 1 & 1 & 1 & -1 & 1 & -1 & 1 & 1 & 1 & 1 & -3 \\
	-1 & 3 & -1 & 1 & -1 & -1 & 1 & -1 & -1 & 1 & -1 & -1 & -1 & 1 & 1 & 1 & 1 & -3 & -1 & 1 & 1 & 1 & -1 & 1 \\
	-1 & -1 & 3 & -1 & 1 & -1 & -1 & 1 & 1 & 1 & -1 & 1 & -1 & 1 & -1 & -1 & -1 & 1 & 1 & -1 & 1 & -3 & 1 & 1 \\
	-1 & 1 & -1 & 3 & -1 & -1 & -1 & 1 & 1 & 1 & 1 & -3 & 1 & -1 & -1 & 1 & -1 & -1 & 1 & -1 & -1 & 1 & 1 & 1 \\
	-1 & -1 & 1 & -1 & 3 & -1 & 1 & -1 & -1 & 1 & 1 & 1 & 1 & -1 & 1 & -3 & 1 & 1 & -1 & 1 & -1 & -1 & -1 & 1 \\
	1 & -1 & -1 & -1 & -1 & 3 & 1 & -1 & 1 & -3 & 1 & 1 & 1 & -1 & -1 & 1 & 1 & 1 & 1 & -1 & -1 & 1 & -1 & -1 \\
	-1 & 1 & -1 & -1 & -1 & 1 & 3 & -1 & -1 & -1 & -1 & 1 & 1 & 1 & -1 & 1 & 1 & -1 & 1 & 1 & -1 & 1 & -3 & 1 \\
	1 & -1 & -1 & 1 & -1 & -1 & -1 & 3 & -1 & 1 & -1 & -1 & 1 & 1 & -1 & 1 & -3 & 1 & 1 & 1 & -1 & 1 & 1 & -1 \\
	-1 & 1 & 1 & 1 & -1 & 1 & -1 & -1 & 3 & -1 & 1 & -1 & -1 & -1 & -1 & 1 & 1 & -1 & 1 & -3 & 1 & -1 & 1 & 1 \\
	-1 & 1 & 1 & 1 & 1 & -3 & -1 & 1 & -1 & 3 & -1 & -1 & -1 & 1 & 1 & -1 & -1 & -1 & -1 & 1 & 1 & -1 & 1 & 1 \\
	1 & -1 & -1 & 1 & 1 & 1 & -1 & -1 & 1 & -1 & 3 & -1 & 1 & -3 & 1 & -1 & 1 & 1 & -1 & -1 & -1 & 1 & 1 & -1 \\
	1 & -1 & 1 & -3 & 1 & 1 & 1 & -1 & -1 & -1 & -1 & 3 & -1 & 1 & 1 & -1 & 1 & 1 & -1 & 1 & 1 & -1 & -1 & -1 \\
	-1 & -1 & -1 & 1 & 1 & -1 & 1 & 1 & -1 & 1 & 1 & -1 & 3 & -1 & -1 & -1 & -1 & 1 & 1 & 1 & -3 & 1 & -1 & 1 \\
	-1 & 1 & 1 & -1 & -1 & -1 & 1 & 1 & -1 & 1 & -3 & 1 & -1 & 3 & -1 & 1 & -1 & -1 & 1 & 1 & 1 & -1 & -1 & 1 \\
	1 & 1 & -1 & 1 & 1 & -1 & -1 & -1 & -1 & 1 & 1 & -1 & -1 & -1 & 3 & -1 & 1 & -1 & -3 & 1 & 1 & 1 & 1 & -1 \\
	1 & 1 & -1 & 1 & -3 & 1 & -1 & 1 & 1 & -1 & -1 & -1 & -1 & 1 & -1 & 3 & -1 & -1 & 1 & -1 & 1 & 1 & 1 & -1 \\
	-1 & 1 & 1 & -1 & 1 & 1 & 1 & -3 & 1 & -1 & 1 & 1 & -1 & -1 & 1 & -1 & 3 & -1 & -1 & -1 & 1 & -1 & -1 & 1 \\
	1 & -3 & 1 & -1 & 1 & 1 & -1 & 1 & 1 & -1 & 1 & 1 & 1 & -1 & -1 & -1 & -1 & 3 & 1 & -1 & -1 & -1 & 1 & -1 \\
	-1 & -1 & 1 & -1 & -1 & 1 & 1 & 1 & 1 & -1 & -1 & 1 & 1 & 1 & -3 & 1 & -1 & 1 & 3 & -1 & -1 & -1 & -1 & 1 \\
	1 & -1 & -1 & -1 & 1 & -1 & 1 & 1 & -3 & 1 & -1 & 1 & 1 & 1 & 1 & -1 & -1 & 1 & -1 & 3 & -1 & 1 & -1 & -1 \\
	1 & 1 & 1 & -1 & -1 & 1 & -1 & -1 & 1 & -1 & -1 & 1 & -3 & 1 & 1 & 1 & 1 & -1 & -1 & -1 & 3 & -1 & 1 & -1 \\
	1 & 1 & -3 & 1 & -1 & 1 & 1 & -1 & -1 & -1 & 1 & -1 & 1 & -1 & 1 & 1 & 1 & -1 & -1 & 1 & -1 & 3 & -1 & -1 \\
	1 & -1 & 1 & 1 & 1 & -1 & -3 & 1 & 1 & 1 & 1 & -1 & -1 & -1 & 1 & -1 & -1 & 1 & -1 & -1 & 1 & -1 & 3 & -1 \\
	-3 & 1 & 1 & 1 & 1 & -1 & 1 & -1 & 1 & 1 & 1 & -1 & 1 & -1 & -1 & -1 & 1 & -1 & 1 & -1 & -1 & -1 & -1 & 3 \\
	\end{array}
	\right)\,.
	\end{align}
\end{minipage}}
	\item[\textbf{4}.] Mixing matrix for $ S=\{1_2,2_2\} $
	
	The mixing matrix for this type of Cweb first appears at three loops, and has the form, 
	\begin{align}
	R(1_2,2_2)=\frac{1}{6} \left(
	\begin{array}{cccc}
	 2 & 2 & -2 & -2 \\
	  2 & 2 & -2 & -2 \\
	  -1 & -1 & 1 & 1 \\
	  -1 & -1 & 1 & 1 \\
	\end{array}
	\right)\,.
	\end{align}
	
	\item[\textbf{5}.] Mixing matrix for $ S=\{1_4,2_8\} $
	
	The mixing matrix for this type of Cweb first appears at four loops, and has the form,
	 
	 \begin{align}
	 	R(1_4,2_8)=\frac{1}{12} \left(
	 	\begin{array}{cccccccccccc}
	 	2 & 1 & -1 & -1 & -1 & 1 & -1 & 2 & -2 & 1 & 1 & -2 \\
	 	2 & 3 & -3 & 1 & -3 & -1 & 1 & 2 & -2 & -1 & 3 & -2 \\
	 	0 & -1 & 1 & -1 & 1 & 1 & -1 & 0 & 0 & 1 & -1 & 0 \\
	 	-2 & 1 & -1 & 3 & -1 & -3 & 3 & -2 & 2 & -3 & 1 & 2 \\
	 	-2 & -3 & 3 & -1 & 3 & 1 & -1 & -2 & 2 & 1 & -3 & 2 \\
	 	2 & -1 & 1 & -3 & 1 & 3 & -3 & 2 & -2 & 3 & -1 & -2 \\
	 	0 & 1 & -1 & 1 & -1 & -1 & 1 & 0 & 0 & -1 & 1 & 0 \\
	 	2 & 1 & -1 & -1 & -1 & 1 & -1 & 2 & -2 & 1 & 1 & -2 \\
	 	-2 & -1 & 1 & 1 & 1 & -1 & 1 & -2 & 2 & -1 & -1 & 2 \\
	 	0 & -1 & 1 & -1 & 1 & 1 & -1 & 0 & 0 & 1 & -1 & 0 \\
	 	0 & 1 & -1 & 1 & -1 & -1 & 1 & 0 & 0 & -1 & 1 & 0 \\
	 	-2 & -1 & 1 & 1 & 1 & -1 & 1 & -2 & 2 & -1 & -1 & 2 \\
	 	\end{array}
	 	\right)\,.
	 \end{align}
	 
	 	\item[\textbf{6}.] Mixing matrix for $ S=\{1_2,2_4,4_2\} $
	
	 The mixing matrix for this type of Cweb first appears at four loops, and has the form,
	 \begin{align}
	 R(1_2,2_4,4_2)=\frac{1}{12}\left(
	 \begin{array}{cccccccc}
	 3 & -3 & -3 & 3 & -3 & 3 & 3 & -3 \\
	 -3 & 3 & 3 & -3 & 3 & -3 & -3 & 3 \\
	 -1 & 1 & 1 & -1 & 1 & -1 & -1 & 1 \\
	 1 & -1 & -1 & 1 & -1 & 1 & 1 & -1 \\
	 -1 & 1 & 1 & -1 & 1 & -1 & -1 & 1 \\
	 1 & -1 & -1 & 1 & -1 & 1 & 1 & -1 \\
	 1 & -1 & -1 & 1 & -1 & 1 & 1 & -1 \\
	 -1 & 1 & 1 & -1 & 1 & -1 & -1 & 1 \\
	 \end{array}
	 \right)
	 \end{align}
\end{itemize}


\bibliographystyle{JHEP}
\bibliography{boom}
\end{document}